\documentclass[11pt,a4paper]{article}
\pdfoutput=1
\usepackage{jheppub}

\usepackage{multirow, graphicx,amssymb,url,mathrsfs,amsmath}
\usepackage{wrapfig,boxedminipage,setspace,subfigure,epsfig}
\usepackage{amsxtra,amstext,latexsym,dsfont,amsfonts}
\usepackage{color,eucal}
\usepackage[dvipsnames]{xcolor}
\usepackage{float}
\usepackage{slashed,comment}
\usepackage{kotex}
\usepackage{contour}

\usepackage{tikz}
\usetikzlibrary{calc,patterns,angles,quotes}
\usetikzlibrary{decorations.pathreplacing,decorations.markings,snakes}

\usepackage{tabularx, array}
\newcolumntype{L}[1]{>{\raggedright\arraybackslash}p{#1}}
\newcolumntype{C}[1]{>{\centering\arraybackslash}p{#1}}
\newcolumntype{R}[1]{>{\raggedleft\arraybackslash}p{#1}}

\newcommand{\R}{\mathbb{R}}

\renewcommand{\d}{\delta}      
\renewcommand{\l}{\lambda}

\newcommand{\dd}{\mathrm{d}}

\newcommand{\grad}{\nabla}

\newcommand{\abs}[1]{\left| #1 \right|}

\newcommand{\tQ}{{\tilde{Q}}}

\usepackage{xparse}
\ExplSyntaxOn
\NewDocumentCommand{\HS}{m}
 {
  \seq_set_split:Nnn \l_tmpa_seq { ~ } { #1 }
  \seq_map_inline:Nn \l_tmpa_seq { \contour{green}{##1} ~ } \unskip
 }
\ExplSyntaxOff

\title{Kasner interiors from analytic hairy black holes}

\author[a,b]{Daniel Are\'an,}
\author[a,b]{Hyun-Sik Jeong,}
\author[a]{Juan F. Pedraza}
\author[a,b]{and Le-Chen Qu}

\emailAdd{daniel.arean@uam.es}
\emailAdd{hyunsik.jeong@csic.es}
\emailAdd{j.pedraza@csic.es}
\emailAdd{lechen.qu@ift.csic.es}

\preprint{\texttt{IFT-UAM/CSIC-24-109}}

\affiliation[a]{Instituto de F\'isica Te\'orica UAM/CSIC, Calle Nicol\'as Cabrera 13-15, 28049 Madrid, Spain}
\affiliation[b]{Departamento de F\'isica Te\'orica, Universidad Aut{\'o}noma de Madrid, 28049 Madrid, Spain}

\abstract{
We conduct an exhaustive study of the interior geometry of a family of asymptotically AdS$_{d+1}$ hairy black holes in an \textit{analytically} controllable setup. The black holes are exact solutions to an Einstein-Maxwell-Dilaton theory and include the well-known Gubser-Rocha model. After reviewing the setup, we scrutinize the geometry beyond the horizon, finding that these backgrounds can exhibit timelike or Kasner singularities. We generalize the no inner-horizon theorem for hairy black holes to accommodate these findings. We then consider observables sensitive to the geometry behind the horizon, such as Complexity = Anything and the thermal $a$-function. 
In the Kasner case, we propose a new variant of complexity that characterizes the late-time rate by the Kasner exponents, extending previous work by Jørstad, Myers and Ruan. Additionally, we elucidate the power-law behavior of the thermal $a$-function near the singularity, directly relating it to the Kasner exponents.
Finally, we introduce axion-like fields in the Gubser-Rocha model to study the impact of translational symmetry breaking on the black hole interior. We show that Kasner singularities occur for both explicit and spontaneous symmetry breaking, with the Kasner exponents depending on the strength of broken translations only in the latter case.
}

\begin{document}
\maketitle

%
\section{Introduction}\label{}

Understanding the internal structure of black holes presents a captivating and fundamental challenge, both theoretically and experimentally. The intricate non-linearity of Einstein's equations often renders the analytical derivation of black hole solutions quite arduous. Consequently, grasping the structure and dynamics of the black hole interior ---especially near the singularity where the spacetime curvature becomes infinitely intense--- remains a significant puzzle.

Despite these challenges, analytical solutions for black holes are known. A prominent example is the neutral Schwarzschild black hole, which is characterized by an event horizon enclosing a spacelike singularity~\cite{Schwarzschild:1916uq}. Other significant solutions include the Reissner-Nordström black hole~\cite{Reissner:1916cle,NKARCITE}, which includes electric charge, and the Kerr black hole~\cite{Kerr:1963ud}, which includes angular momentum. Both of these solutions feature an additional inner Cauchy horizon, where predictability in general relativity breaks down, and they possess a timelike singularity that poses challenges to the strong cosmic censorship conjecture~\cite{Penrose:1969pc}.

On the other hand, holographic duality offers significant promise for unraveling the complexities of black hole interiors. Key approaches include analyzing analytically continued correlation functions~\cite{Fidkowski:2003nf,Festuccia:2005pi,Horowitz:2023ury,Caceres:2023zft}, entanglement entropy~\cite{Hartman:2013qma,Penington:2019npb,Almheiri:2019hni}, and computational complexity~\cite{Stanford:2014jda,Brown:2015bva,Couch:2016exn,Belin:2021bga}. Indeed, many of these efforts have focused on understanding aspects of black hole singularities in terms of CFT data ---see also~\cite{Grinberg:2020fdj,Leutheusser:2021frk,Rodriguez-Gomez:2021pfh,deBoer:2022zps,Jorstad:2023kmq,Ceplak:2024bja,Singhi:2024sdr,Anegawa:2024kdj}.

Among the extensively studied black hole interiors in holography is that of the eternal Schwarzschild-AdS black hole, which characterizes the thermofield double state of the dual CFT~\cite{Maldacena:2001kr}. While the exterior geometry of these black holes is dynamically stable, their interior is known to be unstable: matter fields experience infinite growth as they approach the spacelike singularity, leading to significant backreaction~\cite{Belinsky:1970ew,Belinski:1973zz,Belinsky:1981vdw}. This inherent instability of the Schwarzschild singularity necessitates careful consideration in any holographic investigation of black hole interiors and black hole singularities.

Motivated by this conceptual challenge, Frenkel et al. examined a class of black holes formed by perturbing the dual theory with a relevant scalar operator, revealing a deformation of the Schwarzschild singularity into a more general Kasner singularity \cite{Frenkel:2020ysx}. These singularities align precisely with those discovered by Belinsky-Khalatnikov-Lifshitz (BKL) in the early '70s \cite{Belinsky:1970ew,Belinski:1973zz,Belinsky:1981vdw}. 
Following the pioneering work of \cite{Frenkel:2020ysx}, numerous studies have \textit{numerically} identified Kasner singularities in a variety of holographic models \cite{Hartnoll:2020rwq,Bhattacharya:2020qil,Hartnoll:2020fhc,Sword:2021pfm,Sword:2022oyg,Wang:2020nkd,Caceres:2021fuw,Mansoori:2021wxf,Liu:2021hap,Bhattacharya:2021nqj,Das:2021vjf,Caceres:2022smh,An:2022lvo,Auzzi:2022bfd,Hartnoll:2022rdv,Hartnoll:2022snh,Caceres:2022hei,Liu:2022rsy,Mirjalali:2022wrg,Gao:2023zbd,Caceres:2023zhl,Blacker:2023ezy,Caceres:2023zft,DeClerck:2023fax,Cai:2023igv,Gao:2023rqc,Carballo:2024hem,Caceres:2024edr}. These studies have investigated holographic RG flows induced by scalar or vector fields, transitioning between a UV CFT and a Kasner universe in the trans-IR regime.\footnote{In the holographic framework, the energy scale within the dual field theory corresponds to the bulk radial direction. By analytically continuing the radial direction, the holographic RG flow can extend to the interior of a black hole, where the radial direction becomes timelike, signifying trans-IR behavior.} Within this context, it has been shown that introducing a deformation induced by a relevant operator precludes the existence of a Cauchy horizon.

In this manuscript, we conduct an \textit{analytical} investigation of the black hole interior using a family of asymptotically AdS$_{d+1}$ hairy black holes~\cite{Gouteraux:2014hca,Jeong:2018tua}, which are solutions to a specific Einstein-Maxwell-Dilaton (EMD) theory. Our analysis offers a novel analytic approach to identifying Kasner singularities, complementing previous numerical findings on black hole interiors. Additionally, it extends prior holographic studies in applied holography ~\cite{Ammon:2015wua,Zaanen:2015oix,Natsuume:2014sfa,CasalderreySolana:2011us,Hartnoll:2009sz,Hartnoll:2016apf} by characterizing their corresponding geometries behind the horizon.

The solutions we consider form a family of charged, asymptotically AdS black holes in the presence of a neutral scalar field, characterized by two parameters: the dimension $d$ and a real constant $\delta$ that appears in the potentials of the EMD action. This setup is notable for its generic analytic background solutions. Furthermore, the solution for $\delta = \delta_c \equiv \sqrt{\frac{2}{d(d-1)}}$, known as the Gubser-Rocha model~\cite{Gubser:2009qt}, can be derived from a top-down construction. For $d=4$, this model emerges from the ten-dimensional type IIB string theory as the near-horizon limit of D3-branes~\cite{Gubser:2009qt,Cvetic:1999xp}. Similarly, for $d=3$, it results from a consistent truncation of eleven-dimensional supergravity compactified on AdS$_4 \times S^7$~\cite{Gubser:2009qt}.

A detailed analytical analysis of the black hole interior geometry reveals that these models can exhibit either timelike or Kasner singularities. We provide a concrete analytical computation showing that a deformation induced by a scalar operator may not necessarily destroy the Cauchy horizon of an asymptotically AdS black hole, thereby preserving the existence of a timelike singularity. This generalizes previous proofs regarding the (non-)existence of inner horizons~\cite{Hartnoll:2020rwq,Cai:2020wrp,An:2021plu}. We then investigate various holographic observables to probe the black hole interior, focusing on two intriguing proposals for diagnosing singularities from the CFT perspective: Complexity = Anything~\cite{Jorstad:2023kmq} and the thermal $a$-function~\cite{Caceres:2022smh,Caceres:2022hei}. In both cases we are able to characterize their behavior in terms of near-singularity data. Finally, we explore the effect of translational symmetry breaking on the black hole interior. By incorporating the axion field into our models~\cite{Gouteraux:2014hca,Jeong:2018tua}, we conduct an analytical study of singularities in the context of explicit symmetry breaking and a numerical analysis for cases of spontaneous symmetry breaking.

The structure of this paper is as follows. In Section~\ref{SEC2}, we introduce the holographic Einstein-Maxwell-Dilaton model that forms the basis of this work; we thoroughly review its properties and determine the physically allowed range of parameters. In Section~\ref{sec:interior}, we analyze the structure of the black hole interior. Section~\ref{SEC3} explores two holographic probes of the black hole interior: Complexity = Anything and the thermal $a$-function. Finally, in Section~\ref{sec:transbreaking}, we examine the black hole interior of models dual to theories with explicit or spontaneous translational symmetry breaking. We present our conclusions in Section~\ref{SEC4}.

%
\section{Einstein-Maxwell-Dilaton models}\label{SEC2}

%
\subsection{Models: asymptotically AdS$_{d+1}$ family}
We study the $(d+1)$-dimensional holographic Einstein-Maxwell-Dilaton model~\cite{Gouteraux:2014hca,Jeong:2018tua} described by the action
\begin{equation}\label{GGR}
S=  \int\dd^{d+1}x \,  \sqrt{-g}   \left[   R -  \frac{1}{2}\partial\phi^2 -  \frac{1}{4} Z(\phi) F^2 + V(\phi)  \right] \,,
\end{equation}
that features two matter fields: a neutral scalar field $\phi$, referred to as the dilaton field, and a $U(1)$ gauge field with field strength $F=\dd A$. The terms denoted by $Z(\phi)$ and $V(\phi)$ are the coupling and potential function given as
\begin{align}\label{MODELACTION}
\begin{split}
Z(\phi) = e^{-(d-2)\d\phi}  \,, \qquad 
V(\phi) = V_{1}e^{\frac{\left((d-2)(d-1)\d^{2}-2\right)\phi}{2(d-1)\d}} \,+\, V_{2}e^{\frac{2\phi}{\d-d\d}} \,+\, V_{3}e^{(d-2)\d\phi} \,, \\
\end{split}
\end{align}
with
\begin{align}
\begin{split}
&V_{1}=\frac{8(d-2)(d-1)^3\d^{2}}{\left(2+(d-2)(d-1)\d^2\right)^2}\ , \qquad \quad V_{2}=\frac{(d-2)^{2}(d-1)^2(d(d-1)\d^2-2)\d^{2}}{\left(2+(d-2)(d-1)\d^2\right)^2}\,, \\
&V_{3}=-\frac{2(d-2)^2(d-1)^2\d^2-4d(d-1)}{(2+(d-2)(d-1)\d^2)^2} \,,
\end{split}
\end{align}
where the number of boundary spacetime dimensions $d$ is restricted to the range $d>2$.
The potential $V(\phi)$ has the following asymptotic behavior in the $\phi\rightarrow0$ limit (which is the near AdS boundary limit)
\begin{align}
\begin{split}
V(\phi) = d(d-1) + (d-2)\phi^2 + \mathcal{O}(\phi^3) \,,\label{nbexpansion}
\end{split}
\end{align}
where the first term corresponds to the cosmological constant and the coefficient in the subleading term is related to the mass squared of a scalar field. In particular, (\ref{nbexpansion}) implies $\Lambda=-d(d-1)/2L^2$ (with unit AdS radius, $L=1$) and $m^2=2(2-d)$. Using the formula for the dimension of the dual operator, we find
\begin{equation}
\Delta =  \frac{d}{2} + \sqrt{\frac{d^2}{4} + m^2L^2}  = d-2\,.
\end{equation}
This implies $\phi$ is dual to a relevant operator and, hence, solutions with a scalar profile are expected to induce a non-trivial Renormalization Group (RG) flow.

The equations of motion resulting from the action \eqref{GGR} take the form
\begin{align}\label{BULKEOM}
\begin{split}
R_{\mu\nu} &= \frac{1}{2}\partial_{\mu}\phi\partial_{\nu}\phi+\frac{Z(\phi)}{2}F_{\mu}{^\rho}F_{\nu\rho}-\frac{Z(\phi)F^2}{4(d-1)}g_{\mu\nu}-\frac{V(\phi)}{d-1}g_{\mu\nu} \,, \\
0&=\grad_{\mu}(Z(\phi)F^{\mu\nu}) \,, \\
0&=\square\phi+V'(\phi)-\frac{1}{4}Z'(\phi)F^2 \,.
\end{split}
\end{align}
We are interested in solutions that realize homogeneous charged black hole geometries (that is, solutions that depend only on the radial coordinate). These can be obtained via the ansatz
\begin{equation}\label{OURMETRIC}
\dd s^2=-D(r)\,\dd t^2+B(r)\,\dd r^2+C(r)\,\dd \vec x_{i}^{2} \,, \qquad A=A_t(r) \dd t
 \,, \qquad \phi=\phi(r)\,.
\end{equation}
Indeed, for the potentials~\eqref{MODELACTION} one finds the following analytic
solution of the equations~\eqref{BULKEOM}. The metric functions read
\begin{align}\label{sol11}
\begin{split}
D(r) &= f(r)h(r)^{\frac{-4}{2+(d-2)(d-1)\d^2}}  \,, \qquad  
B(r) = \frac{h(r)^{\frac{4}{(d-2)(2+(d-2)(d-1)\d^2)}}}{f(r)} \,, \\ 
C(r) &= r^2 h(r)^{\frac{4}{(d-2)(2+(d-2)(d-1)\d^2)}} \,,
\end{split}
\end{align}
with
\begin{align}\label{sol11int}
\begin{split}
f(r) &=r^2 \left(h(r)^{\frac{4 (d-1)}{(d-2) \left(2+(d-1)(d-2) \delta ^2\right)}}-\frac{r_h^d}{r^{d}} \,h(r_h)^{\frac{4 (d-1)}{(d-2) \left(2+(d-1)(d-2) \delta ^2\right)}}\right) \,,   \\
 h(r) &=1+\frac{Q}{r^{d-2}} \,,
\end{split}
\end{align}
where $r_{h}$ is the event horizon satisfying $f(r_h)=0$.
While the solutions for the matter fields are given by
\begin{align}\label{sol12}
\begin{split}
A_t(r) &= 2\sqrt{(d-1)Q}\,\frac{\sqrt{(d-2)r_h^{2+d} h(r_h)^{\frac{2 \left(2-(d-2)^2 (d-1) \delta ^2\right)}{(d-2) \left(2+(d-1)(d-2) \delta ^2\right)}} }}{(d-2)r_h^{d-1} h(r)\sqrt{2+(d-2)(d-1) \delta ^2}}\left(1-\frac{r_h^{d-2}}{r^{d-2}}\right),   \\
e^\phi &= h(r)^{\frac{-2(d-1)\delta }{2+(d-2)(d-1)\delta ^2}}\ \,.
\end{split}
\end{align}
It is worth noting that the model~(\ref{GGR})-(\ref{MODELACTION}) depends on two independent parameters:
the dimension $d$, and a real number $\delta$ appearing in the functional forms of the potentials~\eqref{MODELACTION}. The analytic solutions above depend also
on two extra parameters (constants of integration) $r_h$ and $Q$ which, as we will see shortly, determine the temperature and chemical potential of these black hole geometries. We generally require $\d\neq0$ and $Q\neq0$ to obtain hairy black hole solutions within our ansatz. Conversely, for $\d=0$ and $Q\neq0$ we recover the traditional Reissner-Nordström-AdS black hole (see section \ref{appa2}), while for $Q=0$ we recover the Schwarzshcild-AdS black hole (in this case $\delta$ does not feature in the solutions).
{Notice also that one must have $Q\geq 0$ since otherwise the gauge field $A_t$
becomes imaginary.}

Finally, we shall point out that for the particular value of $\delta$
\begin{align}\label{CRITICAL}
\begin{split}
\delta =\delta_c \equiv  \sqrt{\frac{2}{d\,(d-1)}} \,,
\end{split}
\end{align}
our holographic setup becomes the top-down Gubser-Rocha
model~\cite{Gubser:2009qt}.\footnote{See also \cite{Davison:2013txa,Gouteraux:2014hca,Jeong:2018tua} for its extension when the boundary translational symmetry is broken.}  For $d=4$ this background is
a truncation of IIB string theory in AdS$_5 \times S^5$. Similarly, for $d=3$, it results from a consistent truncation of eleven-dimensional supergravity compactified on AdS$_4 \times S^7$~\cite{Gubser:2009qt}. Hereafter, we shall refer to the value of $\d$ in \eqref{CRITICAL} as the critical value.

%
\subsection{IR geometries}
\label{ssec:irgeometries}
In this section, we will study the low-temperature behavior of the black hole geometries
(\ref{OURMETRIC})-(\ref{sol12}). As we will see, these backgrounds feature three different infrared (IR) geometries depending on the value of $\delta$ (at fixed number of dimensions $d$).
For $\delta<\delta_c$, with
$\delta_c$ given in \eqref{CRITICAL},  the extremal geometry becomes
$AdS_2 \times \R^{d-1}$. For $\delta=\delta_c$, the IR geometry is
conformal to $AdS_2\times \R^{d-1}$, featuring a linearly vanishing entropy as $T\to 0$. Finally, for $\delta>\delta_c$, the geometry is gapped: 
the black hole solution ceases to exist below a certain temperature.
We will characterize these three cases by examining the behavior of observables
as the Null Energy Condition, thermal entropy, and butterfly velocity.

We will classify our geometries according to their low-temperature behavior. Thus we first need to read the temperature ($T$) and chemical potential ($\mu$) of the black hole solutions
(\ref{OURMETRIC})-(\ref{sol12}). They are given as
\begin{align}\label{TmuFOR}
\begin{split}
T &\equiv {\frac{1}{4\pi } \left. \frac{D'}{\sqrt{D B}}\right|_{r_{h}}} =  r_{h} \, \frac{2 \left(d(1+\tQ)-\frac{4(d-1)\tQ}{2+(d-2)(d-1)\delta^2}\right)  (1+\tQ)^{\frac{2d-(d-2)^2(d-1)\delta^2}{(d-2)(2+(d-2)(d-1)\delta^2)}}   }{8 \pi (1+\tQ)^{\frac{2(d-1)}{(d-2)(2+(d-2)(d-1)\delta^2)}}    }  \,, \,  \\
\mu &\equiv A_t(\infty) = r_{h} \, \sqrt{\frac{4(d-1) \, \tQ (1+\tQ)^{\frac{4-2(d-2)^2(d-1)\delta^2}{(d-2)(2+(d-2)(d-1)\delta^2)}}}{(d-2)(2+(d-2)(d-1)\delta^2)} }  \,,
\end{split}
\end{align}
where we defined the dimensionless quantity $\tQ$ as
\begin{align}\label{tildeQval}
\begin{split}
\tQ \equiv \frac{Q}{r_{h}^{d-2}} \,.
\end{split}
\end{align}
One can see that $Q$, together with $r_h$, determines the physical quantities $T$ and $\mu$.
In particular, the dimensionless quantity $\tilde Q$ sets the dimensionless ratio $T/\mu$, reflecting the underlying UV conformal symmetry of these theories.
{Recall that, as discussed above, $\tilde Q\geq0$ for our geometries.
Then, we shall check next if an extremal (zero temperature) solution can be found for a value of $\tilde Q$ within that range.}

\paragraph{$\d > \d_c$ : Gapped geometries.} It is easy to check that for these values of $\delta$ there exists a nonzero minimum value of $T/\mu$ given by
\begin{equation}
{T\over\mu}\bigg|_{\text{min}}=
{d-2\over4\pi}\sqrt{{d^2(d-1)\,\delta^2-2d\over d-1}}\,.
\label{eq:tomumin}
\end{equation}
Therefore in this range of parameters, the black hole solutions we study do not have a zero temperature limit. They cease to exist for temperatures below the value above.
That minimum temperature~\eqref{eq:tomumin} vanishes as $\delta$ approaches $\delta_c$ from above. Accordingly, as we see below, for $\delta\leq\delta_c$ our geometries have a well-defined extremal limit.

It is worth noting that two branches of solutions exist for these gapped geometries. 
In Appendix \ref{appB} we compute the free energy of these backgrounds to determine which branch is the thermodynamically stable solution.

\paragraph{$\d < \d_c$ : AdS$_2 \times \R^{d-1}$ extremal geometries.}
In this range of values of $\delta$ one can easily find the value of $\tilde Q$ that makes
$T/\mu$ \eqref{TmuFOR} vanish. It reads
\begin{equation}\label{ZTC1}
\tQ = \frac{d \left( 2+(d-2)(d-1)\d^2 \right)}{(d-2)(2-d(d-1) \d^2)} \,.
\end{equation}
Plugging this condition into our metric \eqref{sol11} and expanding it near the horizon $r_h$ one arrives at
\begin{align}\label{OURMETRIC2}
\begin{split}
\dd s^2 \approx -c_t \,\zeta^2 \,\dd t^2 + \frac{c_\zeta}{\zeta^2} \,\dd \zeta^2 + c_x \,\dd \vec x_{i}^{2} \,,  \qquad \zeta \equiv r-r_h \,.
\end{split}
\end{align}
This extremal solution corresponds to an AdS$_2 \times \R^{d-1}$ geometry with the
three coefficients $(c_t,\,c_\zeta,\,c_x)$ given as
\begin{align}\label{PR1}
\begin{split}
c_t = \frac{d(d-2)}{2}\, \xi^{\frac{2-2\xi+d(d-3+4\xi-d\xi)}{(d-2)(d-1)(\xi-1)}} \,, \qquad
c_\zeta = \frac{2}{d(d-2)}\, \xi^{\frac{1-d-\xi}{(d-1)(\xi-1)}} \,, \qquad
c_x = \xi^{\frac{4}{(d-2)(2+(d-2)(d-1)\d^2)}} \,,
\end{split}
\end{align}
where
\begin{align}\label{PR2}
\begin{split}
\xi \equiv \frac{4(d-1)}{(d-2)(2-d(d-1)\d^2)} \,.
\end{split}
\end{align}
It is easy to check that for $0\leq \d < \d_c$ one has $\xi\geq 0$
and all three coefficients~\eqref{PR1} are positive.

\paragraph{$\d = \d_c$ : conformal to AdS$_2 \times \R^{d-1}$ extremal geometry.}
For $\delta$ equal to the critical value \eqref{CRITICAL} the extremal limit of our geometries is reached for $r_h=0$, which corresponds to $\tQ\to\infty$. In this limit, the near horizon geometry of the metric~\eqref{sol11} takes the form
\begin{align}\label{OURMETRIC3}
\begin{split}
\dd s^2 \approx \tilde{\zeta}^{\frac{2}{d-1}} \left(- \tilde{\zeta}^2 \,\dd t^2 + \frac{1}{\tilde{\zeta}^2} \,\dd \tilde{\zeta}^2 +  \,\dd \vec x_{i}^{2}\right) \,,  \qquad \tilde{\zeta} \equiv r^{\frac{d-2}{2}} \,,
\end{split}
\end{align}
where, for the sake of clarity, we have omitted the metric coefficients. These are dependent on $Q$ and $d$ and are always positive-definite when $Q\geq0$ and $d>2$. 
It is obvious from \eqref{OURMETRIC3} that the extremal geometry for $\delta=\delta_c$ is conformal to AdS$_2 \times \R^{d-1}$. It features a vanishing horizon and thus, as we will see below, a vanishing entropy.

In the remainder of this section, focusing on the extremal geometries above, we will study different observables that will allow us to further characterize these holographic backgrounds. We will look at the Null Energy Condition, thermal entropy, and butterfly velocity of our family of black hole solutions.

\paragraph{Null Energy Condition.} The Null Energy Condition (NEC) states that the matter energy-momentum tensor $T_{\mu\nu}$ obeys
\begin{align}\label{NECFOR}
\begin{split}
T_{\mu\nu}\,n^{\mu}n^{\nu} >0 \,,
\end{split}
\end{align}
for any null vector $n^\mu$, i.e., for any vector satisfying $g_{\mu\nu}\,n^{\mu}n^{\nu}=0$.

For our metric \eqref{OURMETRIC}, one can find the null vector as 
\begin{align}\label{NECFOR2}
\begin{split}
n_{t} = \tilde{c}_t \, \sqrt{D(r)} \,, \qquad n_{r} = \tilde{c}_r \, \sqrt{B(r)} \,, \qquad n_{x} = \tilde{c}_x \, \sqrt{\frac{C(r)}{d-1}} \,,
\end{split}
\end{align}
where ($\tilde{c}_t,\, \tilde{c}_r,\, \tilde{c}_x$) are constants that satisfy
\begin{align}\label{}
\begin{split}
\tilde{c}_r^2 + \tilde{c}_x^2 = \tilde{c}_t^2 \,,
\end{split}
\end{align}
to ensure $g_{\mu\nu}\,n^{\mu}n^{\nu}=0$.

Let us then consider the NEC \eqref{NECFOR} for the extremal geometries~\eqref{OURMETRIC2} of the type AdS$_2\times \R^{d-1}$. One easily finds that the NEC is satisfied provided that
\begin{align}\label{}
\begin{split}
\frac{\tilde{c}_x^2}{c_\zeta} >0 \,,
\end{split}
\end{align}
which is indeed true in the range $\delta<\delta_c$ where those extremal geometries exist.

As for the extremal geometries~\eqref{OURMETRIC3} which are conformal to AdS$_2\times \R^{d-1}$ the NEC is always satisfied.


\paragraph{Thermal entropy.} We shall examine the behavior of the thermal entropy, given by the area of the black hole horizon, for the three different black hole solutions described above. In terms of the metric ansatz~\eqref{OURMETRIC} particularized to the solutions~\eqref{sol11}, the entropy ($s$) is given by
\begin{align}\label{enforre}
\begin{split}
s \equiv 4\pi C(r)^{\frac{d-1}{2}} |_{r_{h}} =  4\pi \, r_h^{d-1} \,  \left( 1 + \tQ \right)^{\frac{2(d-1)}{(d-2)( 2+(d-2)(d-1)\d^2 )}} \,.
\end{split}
\end{align}
For the extremal solutions realizing an AdS$_2\times \mathbb{R}^{d-1}$ geometry, after substituting the zero temperature condition~\eqref{ZTC1}, the entropy reads
\begin{align}\label{entropyT0}
\begin{split}
\frac{s}{\mu^{d-1}} \,=\, 4 \pi \left(\frac{d-2}{d}\right)^{\frac{d-1}{2}} \xi^{\frac{1-d-\xi}{2(\xi-1)}}  \,+\, \mathcal{O}({T}/{\mu}) \,,
\end{split}
\end{align}
where $\xi$, given in \eqref{PR2}, is real and positive in the range $\delta<\delta_c$
where the extremal AdS$_2\times \R^{d-1}$ geometries exist.

As expected, one can check that for $\d = \d_c$ the entropy vanishes as $T$ goes to zero.
Indeed, in the low-temperature limit, one arrives at
\begin{equation}
{s\over\mu^{d-1}} \equiv 4\pi \left(C(r)/\mu^2\right)^{\frac{d-1}{2}}|_{r_h} \,\approx\,
8 \pi^2 (d-2)^{\frac{d-2}{2}} d^{-d/2} \,\, T/\mu \,+\, \dots
\end{equation}

\begin{figure}[]
  \centering
     {\includegraphics[width=7.5cm]{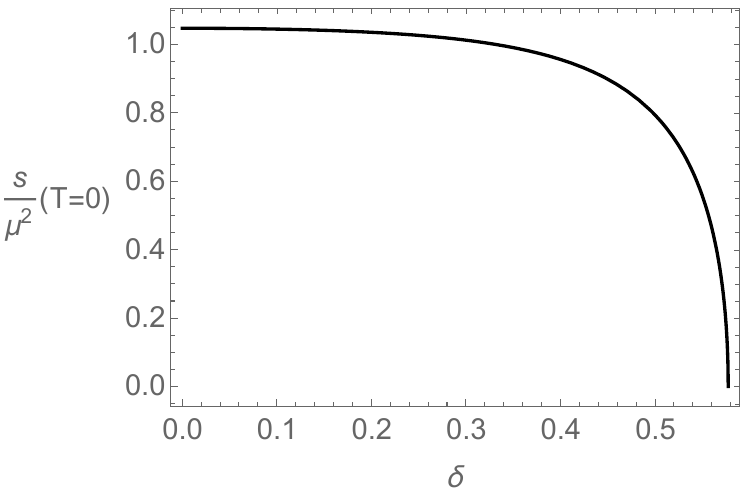} \label{}}
 \quad
     {\includegraphics[width=6.7cm]{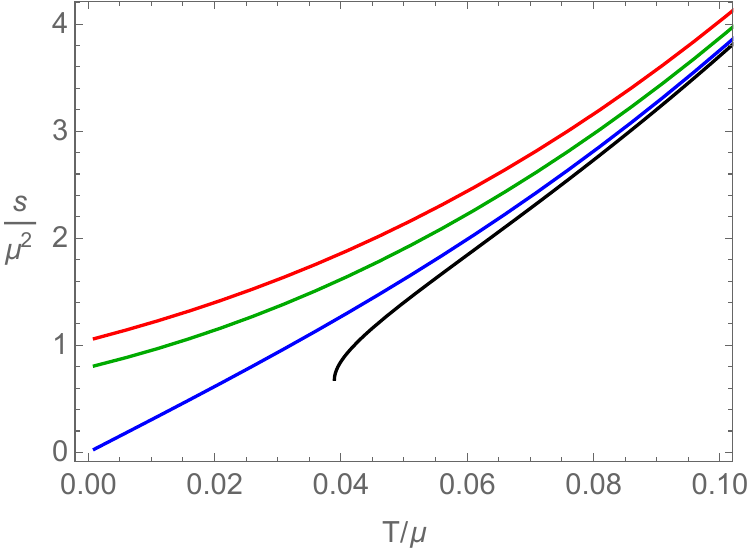} \label{}}
 \caption{Thermal entropy $s/\mu^2$ for $d=3$ and thus with $\d_c = \sqrt{1/3} \approx 0.577$. Left panel: $s/\mu^2$ vs. $\d$ when $T=0$. Right panel: $s/\mu^2$ vs. $T/\mu$ for $\d = 0,\, 0.5,\, \d_c,\, 0.6$ (red, green, blue, black).}\label{ENFIG1}
\end{figure}

We illustrate the behavior of the thermal entropy of these black hole geometries for $d=3$ in Fig.~\ref{ENFIG1}.
In the left panel of Fig.~\ref{ENFIG1} we plot the thermal entropy at $T=0$ as a function of $\delta$. In the right panel of Fig.~\ref{ENFIG1} we show the temperature dependence of the entropy for two different values of $\delta$, one below and another above the critical value.

\paragraph{Butterfly velocity.}
For the metric ansatz~\eqref{OURMETRIC},
the butterfly velocity $v_B$~\cite{Mezei:2016wfz,Blake:2016wvh} is given by (see \cite{Fischler:2018kwt,DiNunno:2021eyf,Eccles:2021zum} for a derivation in slightly more general geometries):
\begin{align}\label{vbforre}
\begin{split}
v_B^2 \equiv{\frac{4\pi T}{(d-1)C'(r)}} \,\, \Bigg|_{r_{h}} = \frac{d(1+\tQ-\xi)(1+\tQ)^{\frac{d(d-3)\xi}{2(d-2)(d-1)(\xi-1)}}}{2d(1+\tQ)+2(1+\tQ)(\xi-1)-d(2+\tQ)\xi} \,,
\end{split}
\end{align}
where the second equality follows after plugging in our solutions~\eqref{sol11}.

In the regime $\delta<\delta_c$ we can study the zero temperature limit of the butterfly velocity. First, one can compute the low $T/\mu$ correction to $\tQ$ in \eqref{ZTC1},
arriving at
\begin{equation}\label{ZTCsmallT}
\tQ = \frac{d \left( 2+(d-2)(d-1)\d^2 \right)}{(d-2)(2-d(d-1) \d^2)} \left[ 1 \,-\, 8\pi \sqrt{\frac{d-1}{d(d-2)^2 \left(2-d(d-1)\d^2\right)}} \, {T}/{\mu} \,+\, \mathcal{O}({T}/{\mu})^2 \right] \,.
\end{equation}
Plugging this expression into~\eqref{vbforre} we arrive at 
\begin{align}\label{LTVB}
\begin{split}
v_B^2 \,=\, 4 \pi \sqrt{\frac{d}{(d-2)^3}} \, \xi^{\frac{2+d(d-3)-2\xi}{2(d-2)(d-1)(\xi-1)}} \,\, {T}/{\mu}  \,+\, \mathcal{O}({T}/{\mu})^2 \,.
\end{split}
\end{align}
For the critical case $\d=\d_c$, the expression for $\tQ$ in \eqref{ZTCsmallT} becomes divergent, indicating the necessity for separate computations. Using \eqref{TmuFOR}
one finds that the extremal limit of the critical case corresponds to
\begin{equation}\label{}
\tQ = \frac{d(d-2)}{16 \pi^2 (T/\mu)^2} \,.
\end{equation}
Therefore, in the limit of low $T/\mu$, \eqref{vbforre} reads
\begin{align}\label{}
\begin{split}
v_B^2 \,=\,   \left({T}/{\mu}\right)^{\frac{d(3-d)}{(d-2)(d-1)}} \left[ d^{\frac{d(d-3)}{2(d-2)(d-1)}}  (d-2)^{\frac{-3d(d-3)-8}{2(d-2)(d-1)}} (4\pi)^{\frac{d(d-3)+4}{(d-2)(d-1)}}       \left({T}/{\mu}\right)^2  \,+\, \mathcal{O}({T}/{\mu})^3 \right] \,.
\end{split}
\end{align}
\begin{figure}[]
  \centering
      {\includegraphics[width=7.1cm]{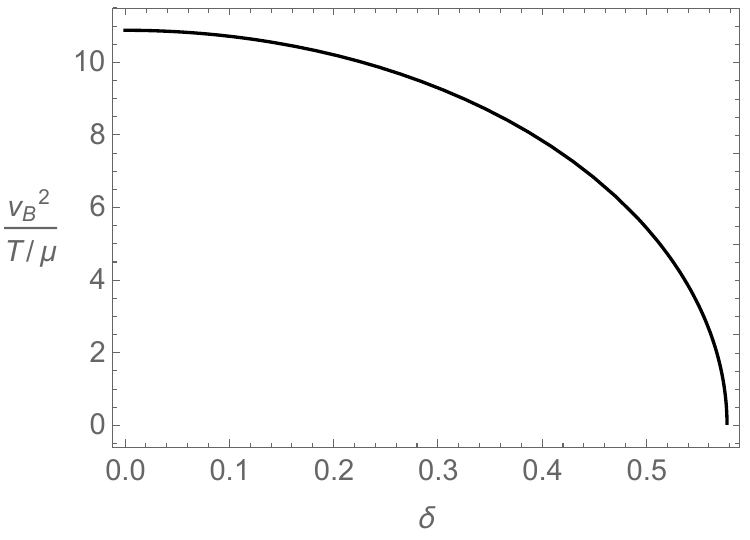} \label{}}
 \quad
      {\includegraphics[width=7.0cm]{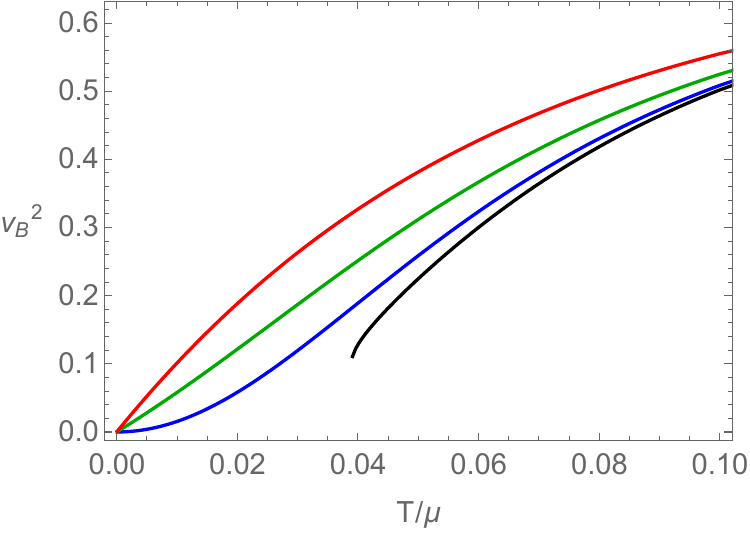} \label{}}
 \caption{Butterfly velocity for $d=3$. Left panel: $v_B^2$ at low temperature \eqref{LTVB}. Right panel: $v_B^2$ at  $\d = 0,\, 0.5,\, \d_c \approx0.577,\, 0.6$ (red, green, blue, black).}\label{VBFIG}
 \end{figure}
In Fig. \ref{VBFIG}, we display the $\d$-dependence and $T$-dependence of $v_B^2$ when $d=3$.

%

\section{Beyond the horizon}\label{sec:interior}
In this section, we will study the interior of the family of black hole solutions
we have presented above.
We will first characterize the type of singularities these models can feature.
Next, we will present a generalization of the no-inner horizon theorem where we show that for some values of the parameters, these black hole geometries with a non-trivial scalar possess an inner horizon.

\subsection{Near singularity limit}\label{sec23}

We shall now study the interior of the black hole geometries (\ref{OURMETRIC})-(\ref{sol12}).
It will be useful to study the singularities of these solutions using the
conventional coordinate system previously employed in the literature, characterized by
\begin{align}\label{ZCO}
\begin{split}
\dd s^2 = \frac{1}{z^2} \left[ -g(z) e^{-\chi(z)} \dd t^2 + \frac{\dd z^2 }{g(z)} + \dd \vec x_{i}^{2} \right]   \,,
\end{split}
\end{align}
that can be obtained from our ansatz~\eqref{OURMETRIC} through the coordinate transformation:
\begin{align}\label{ZCO2}
\begin{split}
z = \frac{1}{\sqrt{C(r)}} \,, \qquad g(z) = \frac{C'(r)^2}{4 B(r) C(r)^2} \,, \qquad \chi(z) = \log \left[\frac{C'(r)^2}{4 D(r) B(r) C(r)}\right] \,.
\end{split}
\end{align}
In the Appendix \ref{appa1} we instead perform the analysis of the singularities employing the original $r$-coordinate \eqref{OURMETRIC}, obtaining results consistent with those in this section.

Taking the near singularity limit $z\to\infty$ (equivalently $r\to0$) in our model~\eqref{sol11},
the coordinate transformation~\eqref{ZCO2} boils down to
\begin{align}\label{ZCO23}
\begin{split}
 r = Q^{-\frac{2}{(d-2)^2 (d-1) \d^2}} \, z^{-1-\frac{2}{(d-2)(d-1)\d^2}} \,,
\end{split}
\end{align}
where it can be verified that $-1-\frac{2}{(d-2)(d-1)\d^2}<0$ when $d>2$ and $\d\geq0$.
Next, plugging \eqref{ZCO23} into our solutions (\ref{OURMETRIC})-(\ref{sol12}),
we arrive at the following approximate solution near the singularity
\begin{align}\label{SLFOR}
\begin{split}
g(z) = & - \left[ \frac{(d-2)(d-1)\d^2}{2+(d-2)(d-1)\d^2 }\right]^2 Q^{\frac{2}{(d-2)(d-1)\d^2}} z^{-\frac{4}{(d-2)(d-1)\d^2}} \\
&\times \left[   r_h^d \left(1+ Q r_h^{2-d}\right)^{\frac{4(d-1)}{(d-2)\left( 2+(d-2)(d-1)\d^2 \right)}} z^{d + \frac{2d}{(d-2)(d-1)\d^2}}  - Q^{\frac{2}{(d-2)(d-1)\d^2}} z^{\frac{4}{(d-2)\d^2}} \right] \,, \\
\chi(z) = & \frac{4}{(d-1)\d^2} \log z  \,, \qquad
\phi(z) =  -\frac{2}{\d} \log z  \,.
\end{split}
\end{align}

Crucially, by comparing the three leading powers of $z$ in this solution for $g(z)$, the geometry
in this limit can be categorized into three distinct classes
\begin{align}\label{CLASSTYPES}
\begin{split}
\d &\neq 0:
\begin{cases}
\,\, d + \frac{2d}{(d-2)(d-1)\d^2} > \frac{4}{(d-2)\d^2}  \qquad\longrightarrow\qquad \d > \d_c \,, \qquad\quad\, (\text{Class I})  \\
\,\, d + \frac{2d}{(d-2)(d-1)\d^2} = \frac{4}{(d-2)\d^2}  \qquad\longrightarrow\qquad \d = \d_c \,,  \qquad\quad\, (\text{Class II}) \\
\,\, d + \frac{2d}{(d-2)(d-1)\d^2} < \frac{4}{(d-2)\d^2} \qquad\longrightarrow\qquad 0< \d < \d_c \,, \quad\,\, (\text{Class III}) 
\end{cases}
\\
\d &= 0 \,,
\end{split}
\end{align}
where the case with $\delta=0$ should be handled separately, as we will demonstrate shortly.
Notice that these three classes correspond to the three different types of IR geometries analyzed in section~\ref{ssec:irgeometries}.
In the following subsection we investigate the singularity across all these classes.

%
\subsection{Singularities}
\label{ssec:singularities}

%
\subsubsection{$\d\neq0$: Kasner or timelike singularity}

\paragraph{Class I ($\d > \d_c$): Kasner singularity.}
This class corresponds to the gapped geometries described around equation~\eqref{eq:tomumin}.
In this Class I \eqref{SLFOR} can be written as
\begin{align}\label{SG1}
\begin{split}
g(z) = g_c \,\, z^{\frac{2+d(d-1)\d^2}{(d-1)\d^2}} \,, \qquad
\chi(z) = \frac{4}{(d-1)\d^2} \log z  \,, \qquad
\phi(z) =  -\frac{2}{\d} \log z  \,, 
\end{split}
\end{align}
where
\begin{align}\label{gcCI}
\begin{split}
g_c \equiv - \left[ \frac{(d-2)(d-1)\d^2}{2+(d-2)(d-1)\d^2 }\right]^2 Q^{\frac{2}{(d-2)(d-1)\d^2}}  \left[   r_h^d \left(1+ Q r_h^{2-d}\right)^{\frac{4(d-1)}{(d-2)\left( 2+(d-2)(d-1)\d^2 \right)}}   \right] < 0 \,.
\end{split}
\end{align}
Performing the additional coordinate transformation
\begin{align}\label{CTAORI}
\begin{split}
 z = \tau^{-\frac{2(d-1)\d^2}{2+d(d-1)\d^2}} \,,
\end{split}
\end{align}
our metric \eqref{ZCO} can be expressed as
\begin{align}\label{singmet1}
\begin{split}
\dd s^2 =   \frac{1}{g_c}\left[\frac{2(d-1)\d^2}{2+d(d-1)\d^2}\right]^2 \dd \tau^2  - g_c \tau^{2 p_t} \dd t^2  + \tau^{2 p_x} \dd \vec x_{i}^{2}  \,, \qquad \phi(\tau) = -\sqrt{2} p_{\phi} \log \tau \,,
\end{split}
\end{align}
with 
\begin{align}\label{KAS1}
\begin{split}
p_t = \frac{2-(d-2)(d-1)\d^2}{2+d(d-1)\d^2} \,, \qquad 
p_x = \frac{2(d-1)\d^2}{2+d(d-1)\d^2} \,, \qquad 
p_\phi = -\frac{2\sqrt{2}(d-1)\d}{2+d(d-1)\d^2} \,.
\end{split}
\end{align}
One can check that these coefficients satisfy the Kasner condition:
\begin{align}\label{eq:kasnercond}
\begin{split}
p_t + (d-1)p_x = p_\phi^2 + p_t^2 + (d-1)p_x^2 = 1 \,.
\end{split}
\end{align}
Therefore, the metric near the singularity in Class I, as given by \eqref{singmet1}, corresponds to the Kasner singularity geometry.
The singularity is spacelike as shown by the inequality~\eqref{gcCI}.

\paragraph{Class II ($\d = \d_c$): Kasner singularity.}
In this class II the geometry in the singularity limit, given by \eqref{SLFOR}, becomes
\begin{align}\label{SG2}
\begin{split}
g(z) = g_c \,\, z^{2d} \,, \qquad
\chi(z) = 2d \log z  \,, \qquad
\phi(z) =  -\sqrt{2d(d-1)} \log z  \,, 
\end{split}
\end{align}
where
\begin{align}\label{}
\begin{split}
g_c \equiv - \left[ \frac{d-2}{2(d-1)} \right]^2 Q^{\frac{d}{d-2}}  \left[   r_h^d \left(1+ Q r_h^{2-d}\right)^{\frac{d}{d-2}} - Q^{\frac{d}{d-2}}   \right] < 0 \,.
\end{split}
\end{align}
As for Class I one can find a coordinate transformation
\begin{align}\label{}
\begin{split}
 z = \tau^{-\frac{1}{d}} \,,
\end{split}
\end{align}
that brings the metric to the usual (spacelike) Kasner singularity metric
\begin{align}\label{new22}
\begin{split}
\dd s^2 =   \frac{1}{g_c} \frac{1}{d^2} \dd \tau^2  - g_c \tau^{2 p_t} \dd t^2  + \tau^{2 p_x} \dd \vec x_{i}^{2}  \,, \qquad \phi(\tau) = -\sqrt{2} p_{\phi} \log \tau \,,
\end{split}
\end{align}
where
\begin{align}\label{KAS2}
\begin{split}
p_t = p_x = \frac{1}{d} \,, \qquad 
p_\phi = -\sqrt{\frac{d-1}{d}} \,,
\end{split}
\end{align}
and they satisfy the Kasner condition~\eqref{eq:kasnercond}.
Note that although the analysis of Class II is conducted separately from that of Class I, the result~\eqref{KAS2} regarding the Kasner exponents can be obtained by setting $\delta=\delta_c$ in \eqref{KAS1}.\\

\paragraph{Class III ($0< \d < \d_c$): timelike singularity.}
For this class of solutions, the near singularity geometry~\eqref{SLFOR} takes the form
\begin{align}\label{SG3}
\begin{split}
g(z) = g_c \,\, z^{\frac{4}{(d-1)\d^2}} \,, \qquad
\chi(z) = \frac{4}{(d-1)\d^2} \log z  \,, \qquad
\phi(z) =  -\frac{2}{\d} \log z  \,, 
\end{split}
\end{align}
where
\begin{align}\label{TLSCON}
\begin{split}
g_c \equiv  \left[ \frac{(d-2)(d-1)\d^2}{2+(d-2)(d-1)\d^2 } \, Q^{\frac{2}{(d-2)(d-1)\d^2}} \right]^2 > 0 \,.
\end{split}
\end{align}
Applying the coordinate transformation
\begin{align}\label{}
\begin{split}
 z = \tau^{-\frac{(d-1)\d^2}{2}} \,,
\end{split}
\end{align}
we get
\begin{align}\label{singmet2}
\begin{split}
\dd s^2 =   \frac{1}{g_c}\left[\frac{(d-1)\d^2}{2}\right]^2 \dd \tau^2  - g_c \tau^{2 \tilde{p}_t} \dd t^2  + \tau^{2 \tilde{p}_x} \dd \vec x_{i}^{2}  \,, \qquad \phi(\tau) = -\sqrt{2} \tilde{p}_{\phi} \log \tau \,,
\end{split}
\end{align}
where
\begin{align}\label{TLSCOMP}
\begin{split}
\tilde{p}_t = \tilde{p}_x = \frac{(d-1)\d^2}{2} \,, \qquad 
\tilde{p}_\phi = -\frac{(d-1)\d}{\sqrt{2}}\,.
\end{split}
\end{align}
Notice that these exponents do not satisfy the Kasner condition~\eqref{eq:kasnercond}; they instead fulfill
\begin{align}\label{eq:class3kas}
\begin{split}
\tilde{p}_t + (d-1)\tilde{p}_x = \frac{d(d-1)\,\d^2}{2}  \,, \qquad \tilde{p}_\phi^2 + \tilde{p}_t^2 + (d-1)\tilde{p}_x^2 =  \frac{(d-1)^2 (2+d\, \d^2)\, \d^2}{4} \,.
\end{split}
\end{align}
Moreover, as indicated by \eqref{TLSCON}, the metric near the singularity for these Class III geometries is timelike. It is worth noting that this timelike singularity is similar to that of Reissner-Nordström black holes. These are the black hole solutions of our model for $\delta=0$ (and $Q\neq0$), and, as we review in section \ref{appa2}, they too feature a timelike singularity that can be expressed in the Kasner metric form without satisfying the Kasner condition.

We can summarize the type of singularities of the family of black hole geometries~(\ref{OURMETRIC})-(\ref{sol12}) as follows
\begin{align}\label{FINRESUM}
\begin{split}
\d &\neq 0\quad\&\quad Q\neq 0:
\begin{cases}
\,\, \text{Kasner singularity} \,, \qquad\quad\,\,\,\, (\d > \d_c)  \\
\,\, \text{Kasner singularity} \,,  \qquad\quad\,\,\,\, (\d = \d_c) \\
\,\, \text{Timelike singularity} \,, \qquad\quad (0< \d < \d_c)
\end{cases}
\\
\d &= 0\quad\&\quad Q\neq 0:\,\, \text{Timelike singularity (Reissner-Nordstr\"om)}\,, \\
Q&=0:\,\, \text{Kasner singularity (Schwarzschild)} \,, 
\end{split}
\end{align}
where the last two cases (Reissner-Nordstr\"om and Schwarzschild) will be shown below.
Notice that we have found an instance (our solutions with
$0<\delta<\delta_c$) 
of black holes that in the presence of a nontrivial scalar exhibit
a timelike singularity.
This points to the possibility, which we will confirm shortly, 
that for those solutions, the deformation introduced by the scalar has not prevented the presence of an inner (Cauchy) horizon. This will prompt us to generalize the theorems that established how certain scalar deformations impeded the formation of inner
horizons~\cite{Hartnoll:2020rwq,Cai:2020wrp,An:2021plu}.

%
\subsubsection{$\d=0$: Timelike singularity}\label{appa2}

Next, we examine the case where the parameter $\d=0$ while having a finite $Q$, leading to Reissner-Nordström black holes. Specifically, implementing the following coordinate transformation into \eqref{sol11},   
\begin{align}\label{}
\begin{split}
 \tilde{r} = r \left( 1 + Q \, r^{2-d} \right)^{\frac{1}{d-2}} \,,  \qquad \tilde{\mu} = \sqrt{\frac{2(d-1)}{d-2}Q \, r_h^{4-d} \left( 1 + Q \, r_h^{2-d} \right)^{\frac{2}{d-2}}  } \,,
\end{split}
\end{align}
we obtain the familiar Reissner-Nordström black hole metric
\begin{align}\label{RNMET11}
\begin{split}
\dd s^2 &=  -f(\tilde{r}) \dd t^2 + \frac{\dd \tilde{r}^2}{f(\tilde{r})} + \tilde{r}^2 \dd \vec x_{i}^{2}   \\ 
&=    -\frac{d-2}{2(d-1)} \left(\frac{\tilde{r}_h}{\tilde{r}}\right)^{2d-4} \tilde{\mu}^2 \, \dd t^2 + \left[\frac{d-2}{2(d-1)} \left(\frac{\tilde{r}_h}{\tilde{r}}\right)^{2d-4} \tilde{\mu}^2 \right]^{-1} \dd \tilde{r}^2 + \tilde{r}^2 \dd \vec x_{i}^{2} \,.
\end{split}
\end{align}
where the singularity limit $\tilde{r}\rightarrow0$ has been taken in the second equality together with 
\begin{align}\label{}
\begin{split}
f(\tilde{r}) = \tilde{r}^2 \left[ 1 - \left(\frac{\tilde{r}_h}{\tilde{r}}\right)^d -\frac{d-2}{2(d-1)} \left\{ \left( \frac{\tilde{r}_h}{\tilde{r}} \right)^d - \left( \frac{\tilde{r}_h}{\tilde{r}} \right)^{2(d-1)}   \right\} \frac{\tilde{\mu}^2}{\tilde{r}_h^2}  \right] \,.
\end{split}
\end{align}
Here $\tilde{\mu}$ is the chemical potential given by the asymptotic value of the gauge field $A_t$.

Next, using the coordinate transformation
\begin{align}\label{}
\begin{split}
 \tilde{r} = \tau^{\frac{1}{d-1}} \,,
\end{split}
\end{align}
the metric \eqref{RNMET11} becomes
\begin{align}\label{RNMET12}
\begin{split}
\dd s^2 =   \frac{2 \tilde{r}_h^{4-2d}}{(d-2)(d-1)\tilde{\mu}^2} \dd \tau^2  - \frac{(d-2)\tilde{r}_h^{2d-4}\tilde{\mu}^2}{2(d-1)} \tau^{2 \tilde{p}_t} \dd t^2  + \tau^{2 \tilde{p}_x} \dd \vec x_{i}^{2}  \,, 
\end{split}
\end{align}
where
\begin{align}\label{}
\begin{split}
\tilde{p}_t = \frac{1}{d-1}-1 \,, \qquad 
\tilde{p}_x = \frac{1}{d-1} \,. 
\end{split}
\end{align}
{Thus one finds that the Reissner-Nordström black holes feature the timelike singularity \eqref{RNMET12} that does not satisfy the Kasner conditions \eqref{eq:kasnercond} since we have}
\begin{align}\label{}
\begin{split}
\tilde{p}_t + (d-1)\tilde{p}_x = \frac{1}{d-1} \,, \qquad \tilde{p}_t^2 + (d-1)\tilde{p}_x^2 = \frac{3+d(d-3)}{(d-1)^2} \,.
\end{split}
\end{align}
%

%
\subsubsection{$Q=0$: Schwarzschild singularity}\label{se242new}

We close this subsection by exploring the remaining case where $Q=0$. As evident from \eqref{sol11int}-\eqref{sol12}, when $Q=0$ both the gauge field and the scalar vanish. Consequently, the geometry becomes that of a Schwarzschild black hole in the absence of a scalar field, and $\d$ is not present in the action
as the potential in \eqref{MODELACTION} becomes $V(0)=d(d-1)$.

In this scenario, the singularity limit is not attained via~ \eqref{ZCO23}. We start from the following usual form of the
Schwarzschild black hole metric
\begin{align}\label{DELTA0EQ1}
\begin{split}
\dd s^2 &= \frac{1}{z^2} \left[ - \left(1- \frac{z^d}{z_h^d}\right)  \dd t^2 + \frac{\dd z^2 }{\left(1- \frac{z^d}{z_h^d}\right)} + \dd \vec x_{i}^{2} \right]   \\ 
&=   \frac{z^{d-2}}{z_h^d}  \dd t^2 -\frac{z_h^d}{z^{d+2}} \dd z^2 + \dd \vec x_{i}^{2} \,,
\end{split}
\end{align}
where $z\rightarrow\infty$ is used in the second equality. 
Then, with the coordinate transformation
\begin{align}\label{}
\begin{split}
 z = \tau^{-\frac{2}{d}} \,,
\end{split}
\end{align}
we have
\begin{align}\label{swakas}
\begin{split}
\dd s^2 =   -\frac{4 z_h^{d}}{d^2} \dd \tau^2  + z_{h}^{-d} \tau^{2 p_t} \dd t^2  + \tau^{2 p_x} \dd \vec x_{i}^{2}  \,, 
\end{split}
\end{align}
where
\begin{align}\label{}
\begin{split}
p_t = \frac{2-d}{d} \,, \qquad 
p_x = \frac{2}{d} \,.
\end{split}
\end{align}
This is the Schwarzschild singularity that satisfies
\begin{align}\label{}
\begin{split}
p_t + (d-1)p_x =  p_t^2 + (d-1)p_x^2 = 1 \,.
\end{split}
\end{align}
%

%
\subsection{More on the interior structure}
So far we have determined the singularity type of the different black hole solutions we study. Next, we will further analyze the geometry behind the horizon, characterizing the whole holographic flow from the horizon to the singularity. In particular, we will try and determine the existence of inner horizons for our solutions.

%

\subsubsection{Proof for the existence of inner horizon or lack thereof}
\begin{figure}[]
 \centering
     {\includegraphics[width=14.0cm]{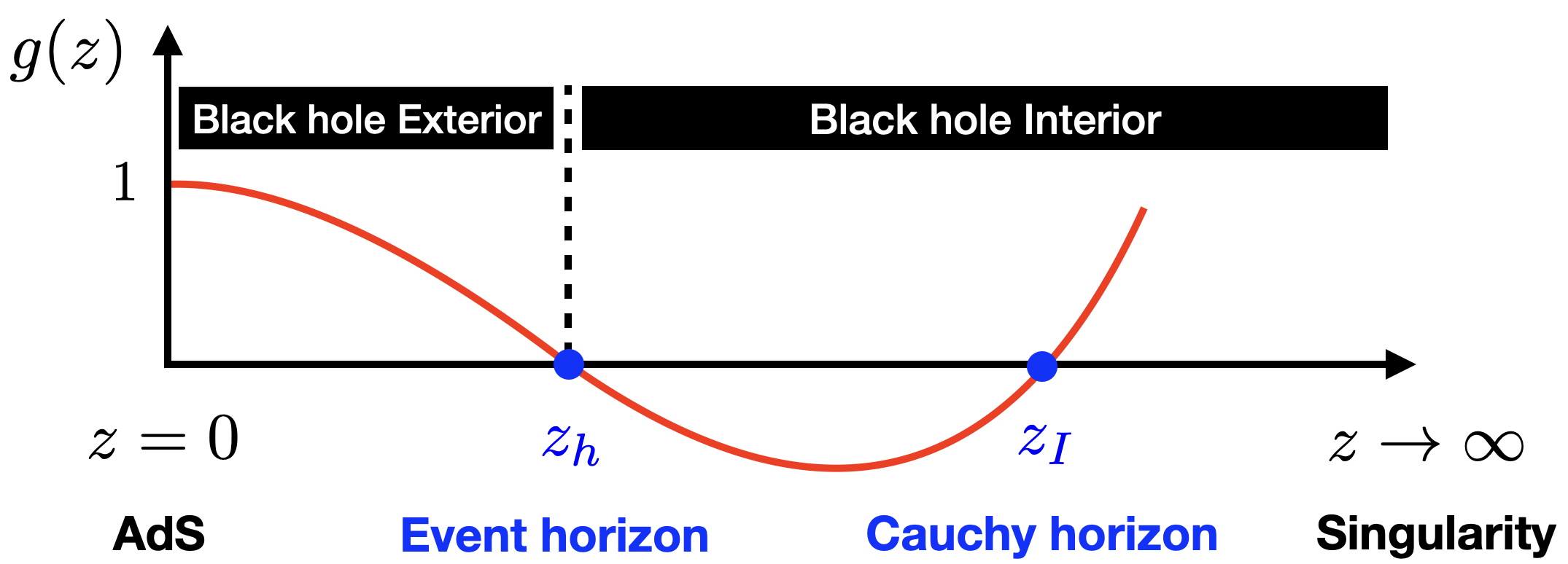} \label{}}
\caption{Schematic structure of the blackening factor $g(z)$ from the AdS boundary to the singularity where $g(z)$ vanishes both at the event horizon $z_h$ and Cauchy horizon $z_I$.}\label{SCP}
\end{figure}

In the previous subsection, we have proved that the solutions
with $0<\d<\d_c$ posses timelike singularities.
This implies the existence of two horizons: an event (outer) horizon at $z_h$, and a Cauchy (inner) horizon at $z_I$, where $g(z_h)=g(z_I)=0$.
In Fig.~\ref{SCP} we illustrate this fact through a depiction of the behavior of the blackening factor $g(z)$ for those solutions with a timelike singularity.

Here, following \cite{Hartnoll:2020rwq}, we prove that the inner horizon $z_I$ can only appear when $0<\d<\d_c$, which is consistent with our finding regarding timelike singularities.
We employ the analytical solutions~(\ref{sol11})-(\ref{sol12})
of the generic metric ansatz~\eqref{OURMETRIC}
wherein the two horizons are denoted as $r_h$ and $r_I$.
We start by rewriting the equation of motion of the dilaton field \eqref{BULKEOM} as
\begin{align}\label{EFFEOM}
\begin{split}
 \frac{1}{\sqrt{D B C^{d-1}}} \left( \sqrt{\frac{D \, C^{d-1}}{B}} \phi^{'} \right)^{'} \,=\, -\dot{V} - \frac{\dot{Z}}{2} \frac{A_t'^2}{DB} \,=:\, V_{\text{eff}} \,,
\end{split}
\end{align}
where a prime (\,$'$\,) denotes differentiation with respect to $r$, and a dot (\,$\dot{}$\,) with respect to $\phi$. 
Furthermore, it is useful to rewrite this equation as
\begin{align}\label{INTMEQ}
\begin{split}
 \left( \sqrt{\frac{D \, C^{d-1}}{B}} \, \phi^{'} \phi \right)^{'} \,=\, \sqrt{D B C^{d-1}} \left( \phi \,V_{\text{eff}} \,+\, \frac{\phi'^2}{B} \right)\,,
\end{split}
\end{align}
using $\left( F \, \phi \right)^{'} \,=\, \phi \, F' + F \, \phi'$, where $F\equiv\sqrt{\frac{D \, C^{d-1}}{B}} \, \phi^{'}$.

Now, assuming the existence of an inner horizon at $r_I$, we can integrate both sides of \eqref{INTMEQ}, arriving at
\begin{align}\label{}
\begin{split}
  \sqrt{\frac{D \, C^{d-1}}{B}} \, \phi^{'} \phi \, \bigg|_{r_h}^{r_I}  \,=\, \int_{r_h}^{r_I} \sqrt{D B C^{d-1}} \left( \phi \,V_{\text{eff}} \,+\, \frac{\phi'^2}{B} \right) \dd r \,.
\end{split}
\end{align}
Plugging in our analytic solutions~(\ref{OURMETRIC})-(\ref{sol12})
we obtain
\begin{align}\label{PFEQ}
\begin{split}
  r^{d-1} f \, \phi^{'} \phi \, \bigg|_{r_h}^{r_I}  \,=\, \int_{r_h}^{r_I} r^{d-1} h^{\frac{4}{(d-2)\left(2+(d-2)(d-1)\d^2\right)}} \left[ \phi \,V_{\text{eff}} \,+\, \underbrace{f h^{-\frac{4}{(d-2)\left(2+(d-2)(d-1)\d^2\right)}}{\phi'^2}}_{ <\, 0 } \right] \dd r \,.
\end{split}
\end{align}
The left-hand side of this equation vanishes if an inner horizon indeed exists (recall that $f(r_h) = f(r_I)=0$).
Then, examining the sign of the integrand on the right hand side, one can ascertain the presence of the inner horizon.
As $f(r)<0$ between the two horizons, it is evident that the second term on the right-hand side is negative. 
Thus, in order to have an inner horizon, the following condition must hold
\begin{align}\label{}
\begin{split}
  \phi \, V_{\text{eff}} > 0 \,,
\end{split}
\end{align}
which can be simplified to
\begin{align}\label{CRUCON}
\begin{split}
  V_{\text{eff}} <0 \,.
\end{split}
\end{align}
Plugging in our analytic solutions~\eqref{sol11}-\eqref{sol12},
this inequality becomes
\begin{align}\label{CRUCON2}
\begin{split}
 \left( d(d-1)\d^2 -2 \right) h \,<\, (d-2) (d-1)\d^2 - \frac{2d}{d-2} \,.
\end{split}
\end{align}
This result, together with the fact, guaranteed by \eqref{sol11int}, that the metric component $h>1$, results in the following condition for the existence of an inner horizon 
\begin{align}\label{CRUCON3}
\begin{split}
 \exists \, \text{Inner horizon}: \quad  0 < \delta < \d_c \,,
\end{split}
\end{align}
which agrees with our finding that indeed a timelike singularity appears in this range.

%
\subsubsection{Holographic flows to singularities}
We shall next analyze the holographic flows corresponding to the hairy ($\delta\neq0$) black hole geometries we are studying.
We will present our results in terms of the ansatz
\eqref{ZCO} to allow for easier comparison with the singularity analysis of section~\ref{ssec:singularities}.
In order to illustrate the main features of these backgrounds we will plot and discuss the following two solutions with
$d=3$ at $T/\mu=1$:
\begin{align}\label{TWOEX}
\begin{split}
 \d =\frac{\d_c}{3}  \,, \qquad \d = \d_c = \sqrt{\frac{1}{3}} \,.
\end{split}
\end{align}
The first case falls in class III and features a timelike singularity, while the second one, for $\delta$ set to its critical value, presents a Kasner singularity.

\begin{figure}[]
 \centering
     \subfigure[$g(z)$ at $\d ={\d_c}/{3}$]
     {\includegraphics[width=7.0cm]{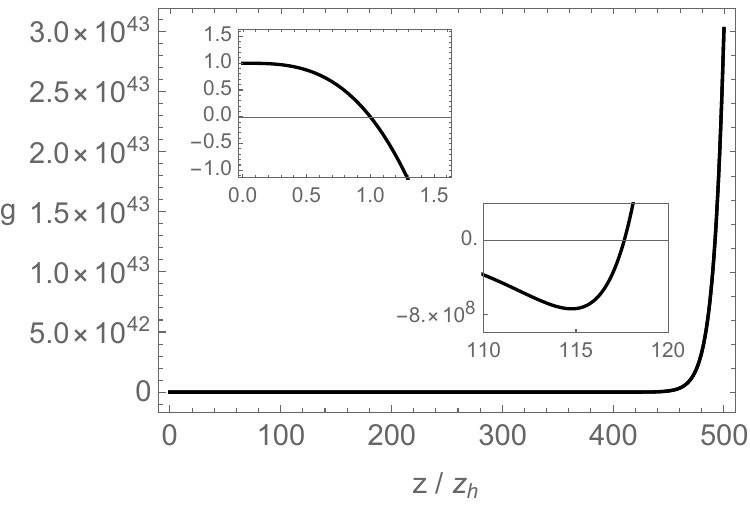} \label{}}
\quad
     \subfigure[$g(z)$ at $\d =\d_c$]
     {\includegraphics[width=7.0cm]{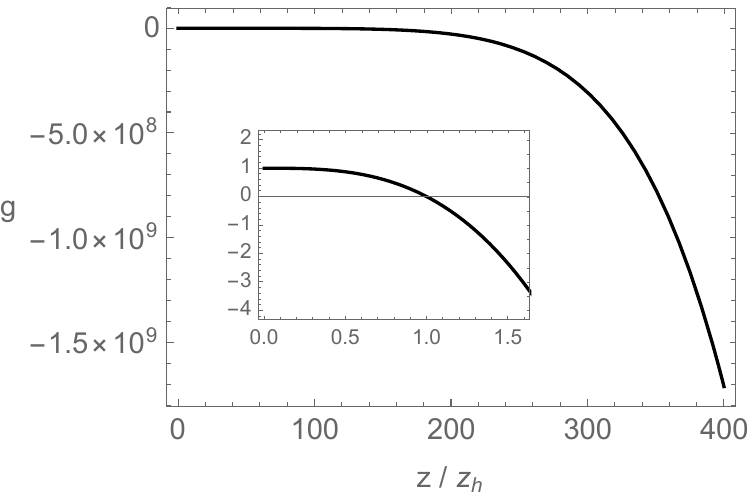} \label{}}
     \subfigure[$A_t(z)$ at $\d ={\d_c}/{3}$]
     {\includegraphics[width=6.5cm]{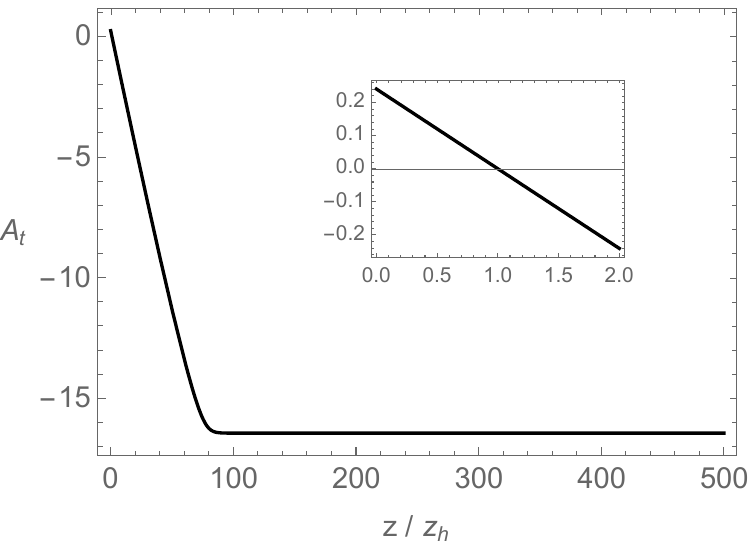} \label{}}
\quad
     \subfigure[$A_t(z)$ at $\d =\d_c$]
     {\includegraphics[width=6.5cm]{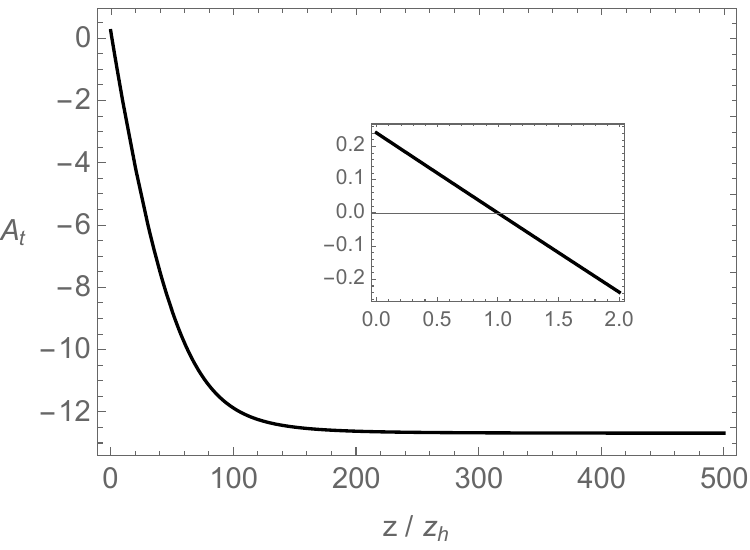} \label{}}
\caption{Holographic flows of the blackening factor $g(z)$ and gauge field $A_t(z)$ for $d=3$. The left column ((a), (c)) corresponds to geometries with $\d ={\d_c}/{3}$, while the right one ((b), (d)) to $\d =\d_c$. Here $\d_c = \sqrt{1/3}$. The insets show the behavior near the horizon.}\label{NMFIG1}
\end{figure}

\begin{figure}[]
 \centering
     \subfigure[$z \, \dd (\log g_{tt}')/\dd z$ at $\d ={\d_c}/{3}$]
     {\includegraphics[width=7.0cm]{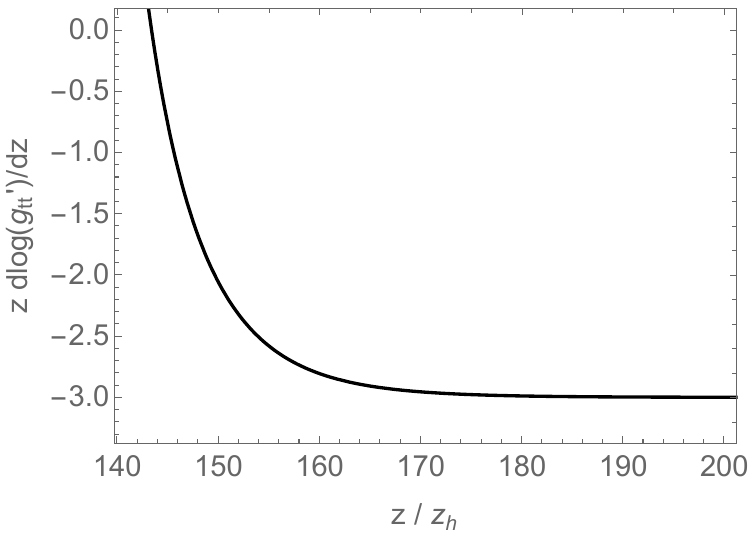} \label{}}
\quad
     \subfigure[$z \, \dd (\log g_{tt}')/\dd z$ at $\d ={\d_c}$]
     {\includegraphics[width=7.0cm]{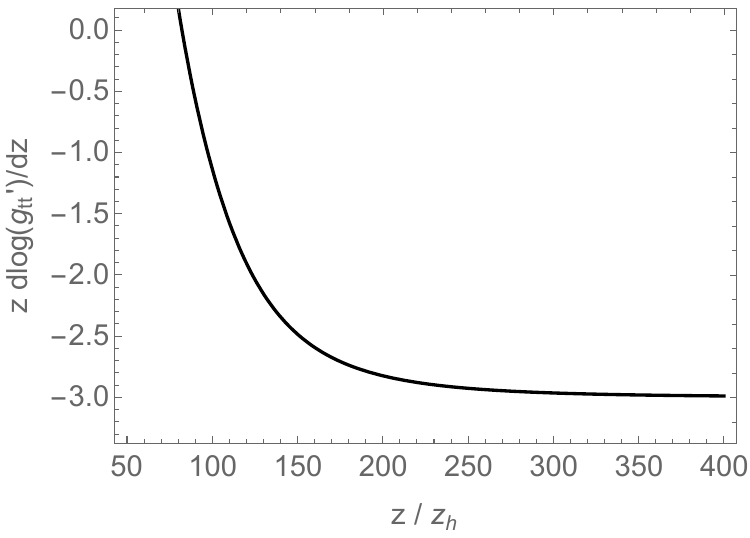} \label{}}
\quad
     \subfigure[$z \, \dd \chi/\dd z$ at $\d ={\d_c}/{3}$]
     {\includegraphics[width=7.0cm]{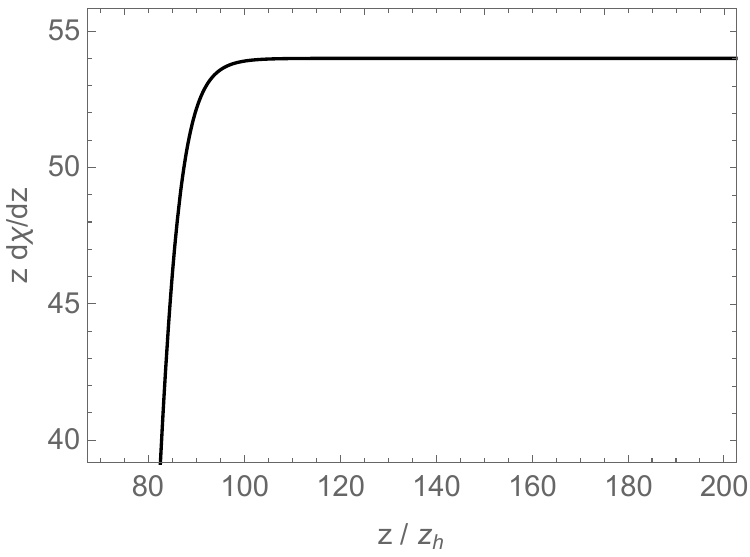} \label{}}
     \subfigure[$z \, \dd \chi/\dd z$ at $\d ={\d_c}$]
     {\includegraphics[width=7.0cm]{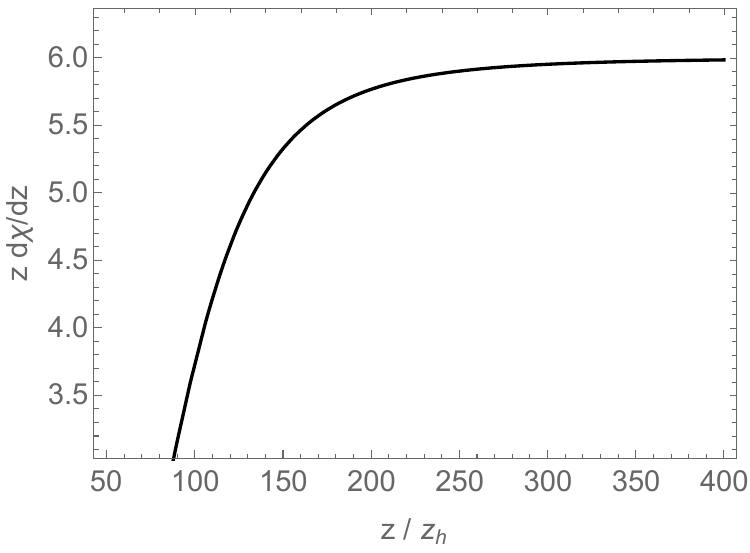} \label{}}
\quad
     \subfigure[$z \, \dd \phi/\dd z$ at $\d ={\d_c}/{3}$]
     {\includegraphics[width=7.0cm]{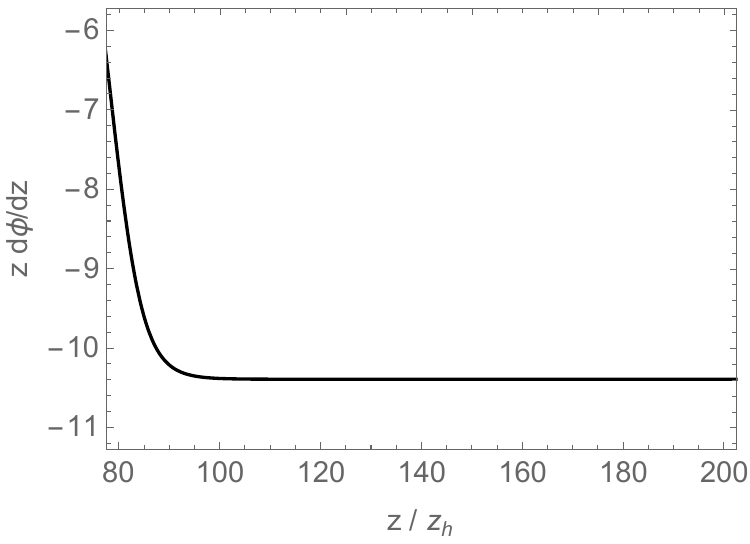} \label{}}
\quad
     \subfigure[$z \, \dd \phi/\dd z$ at $\d ={\d_c}$]
     {\includegraphics[width=7.0cm]{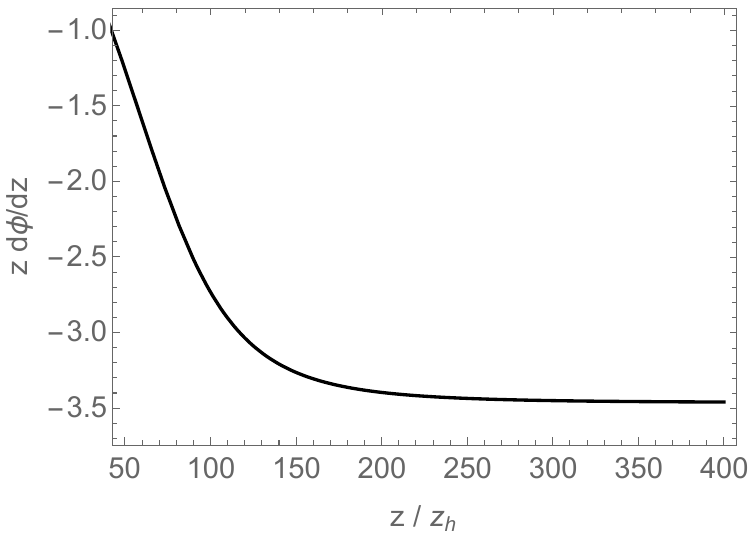} \label{}}
\caption{The functions $z \, \dd X/\dd z$ when $X = \left\{ \log g_{tt}',\, \chi,\, \phi \right\}$. The left column ((a), (c), (e)) is at $\d ={\d_c}/{3}$, while the right one ((b), (d), (f)) is at $\d =\d_c$. Here $\d_c = \sqrt{1/3}$. The saturated value near the singularity ($z\rightarrow\infty$) is \eqref{APPVAL}.}\label{NMFIG2}
\end{figure}

First, in Fig. \ref{NMFIG1} we plot the blackening factor $g(z)$ and the gauge field $A_t$. Additionally,
the metric function $\chi(z)$ and the dilaton field $\phi(z)$
are shown in Fig.~\ref{NMFIG2}.
The plots of $g(z)$ clearly show the existence of an inner horizon for the case $\d ={\d_c}/{3}$, as expected since this value falls in the range \eqref{CRUCON3}. Instead, for $\d = \d_c$ only the event horizon is present. One can also check in Fig.~\ref{NMFIG1} that
the gauge field $A_t$ vanishes only at the event horizon.\footnote{It is worth noting that, unlike our neutral scalar field theory, in scenarios involving non-trivial charged scalar hairs, the gauge field can also vanish at the inner horizon. This phenomenon is demonstrated in studies such as~\cite{Cai:2020wrp,An:2021plu}. For the same reason, Josephson oscillations do not occur in our dilaton fields \cite{Mansoori:2021wxf}, as illustrated in Fig. \ref{NMFIG2}.}

We shall next present results for three functions useful for characterizing the geometry near the singularity:
\begin{align}\label{THF}
\begin{split}
  z \frac{\dd}{\dd z} \log (g_{tt}')  \,, \qquad z \frac{\dd}{\dd z} \chi \,, \qquad  z \frac{\dd}{\dd z} \phi \,,
\end{split}
\end{align}
where
\begin{align}\label{}
\begin{split}
  g_{tt}(z) = -\frac{g(z) e^{-\chi(z)}}{z^2} \,.
\end{split}
\end{align}
The near singularity analysis of section~\ref{ssec:singularities} (see eqs.
\eqref{SG1}, \eqref{SG2}, and \eqref{SG3}) implies that these
functions \eqref{THF} approach a constant towards the singularity
\begin{align}\label{APPVAL}
\begin{split}
   z \frac{\dd}{\dd z} \log (g_{tt}') & \approx 
\begin{cases}
d-3 - \frac{2}{(d-1)\d^2} \,, \qquad\, (\text{Class I})  \\
-3  \qquad\qquad\qquad\qquad (\text{Class II \,\&\, III}) \,,
\end{cases}
\\
z \frac{\dd}{\dd z} \chi &\approx \frac{4}{(d-1)\d^2}\,, \qquad  z \frac{\dd}{\dd z} \phi \approx -\frac{2}{\d} \,.
\end{split}
\end{align}
Note that for the case of the Kasner singularity, these constants can be associated with the Kasner exponent via \eqref{KAS1} and \eqref{KAS2}.
In Fig. \ref{NMFIG2} we display the functions \eqref{THF} inside the event horizon.
As expected, all three functions approach the constant values~\eqref{APPVAL} towards the singularity.

\begin{figure}[]
 \centering
     \subfigure[$g_{tt}$ vs. $z$ at $\d ={\d_c}/{3}$]
     {\includegraphics[width=7.1cm]{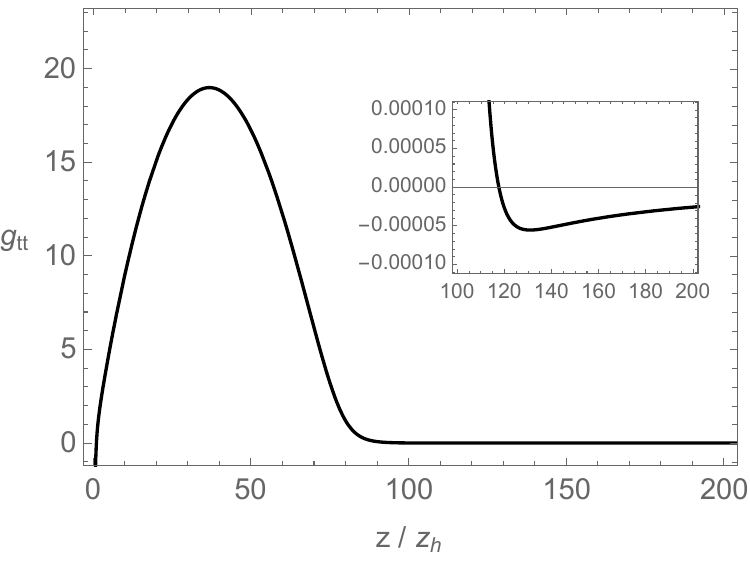} \label{}}
\quad
     \subfigure[$g_{tt}$ vs. $z$ at $\d ={\d_c}$]
     {\includegraphics[width=7.0cm]{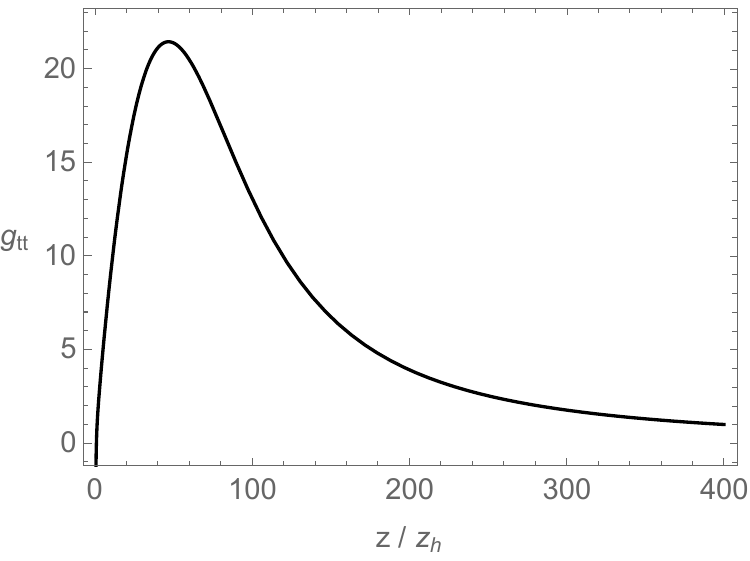} \label{}}
\caption{The plot of $g_{tt}$ when $\d = \left({\d_c}/{3},\, \d_c \right)$ (left, right). The left panel shows the inner horizon in its inset, which can be compared with the plot of the blackening factor in Fig. \ref{NMFIG1} (a).}\label{NMFIG3}
\end{figure}

We close this section with the plot of $g_{tt}(z)$ in Fig.~\ref{NMFIG3}.
For both cases in \eqref{TWOEX} 
we find that $g_{tt}$ reaches a maximum value within the interior region and subsequently diminishes.\footnote{A similar behavior of $g_{tt}$ is found in other models. 
For instance, see \cite{DeClerck:2023fax} for the case of massive vector fields.}
Moreover, as expected, an inner horizon appears only when $\d < \d_c$ (left plot).
Accordingly, on the right plot of Fig.~\ref{NMFIG3}, $g_{tt}$ does not vanish (indicating no inner horizon) and instead remains very small.
This phenomenon can be associated with the collapse of the Einstein-Rosen bridge~\cite{Hartnoll:2020rwq}. Recall that, in the interior of a black hole, $g_{tt}$ serves as a measure for the spatial coordinate $t$ along the wormhole that connects the two exterior regions of the black hole, i.e., the Einstein-Rosen bridge. A rapid decrease in $g_{tt}$ may therefore signify a collapse of the Einstein-Rosen bridge.

%
\section{Probes of the black hole interior}\label{SEC3}

In this section, we now investigate gravitational observables that can probe the interior of the black holes. We focus on two recent proposals for diagnosing the black hole singularity: the ``Complexity = Anything'' proposal~\cite{Jorstad:2023kmq} and the thermal $a$-function~\cite{Caceres:2022smh}.

%
\subsection{Complexity = Anything}

Quantum complexity is an interesting entry of the holographic dictionary, which has been highlighted by recent research in the intersection of quantum information theory and quantum gravity~\cite{Susskind:2014moa,Susskind:2018pmk}. Initially, two main conjectures were proposed for its holographic dual: ``Complexity = Volume'' \cite{Susskind:2014rva,Stanford:2014jda} and ``Complexity = Action'' \cite{Brown:2015bva,Brown:2015lvg}. Recently, a new infinite family of gravitational observables, coined ``Complexity = Anything'' \cite{Belin:2021bga,Belin:2022xmt}, has been introduced, encompassing the previous conjectures as specific cases.

A major motivation for studying quantum complexity within holography is to enhance our understanding of the emergence of spacetime through quantum information \cite{Czech:2017ryf,Susskind:2019ddc,Pedraza:2021mkh,Pedraza:2021fgp,Pedraza:2022dqi,Carrasco:2023fcj}, in particular black hole interiors. Recent work \cite{Jorstad:2023kmq} has shown that the ``Complexity = Anything'' approach \cite{Belin:2021bga,Belin:2022xmt} provides a framework for examining the geometric properties of black hole singularities. For a recent review, see \cite{Myers:2024vve}. Specifically, by focusing on the late-time growth rate of complexity, the authors of \cite{Jorstad:2023kmq} demonstrated that, for a particular class of complexity functionals, the extremal surfaces relevant to the computation of complexity can probe the black hole singularity. This suggests that the geometric structure of the singularities may determine certain properties of these surfaces as well as the associated dual complexities.

In this paper, following the methodology of \cite{Jorstad:2023kmq}, we investigate holographic complexity as a tool for probing black hole singularities. Our analysis provides novel insights in two key areas. First, we extend the formalism of \cite{Jorstad:2023kmq} to a generic metric of the form \eqref{OURMETRIC}; when $D(r)=1/B(r)=f(r)$ and $C(r)=r^2$, our results align with the simpler setting in \cite{Jorstad:2023kmq}. Second, we extend their findings to include black holes with scalar hair, which is an essential ingredient for obtaining more generic and intricate black hole interiors. In particular, using our analytic AdS black holes \eqref{GGR}, we provide explicit analytic examples illustrating the relationship between holographic complexity and the structure of general Kasner singularities.

%
\subsubsection{Formalism and general results}

In this section, we extend the ``Complexity = Anything'' formalism from \cite{Jorstad:2023kmq} to accommodate our generic metric \eqref{OURMETRIC}. This will allow us to probe a generic Kasner singularity in an analytically controllable setup.

Let us first summarize the main ingredients of the ``Complexity = Anything'' proposal \cite{Belin:2021bga,Belin:2022xmt}. The idea is to study a family of codimension-zero or codimension-one observables with certain universal properties that are expected for complexity; that is, linear growth of complexity at late times and switchback effect. Without loss of generality, we will focus on the codimension-zero case, as codimension-one proposals can be obtained from particular limiting cases of the former. 

The definition of complexity follows a two-step procedure. First, one needs to specify a bulk codimension-zero region $\mathcal{M}$. To do so, one defines a functional of the form
\begin{equation}
W(\mathcal{M})=\int_{\Sigma_{+}}\!\! \dd^d \sigma \sqrt{h}\,F_{+}(g_{\mu\nu},X_{+}^{\mu}) + \int_{\Sigma_{-}}\!\!  \dd^d \sigma \sqrt{h}\,F_{-}(g_{\mu\nu},X_{-}^{\mu}) + \int_{\mathcal{M}}\!\!  \dd^{d+1} x \sqrt{g}\,G(g_{\mu\nu})\,,\label{defcalM}
\end{equation}
where $\Sigma_{\pm}$ are future and past boundaries of $\mathcal{M}$ respectively ---see Fig. \ref{CMCFIG1} below,--- $F_{\pm}$ are general functionals of the metric $g_{\mu\nu}$ and embedding functions $X^\mu_\pm$, and $G$ is a function of $g_{\mu\nu}$. The region $\mathcal{M}$ can then be found by extremizing such a functional, such that\footnote{Using stokes theorem, it can be shown that the above extremization can always be cast as independent extremizations of the functions $F_{\pm}\pm \tilde{G}$, where $\tilde{G}$ is the primitive of $G$.}
\begin{figure}[]
 \centering
     {\includegraphics[width=7.8cm]{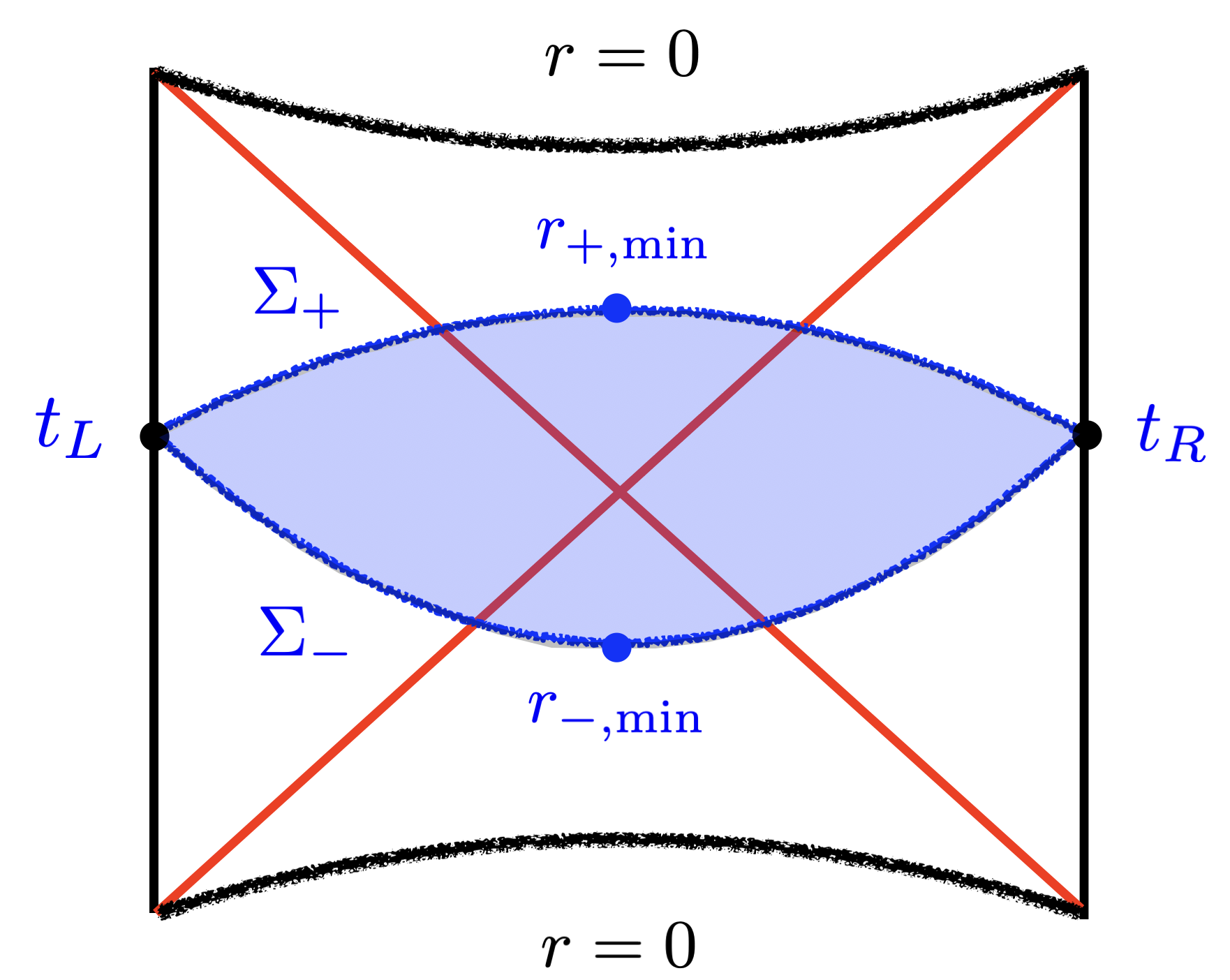} \label{}}
\caption{A sketch of the codimension-zero region $\mathcal{M}$ (blue shaded region) together with future and past boundaries $\Sigma_{\pm}$ (blue lines). Below, we will specialize in a set of proposals where $\Sigma_{\pm}$ are taken to be constant
mean curvature slices. The parameters $r_{\pm, \text{min}}$ represent the minimum values attained by $r_{\pm}$, while $t_{L,\,R}$ denote the left/right boundary times.}\label{CMCFIG1}
\end{figure}
\begin{equation}
\delta[W(\mathcal{M})]=0\,.
\end{equation}
To solve this extremization problem, one should impose boundary conditions such that $\mathcal{M}$ is anchored to the relevant boundary Cauchy slice $\sigma_{\text{CFT}}$. This is guaranteed by requiring that $\partial\Sigma_\pm=\sigma_{\text{CFT}}$. In a two-sided black hole $\sigma_{\text{CFT}}$ has two components so that $\sigma_{\text{CFT}}=\sigma_{L}\cup \sigma_{R}$, and typically $\sigma_{L,R}$ are taken to be constant-$t$ slices: $t=t_L$ and $t=t_R$, respectively.

The second step amounts to evaluating a ``complexity observable'' on region $\mathcal{M}$, so that
\begin{equation}
\mathcal{C}(\sigma_{\text{CFT}})=\int_{\Sigma_{+}}\!\! \dd^d \sigma \sqrt{h}\,\mathcal{F}_{+}(g_{\mu\nu},X_{+}^{\mu}) + \int_{\Sigma_{-}}\!\!  \dd^d \sigma \sqrt{h}\,\mathcal{F}_{-}(g_{\mu\nu},X_{-}^{\mu}) + \int_{\mathcal{M}}\!\!  \dd^{d+1} x \sqrt{g}\,\mathcal{G}(g_{\mu\nu})\,.
\end{equation}
Here, $\mathcal{F}_{\pm}$ are scalar functionals that may be constructed from the bulk intrinsic curvature and/or the extrinsic curvatures of $\Sigma_\pm$. Similarly, $\mathcal{G}$, is constructed from the bulk intrinsic curvature. In
general, $\mathcal{F}_{\pm}$ and $\mathcal{G}$ are independent and may be chosen to be completely different from the functionals appearing in eq. (\ref{defcalM}).

The above prescription defines an infinite family of gravitational observables that probe the black hole's interior. However, this two-step process does not ensure the results meet the expected properties for complexity. On general grounds, we expect a ``good'' measure of complexity must exhibit linear growth at late times and the switchback effect \cite{Belin:2021bga,Belin:2022xmt}, but these properties must be verified on a case-by-case basis. Even with these constraints, there remains a large family of proposals from the bulk perspective that capture the main features of complexity. This flexibility in defining holographic complexity should be viewed as a feature, rather than a drawback, reflecting the inherent ambiguities in defining complexity from the field theory side.

Below, we will examine specific complexity proposals that can probe the black hole's singularity. Since $\mathcal{M}$ includes regions outside and inside the event horizon, using Eddington-Finkelstein coordinates is advantageous. The metric in these coordinates is given by:
\begin{equation}\label{EddiE}
\dd s^2 = -D(r) \, \dd v^2 + 2\sqrt{D(r)B(r)} \, \dd v \, \dd r  + C(r)\dd \vec x_{i}^{2} \,,
\end{equation}
where the infalling coordinate $v$ is defined as
\begin{equation}\label{IFCC}
v = t + r_*(r) \,, \qquad r_*(r) = \int_{\infty}^{r} \sqrt{\frac{B(\tilde{r})}{D(\tilde{r})}} \, \dd \tilde{r} \,.
\end{equation}
We will eventually restore the original coordinates using this relation to investigate the evolution of complexity in terms of the (physical) boundary time $t$.

\paragraph{Observables on constant mean curvature slices.}
Following \cite{Jorstad:2023kmq}, we now focus on a codimension-one variant of Complexity = Anything, which can be obtained as a limiting case of the more general codimension-zero case outlined above. To define the gravitational observable, we first need to define the region of interest $\mathcal{M}$. This is obtained by extremizing the functional \cite{Belin:2022xmt}
\begin{equation}\label{CCMC}
W_{\text {CMC }} = \frac{1}{G_{\mathrm{N}} L}\left[\alpha_{+} \int_{\Sigma_{+}}\!\! \dd^d \sigma \sqrt{h} + \alpha_{-} \int_{\Sigma_{-}}\!\! \dd^d \sigma \sqrt{h} + \frac{\alpha_{\mathrm{B}}}{L} \int_{\mathcal{M}}\! \dd^{d+1} x \sqrt{g}\right] \,,
\end{equation}
where $G_N$ is the Newton constant, $L$ the AdS radius, and $\alpha_{\pm}$ and $\alpha_{B}$ some dimensionless constants. The embedding functions of the future and past hypersurfaces bounding $\mathcal{M}$ are taken to be $X^\mu_\pm=\{v_\pm(\sigma),r_\pm(\sigma),x^i\}$ and the induced metric on $\Sigma_{\pm}$ is
\begin{equation}\label{EddiE2}
    \dd s_{h}^2 = \left(-D(r_{\pm})\dot{v}_{\pm}^2 + 2 \sqrt{D(r_{\pm})B(r_{\pm})} \dot{v}_{\pm}\dot{r}_{\pm}\right) \, \dd \sigma^2+C(r_{\pm}) \, \dd \vec x_{i}^{2} \,.
\end{equation}
By extremizing the above functional, the resulting hypersurfaces $\Sigma_{\pm}$ are found to be Constant Mean Curvature (CMC) slices \cite{Belin:2022xmt,Jorstad:2023kmq}, with extrinsic curvatures given by
\begin{equation}\label{KSFOR}
K_{\Sigma_{+}}=-\frac{\alpha_{\mathrm{B}}}{\alpha_{+} L}, \qquad K_{\Sigma_{-}}=\frac{\alpha_{\mathrm{B}}}{\alpha_{-} L} \,.
\end{equation}
To define the codimension-one observable, we then set $\mathcal{G}$ and either $\mathcal{F}_+$ or $\mathcal{F}_-$ to zero. This results in the following complexity observable, evaluated on a single hypersurface:
\begin{align}\label{CANYE}
\begin{split}
\mathcal{C}^{\pm} = \frac{1}{G_NL} \int_{\Sigma_{\pm}}\!\! \mathrm{d}^d \sigma    \sqrt{h}  \,  \mathcal{F}(\sigma)  \,,
\end{split}
\end{align}
where $\mathcal{F}(\sigma)$ is a dimensionless scalar function.

\paragraph{The maximization of CMC slices.}
Next, we provide an algorithm to obtain the CMC slices given a metric of the form (\ref{EddiE}). We start by rewriting the bulk term in \eqref{CCMC} as
\begin{equation}\label{THRTERM}
\begin{aligned}
\frac{\alpha_{\mathrm{B}}}{L} \int_{\mathcal{M}}\dd^{d+1} x \sqrt{g} & =\frac{\alpha_{\mathrm{B}}}{L} \int \dd^{d-1} x \int \dd v \int_{r_{+}}^{r_{-}} \dd r\sqrt{D(r)B(r)C(r)^{d-1}}   \\
& ={\alpha_{\mathrm{B}} V_{d-1}} \left[\int_{\Sigma_{-}} \dd \sigma \, \dot{v}_{-}(\sigma) \, b(r_-) \,-\, \int_{\Sigma_{+}} \dd \sigma \, \dot{v}_{+}(\sigma) \, b(r_+) \right] \,,
\end{aligned}
\end{equation}
where we defined a function $b(r)$ and the spatial volume $V_{d-1}$ such that
\begin{equation}\label{REDBV}
\frac{\partial b(r)}{\partial r} = \frac{\sqrt{D(r)B(r)C(r)^{d-1}}}{L}\,, \qquad V_{d-1} = \int \dd^{d-1} x \,.
\end{equation}
Then, plugging \eqref{THRTERM} into \eqref{CCMC}, we obtain 
\begin{equation}\label{MAXSIG}
W_{\text {CMC}} =  \frac{V_{d-1}}{G_{\mathrm{N}}L} \sum_{\varepsilon=+,-} \int_{\Sigma_{\varepsilon}} \dd \sigma 
\, \mathcal{L}_{\varepsilon}  \,,
\end{equation}
where
\begin{equation}
\mathcal{L}_{\varepsilon} =\alpha_\varepsilon \, C(r_{\pm})^{\frac{d-1}{2}} \sqrt{-D(r_{\pm}) \, \dot{v}_{\pm}^2+2 \sqrt{D(r_{\pm})B(r_{\pm})} \, \dot{v}_{\pm} \dot{r}_{\pm} }   -\varepsilon \, \alpha_{\mathrm{B}} \, \dot{v}_{\pm} \, b\left(r_{\pm} \right)\,.
\end{equation}
Thus, finding the corresponding CMC slices (i.e., extremal surfaces of the functional $W_{\text{CMC}}$) is equivalent to solving the equations of motion derived from the ``Lagrangian'' $\mathcal{L}_{\varepsilon}$ in \eqref{MAXSIG}.
To achieve this, it is useful to fix the gauge by choosing
\begin{equation}\label{gaugeeq}
\sqrt{-D(r_{\pm})\dot{v}_{\pm}^2+2 \sqrt{D(r_{\pm})B(r_{\pm})} \dot{v}_{\pm}\dot{r}_{\pm}}=C(r_{\pm})^{\frac{d-1}{2}}\sqrt{B(r_{\pm})D(r_{\pm})} \,.
\end{equation}
This is possible because the functional $W_{\text{CMC}}$ is diffeomorphism invariant under the transformation $\sigma \rightarrow \tilde{\sigma}(\sigma)$~\cite{Jorstad:2023kmq}. By using \eqref{gaugeeq} in the Lagrangian $\mathcal{L}_{\varepsilon}$, one can derive the equation of motion for the radial profile $r_{\pm}(\sigma)$ as:
\begin{equation}\label{emoeq}
\dot{r}_{\pm}^2+\mathcal{U}\left(P_v^{ \pm}, r_{\pm} \right)=0 \,,
\end{equation}
where the ``potential'' $\mathcal{U}$ is given by
\begin{equation}\label{Ufunction}
\mathcal{U}\left(P_v^{ \pm}, r_{\pm}\right) = -D(r_{\pm})C(r_{\pm})^{d-1}-\left(P_v^{ \pm}\pm\frac{\alpha_{\mathrm{B}}}{\alpha_\pm} \, b(r_{\pm})\right)^2 \,,
\end{equation}
and the conjugate conserved momenta $P_v^\pm$ are
\begin{equation}\label{consereq}
P_v^{ \pm} = \frac{\partial \mathcal{L}_{ \pm}}{\partial \dot{v}_{\pm}}=\dot{r}_{\pm}-\sqrt{\frac{D(r_{\pm})}{B(r_{\pm})}} \dot{v}_{\pm} \mp \frac{\alpha_{\mathrm{B}}}{\alpha_\pm} \, b(r_{\pm}) \,.
\end{equation}
Finally, by solving \eqref{emoeq} for a given value of $P_v^{\pm}$, we can determine $r_{\pm}$. Substituting this function back into \eqref{consereq}, we can also derive $v_{\pm}$. Consequently, we obtain the full embedding functions $\{r_{\pm}(\sigma)\,, v_{\pm}(\sigma)\}$ for the CMC surfaces $\Sigma_{\pm}$ that maximize \eqref{CCMC}.

\begin{figure}[t]
 \centering
     {\includegraphics[width=12.0cm]{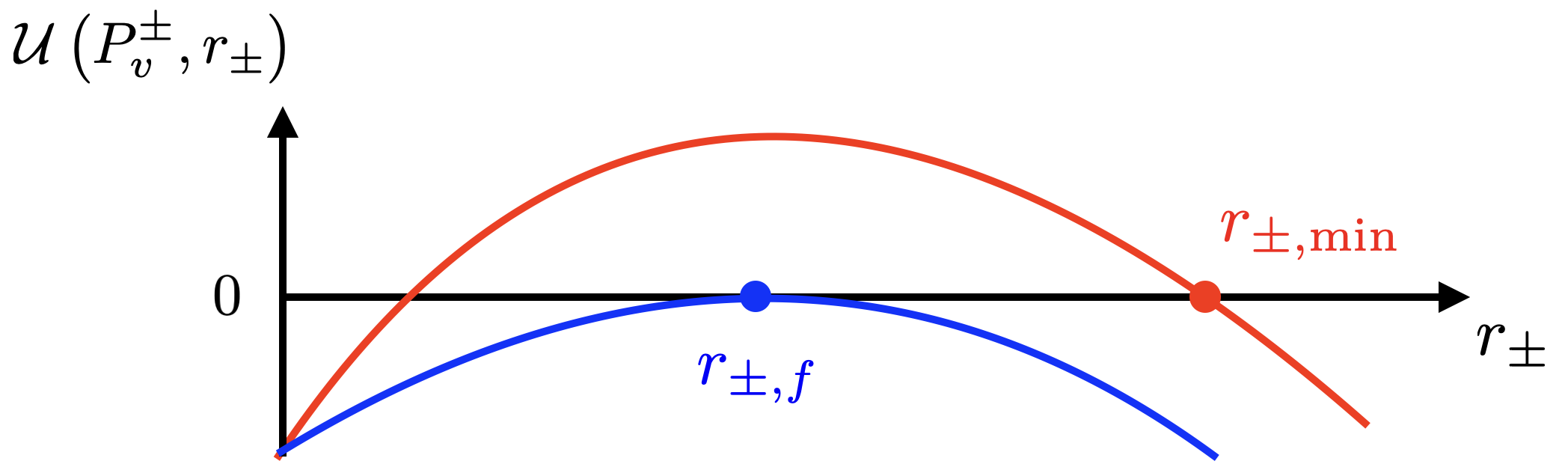} \label{}}
\caption{A sketch of the potential $\mathcal{U}\left(P_v^{ \pm}, r_{\pm}\right)$ in \eqref{Ufunction}. Different colors denote the distinct values of the momentum $P_v^{ \pm}$, and $r_{\pm, \text{min}}$ are the zeros furthest away from the singularity.}\label{SKETU}
\end{figure}

\paragraph{Boundary time and conserved momentum.} 
Let us now discuss the physical meaning of the momenta $P_v^{ \pm}$. First, note that the potential can exhibit zeros at a fixed $P_v^{ \pm}$, i.e.,
\begin{equation}\label{RMINFOR}
\mathcal{U}\left(P_v^{ \pm}, r_{\pm}\right) = 0 \,.
\end{equation}
We denote $r_{\pm, \text{min}}$ as the zeros that are furthest away from the singularity ---see Fig. \ref{SKETU}.\footnote{Note that the minimal radius $r_{\pm, \text{min}}$ is defined as the point at which trajectories with ``zero-energy'' reflect off the potential~\cite{Belin:2021bga,Belin:2022xmt,Jorstad:2023kmq}.}%
Additionally, note that when $\mathcal{U}=0$, \eqref{emoeq} simplifies to $\dot{r}{_\pm}^2=0$. This indicates that the values of $r_{\pm}$ correspond to the minima $r_{\pm, \text{min}}$ as shown in Fig. \ref{CMCFIG1}. Furthermore, it has been shown \cite{Jorstad:2023kmq} that fixing the momentum is equivalent to fixing a boundary time. To illustrate this, we rewrite equation \eqref{IFCC} as follows:
\begin{align}
\begin{split}
\dd t &=\frac{\dot{v}_{\pm}}{\dot{r}_{\pm}} \, \dd r_{\pm}-\sqrt{\frac{B(r_{\pm})}{D(r_{\pm})}} \, \dd r_{\pm} \\
&=-\frac{\sqrt{B(r_{\pm})}\left(P_v \pm \frac{\alpha_B}{\alpha_\pm} b(r_{\pm})-\dot{r}_{\pm}\right)}{\sqrt{D(r_{\pm})}\dot{r}_{\pm}} \, \dd r_{\pm}-\sqrt{\frac{B(r_{\pm})}{D(r_{\pm})}} \, \dd r_{\pm} \\
&=-\frac{\sqrt{B(r_{\pm})}\left(P_v \pm \frac{\alpha_B}{\alpha_\pm} b(r_{\pm})\right)}{\sqrt{D(r_{\pm})}\sqrt{-\mathcal{U}\left(P_v^{ \pm}, r_{\pm}\right)}} \, \dd r_{\pm} \,,
\end{split}
\end{align}
where we used both \eqref{emoeq} and \eqref{consereq}.
Then, considering left-right
symmetric time evolution, $t_R=t_L\equiv t_b/2$, the corresponding boundary time $t_b$ yields
\begin{equation}\label{tau}
t_b = 2 \int_{\infty}^{r_{\pm,\min }} \frac{\sqrt{B(r_{\pm})}\left(P_v^\pm \pm \frac{\alpha_B}{\alpha_\pm} b(r_{\pm})\right)}{\sqrt{D(r_{\pm})}\sqrt{-\mathcal{U}\left(P_v^{ \pm}, r_{\pm}\right)}}\dd r_{\pm} \,.
\end{equation}
Thus, fixing momentum is equivalent to fixing the boundary time.

\paragraph{The late-time limit and large mean curvature.}
A key concept in investigating singularities through Complexity = Anything involves analyzing the late-time behavior of complexity using equation \eqref{tau} \cite{Jorstad:2023kmq}. To determine the behavior at late times ($t_b \to \infty$), it is essential to discuss the expansion of the potential near $r_{\pm, \text{min}}$,
\begin{align}\label{UEXPFOR}
\begin{split}
\mathcal{U}\left(P_v^\pm, r_{\pm} \right) = \,  
&\, \partial_{r_{\pm}} \mathcal{U}\left(P_v^\pm, r_{\pm}\right) \big|_{r_{\pm}=r_{\pm, \text{min}}} \left(r_{\pm}-r_{\pm, \text{min}}\right) \\
&\,+\, \frac{1}{2} \partial_{r_{\pm}}^{2} \mathcal{U}\left(P_v^\pm, r_{\pm}\right) \big|_{r_{\pm}=r_{\pm, \text{min}}} \left(r_{\pm}-r_{\pm, \text{min}}\right)^2 \,+\, \,\cdots\,  \,,
\end{split}
\end{align}
where we used that $\mathcal{U}\left(P_v^\pm, r_{\pm, \text{min}}\right) = 0$.
Next, we identify the `final' zero, $r_{\pm, f}$, such that
\begin{equation}\label{RPMFFOR}
\mathcal{U}\left(P_v^\pm, r_{\pm,f}\right)
\,=\,
\left.\partial_{r_{\pm}} \mathcal{U}\left(P_v^\pm, r_{\pm}\right)\right|_{{r_{\pm}}=r_{\pm,f}}
\,=\, 0\,.
\end{equation}
Then, due to the quadratic term in \eqref{UEXPFOR}, the integral in \eqref{tau} may become divergent, resulting in an infinite $t_b$ ~\cite{Belin:2021bga,Belin:2022xmt,Jorstad:2023kmq}. See Fig. \ref{SKETU} for a sketch of $r_{\pm, f}$. Therefore, determining $r_{\pm, f}$ is crucial for understanding the late-time behavior of complexity.

The equation to determine $r_{\pm, f}$ can be derived by combining \eqref{Ufunction} and \eqref{RPMFFOR}, yielding:
\begin{equation}\label{finaleq}
K_{\Sigma_{\pm}}^2 + \frac{\left[C(r_{\pm,f}) D^{\prime}(r_{\pm,f}) + (d-1) D(r_{\pm,f}) C^{\prime}(r_{\pm,f})\right]^2}{4 B(r_{\pm,f}) D(r_{\pm,f})^2 C(r_{\pm,f})^2}=0 \,.
\end{equation}
Alternatively, this equation can be derived from the formula for the extrinsic curvature of a constant $r$-slice:
\begin{equation}\label{extcureq}
\begin{aligned}
     K&=\abs{\frac{C(r) D^{\prime}(r)+(d-1) D(r) C^{\prime}(r)}{2 \sqrt{-B(r)} D(r)  C(r)}},
    \end{aligned}
\end{equation}
evaluated at $r = r_{\pm,f}$. 

An intriguing insight into the black hole singularity emerged from the work of \cite{Jorstad:2023kmq}. In particular, in the limit of large mean curvature, $
K_{\Sigma_{\pm}} \,\,\rightarrow\,\,  \pm \, \infty \,,
$
one can show that 
\begin{align}\label{DFINALRELA}
\begin{split}
   r_{\pm,f} \,\rightarrow\,
\begin{cases}
\,\,0 \,, \quad\,\,  (r_{\pm,f} \text{ is the singularity})\,,  \\
\,\,r_h  \,, \quad  (r_{\pm,f} \text{ is the horizon}) \,.
\end{cases}
\end{split}
\end{align}
In other words, in the large mean curvature and late-time limits, the CMC slices $\Sigma_{\pm}$ asymptotically approach a constant $r$-surface where $r_{\pm}(\sigma) = r_{\pm,f}$, coinciding with either the singularity or the horizon.
\begin{figure}[]
 \centering
     \subfigure[At a fixed finite time]
     {\includegraphics[width=6.5cm]{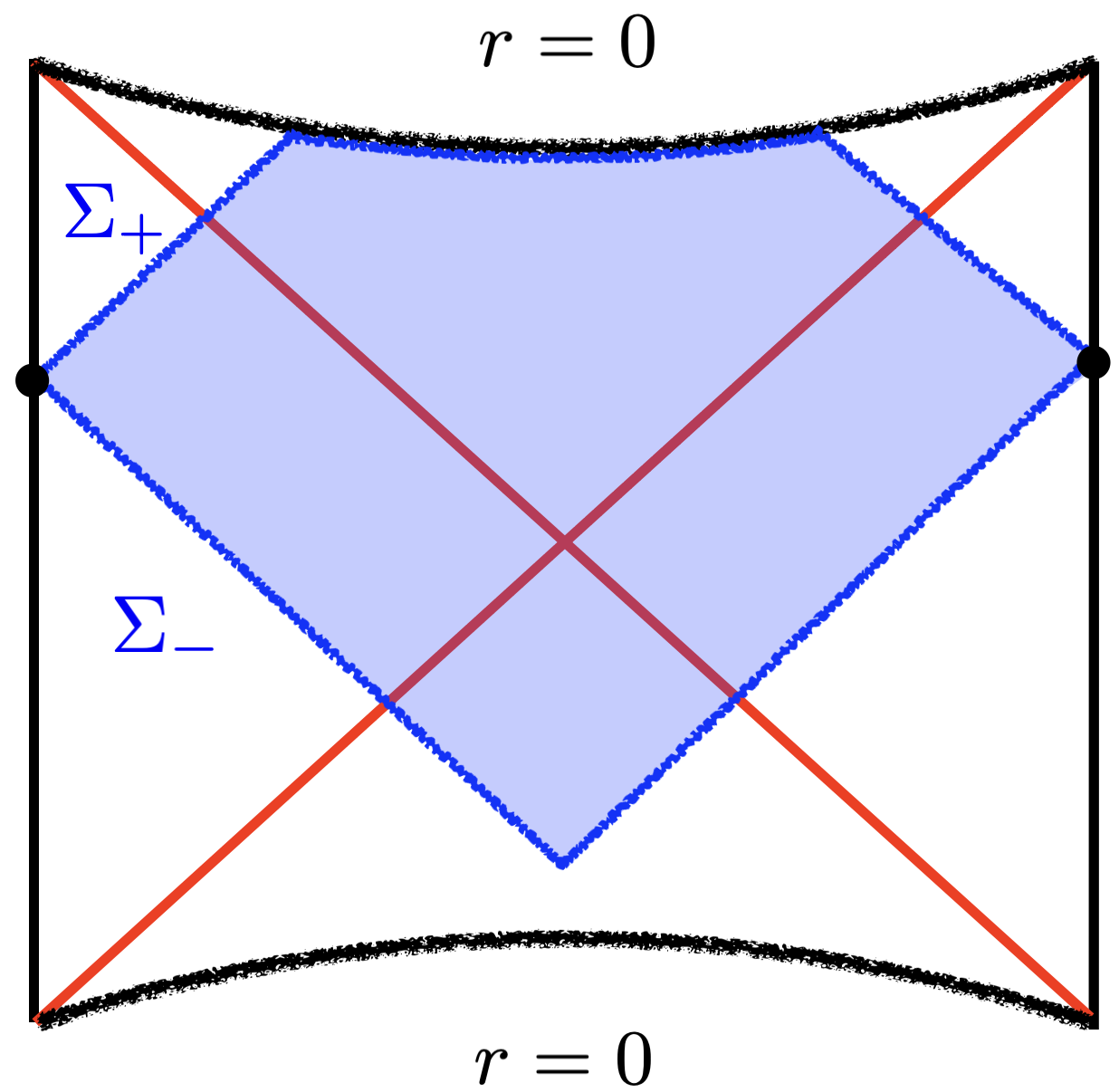} \label{}}
\qquad\quad
     \subfigure[At the late-time limit]
     {\includegraphics[width=6.68cm]{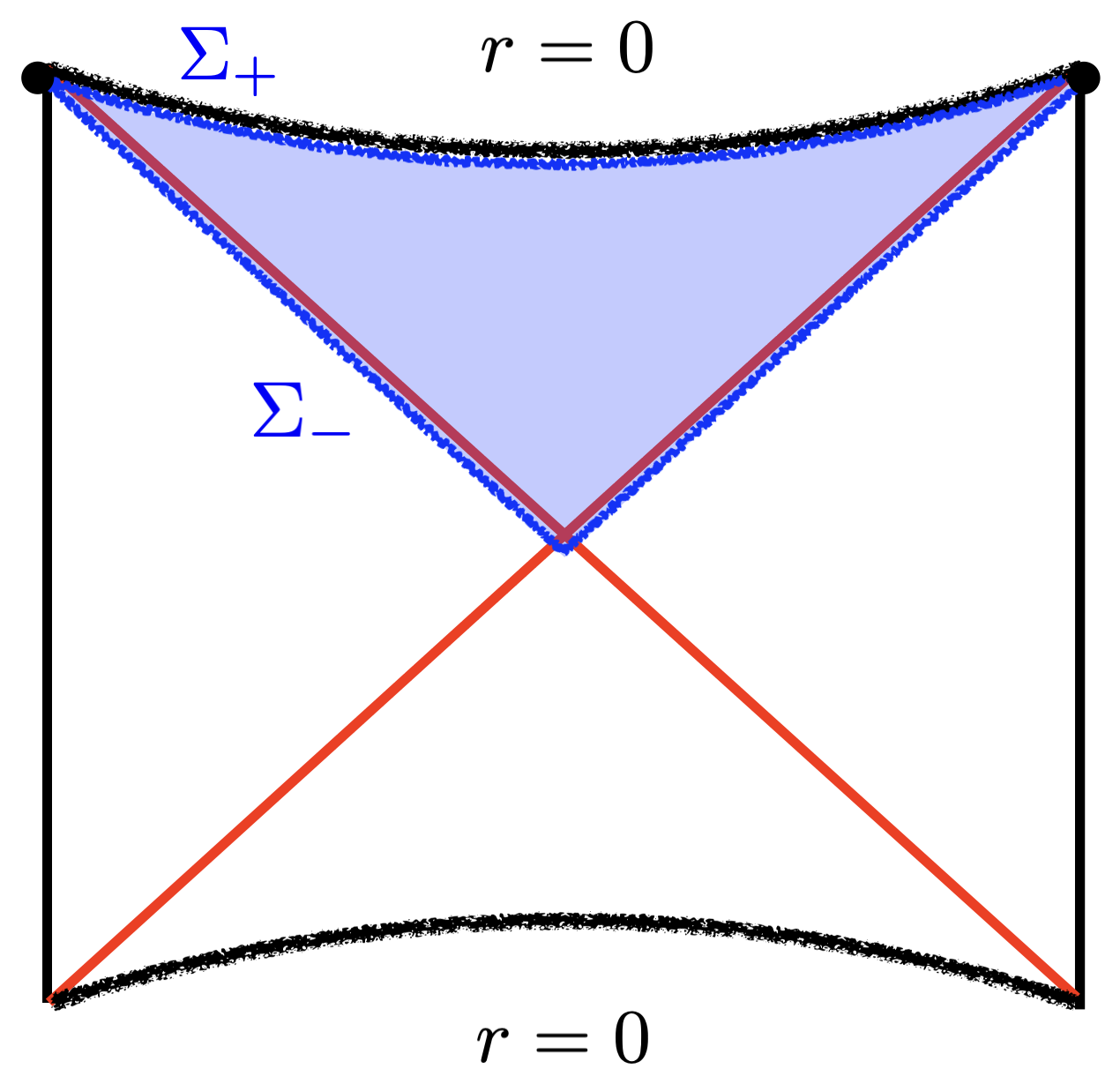} \label{}}
\caption{Sketch of the codimension-zero region $\mathcal{M}$ when the mean curvature is large. In panel (a), at a fixed finite time, the boundaries $\Sigma_{\pm}$ become null, and the region becomes the Wheeler-DeWitt (WDW) patch. Panel (b) illustrates the same scenario in the late-time limit.}\label{LKPENROSE}
\end{figure}
Specifically, considering that the CMC slice with a large mean curvature tends to approach the boundary of the WDW patch~\cite{Belin:2022xmt}, we observe the scenario depicted in the left panel of Fig. \ref{LKPENROSE} at a fixed finite time. In the late-time limit, the past boundary $\Sigma_-$ approaches the event horizon, while the future boundary $\Sigma_+$ converges toward either the spacelike singularity or an inner horizon (if there is one), as illustrated in the right panel of Fig. \ref{LKPENROSE}.

\paragraph{Complexity in the late-time limit.}
Building upon the previous discussion, we now investigate the late-time behavior of the complexity defined in \eqref{CANYE}, evaluated on the previously extremized surface $\Sigma_\pm$:
\begin{align}\label{COMFIN}
\begin{split}
\mathcal{C}^{\pm} &= \frac{1}{G_NL} \int_{\Sigma_{\pm}} \mathrm{d}^d \sigma  \,  \sqrt{h}  \,\,  \mathcal{F}(\sigma)  \, \\
&= \frac{V_{d-1}}{G_{\mathrm{N}}L}  \int_{\Sigma_\pm} \dd \sigma \, C(r_{\pm})^{\frac{d-1}{2}} \,\,\mathcal{F}(r_{\pm})\, \sqrt{-D(r_{\pm})\dot{v}_{\pm}^2+2 \sqrt{D(r_{\pm})B(r_{\pm})} \, \dot{v}_{\pm}\dot{r}_{\pm}} \,,
\end{split}
\end{align}
where we have used \eqref{EddiE2} and \eqref{REDBV}. Moreover, recall that $\mathcal{F}(\sigma)$ can be any arbitrary scalar functional of the background metric and extrinsic curvature. Assuming the homogeneity of the dual quantum circuit, we have taken it to depend solely on the radial coordinate $r$.

By applying both the gauge condition \eqref{gaugeeq} and the equation of motion \eqref{emoeq}, the complexity formula \eqref{COMFIN} can be further simplified to
\begin{equation}\label{COMFIN22}
\begin{aligned}
\mathcal{C}^\pm = \frac{2V_{d-1}}{G_{\mathrm{N}}L}  \int_{r_{\pm,\text{min}}}^\infty \dd r_{\pm} \,\, C(r_{\pm})^{d-1} \,\mathcal{F}(r_{\pm})\, \sqrt{\frac{D(r_{\pm})B(r_{\pm})}{-\mathcal{U}(P^\pm_v,r_{\pm})}} \,.
\end{aligned}
\end{equation}
Given the implicit dependence on boundary time \eqref{tau}, which is encoded through $r_{\pm}$ and $P^\pm_v$, its growth rate can be formulated as
\begin{align}\label{speedeqc}
\begin{split}
\frac{\dd \mathcal{C}^{\pm}}{\dd t_b} = &\,\frac{2V_{p-1}}{G_{\mathrm{N}}L}\Bigg[
C(r_{\pm,\min})^{d-1} \,\, \mathcal{F}(r_{\pm,\min}) \, \sqrt{\frac{D(r_{\pm,\min})B(r_{\pm,\min})}{-\mathcal{U}(P^\pm_v,r_{\pm,\min})}} \,\, \frac{\dd r_{\pm,\min }}{\dd t_b} \\
&  +\, \frac{\dd P^\pm_v}{\dd t_b} \int_{r_{\pm,\min }}^{\infty} \dd r_{\pm} \,\, \Bigg\{ \mathcal{F}(r_{\pm})\, \sqrt{D(r_{\pm})B(r_{\pm})}C(r_{\pm})^{d-1} \frac{ P_v^\pm \pm\frac{\alpha_B}{\alpha_\pm} b(r_{\pm})}{\left(-\mathcal{U}\left(P_v^\pm, r_{\pm}\right)\right)^{\frac{3}{2}}} \Bigg\} \Bigg]\,,
\end{split}
\end{align}
where we used \eqref{Ufunction} to evaluate $\partial_{P_v^\pm} \mathcal{U}\left(P_v^\pm, r_{\pm}\right)$. It is worth noting that the expression \eqref{speedeqc} may initially seem ill-defined due to \eqref{RMINFOR}. However, it can be shown that this apparent issue is resolved; specifically, the term $\mathcal{U}(P^\pm_v,r_{\pm,\min})$ is canceled out by the contribution from ${\dd r_{\pm,\min }}/{\dd t_b}$. To demonstrate this, we first write
\begin{align}\label{differeq}
\begin{split}
\frac{\dd r_{\pm,\min}}{\dd t_b} = 
& \left[\left(P_v^\pm \pm \frac{\alpha_B}{\alpha_\pm} b(r_{\pm,\min})\right)^{-1} \sqrt{\frac{ -\mathcal{U}\left(P_v^\pm, r_{\pm,\min}\right)D(r_{\pm,\min})}{ B(r_{\pm,\min}) }}\right] \\
& \quad \times \left[\frac{1}{2} + \frac{\dd P_v^\pm}{\dd t_b} \int_{r_{\pm,\min }}^{\infty} \,\dd r_{\pm}\, \frac{\sqrt{D(r_{\pm})B(r_{\pm})}C(r_{\pm})^{d-1}}{\left(-\mathcal{U}\left(P_v^\pm, r_{\pm}\right)\right)^{\frac{3}{2}}} \right]  \,,
\end{split}
\end{align}
which can be obtained by differentiating \eqref{tau}. Then, plugging \eqref{differeq} into \eqref{speedeqc}, we obtain
\begin{align}\label{differeq22}
\begin{split}
\frac{\dd \mathcal{C}^{\pm}}{\dd t_b} = &\, \frac{V_{p-1}}{G_{\mathrm{N}}L}\Bigg[
\frac{C(r_{\pm,\min})^{d-1} \, \mathcal{F}(r_{\pm,\min}) \, D(r_{\pm,\min})}{P_v^\pm \pm \frac{\alpha_B}{\alpha_\pm} b(r_{\pm,\min})} \,\,  \\
& +\, \frac{\dd P^\pm_v}{\dd t_b} \int_{r_{\pm,\min }}^{\infty} \dd r_{\pm} \,\, \Bigg\{ 2 \mathcal{F}(r_{\pm})\, \sqrt{D(r_{\pm})B(r_{\pm})}C(r_{\pm})^{d-1} \frac{ P_v^\pm \pm\frac{\alpha_B}{\alpha_\pm} b(r_{\pm})}{\left(-\mathcal{U}\left(P_v^\pm, r_{\pm}\right)\right)^{\frac{3}{2}}} \\
&  - 2 \, \mathcal{F}(r_{\pm,\min}) \sqrt{D(r_{\pm})B(r_{\pm})}C(r_{\pm})^{d-1}\frac{P_v^\pm \pm \frac{\alpha_B}{\alpha_\pm} b(r_{\pm,\min})}{\left(-\mathcal{U}\left(P_v^\pm, r_{\pm}\right)\right)^{\frac{3}{2}}} \, \Bigg\} \Bigg] \,.
\end{split}
\end{align}
Note that the integrand in the second term vanishes when $r_{\pm} \rightarrow r_{\pm,\min}$.

Finally, following \cite{Jorstad:2023kmq}, we consider the limit of large mean curvature. In this scenario, the integral term of \eqref{differeq22} tends towards zero as the CMC slices approach the constant $r$-surface ($r_{\pm,\min} = r_{\pm,f}$), as indicated by \eqref{DFINALRELA}. Consequently, the integration range of $r$ is confined to remain at a constant $r_{\pm,f}$ (which is a set of measure zero), leading to the integral evaluating to zero. We are thus left with
\begin{align}\label{dcdtaue}
\begin{split}
(K_{\Sigma_{\pm}} \,\rightarrow\,  \pm \, \infty): \quad \lim_{t_b \rightarrow \infty}\frac{\dd \mathcal{C}^{\pm}}{\dd t_b} \,=\, \frac{V_{d-1}}{G_{\mathrm{N}}L} \,\mathcal{F}(r_{\pm,f}) \, C(r_{\pm,f})^{\frac{d-1}{2}}\sqrt{-D(r_{\pm,f})} \,,
\end{split}
\end{align}
where we used both \eqref{Ufunction} and \eqref{RMINFOR}.

Two remarks are in order.
Firstly, the final expression \eqref{dcdtaue} can be interpreted as the volume measure of the induced metric on the ultimate slice $r = r_{\pm,f}$, \eqref{EddiE}, multiplied by the geometric factor $\mathcal{F} (r_{\pm,f})$.\footnote{At late times, the portions of the CMC slice extending towards the asymptotic boundary do not affect equation \eqref{dcdtaue}~\cite{Jorstad:2023kmq} because their volume approaches a constant value.}
Secondly, it is worth recalling that \eqref{COMFIN22} can be considered a `good' candidate for the holographic dual of complexity due to its potential to exhibit linear growth at late times~\cite{Jorstad:2023kmq}. In this limit, \eqref{dcdtaue} indeed approaches a constant, consistent with the expected behavior for the circuit complexity associated with the dual thermofield double state~\cite{Susskind:2018pmk,Brown:2017jil,Haferkamp:2021uxo}.

%
\subsubsection{Diagnosing the singularity}

In this section, we aim to further investigate the late-time linear growth of complexity in the limit of large mean curvature. Namely, by exploring different choices of $\mathcal{F}$, we will examine the properties of \eqref{dcdtaue} to try to uncover information about the singularities allowed by our generalized metric. We will separately investigate the complexity associated with the past boundary ($\mathcal{C}^{-}$) and the future boundary ($\mathcal{C}^{+}$). Following \cite{Jorstad:2023kmq}, we will first consider
\begin{equation}\label{Fchoices}
\mathcal{F}=1  \qquad \text{or} \qquad \mathcal{F} = \abs{L K}\,,
\end{equation}
with $K$ given in \eqref{extcureq}. As expected, $\mathcal{C}^{-}$ does not probe the singularity and thus remains unaffected by it. In contrast, for $\mathcal{C}^{+}$, the extremal surfaces do reach the singularity (in cases where we do not have an inner horizon) enabling us to diagnose it. For the specific choices of $\mathcal{F}$ presented in (\ref{Fchoices}), we find that the linear growth rate is either zero or a constant, independent of the Kasner exponents. This observation prompts us to investigate more general choices of $\mathcal{F}$. Notably, we find that a third choice of the form
\begin{equation}\label{Fchoices2}
\mathcal{F}=L\sqrt{K_{\mu \nu}K^{\mu \nu}} \,,
\end{equation}
yields a linear growth rate that explicitly depends on the Kasner exponents.

\paragraph{Complexity with the past boundary.}
Let us first discuss $\mathcal{C}^{-}$. As previously mentioned, in this case, the past boundary $\Sigma_-$ approaches the event horizon $r_+$ (see Fig. \ref{LKPENROSE}).
Starting with the case where $\mathcal{F}(r)=1$, \eqref{dcdtaue} evaluates to
\begin{equation}
\mathcal{F}=1 : \quad \lim_{t_b \rightarrow \infty}\frac{\dd \mathcal{C}^{-}}{\dd t_b} 
= \frac{V_{d-1}}{G_{\mathrm{N}}L} C(r_+)^{\frac{d-1}{2}}\sqrt{-D(r_+)} = 0 \,,
\end{equation}
where we have used the fact that the blackening function $D(r_+)$ vanishes at $r = r_+$. This result is expected, as in this limit, the CMC slice converges towards the past boundary of the WdW patch, which is a null surface.

Next, considering the case $\mathcal{F}=\abs{LK}$, we have 
\begin{align}\label{dcdtlkfeq}
\begin{split}
\mathcal{F}= \abs{LK} : \quad \lim_{t_b \rightarrow \infty}\frac{\dd \mathcal{C}^{-}}{\dd t_b} 
&= \left.\frac{V_{d-1}}{G_{\mathrm{N}}} \abs{\frac{C(r) D^{\prime}(r)+(d-1) D(r) C^{\prime}(r)}{2 \sqrt{B(r)D(r)}}}C(r)^{\frac{d-3}{2}}\right|_{r=r_{+}} \\
&= 8 \pi \, T_+S_+ \,,
\end{split}
\end{align}
where we also used $D(r_{+})=0$, and the explicit expressions for the Hawking temperature and Bekenstein-Hawking entropy:
\begin{equation}\label{TSFOR}
T_{\pm} = \left.\frac{1}{4\pi } \frac{\abs{D(r)'}}{\sqrt{D(r) B(r)}} \right|_{r=r_{\pm}} \,, \quad 
S_{\pm}=\left. \frac{V_{d-1}}{4G_N}C(r)^{\frac{d-1}{2}} \right|_{r=r_{\pm}} \,.
\end{equation}
Our result in \eqref{dcdtlkfeq} indicates that, for the given choice of $\mathcal{F}$, the late-time growth of $\mathcal{C}^{-}$ is proportional to the product $T\,S$ for the broader class of metrics we are studying, thus, extending the findings of \cite{Jorstad:2023kmq}.

\paragraph{Complexity with the future boundary.} 
Up to this point, we have focused on $\mathcal{C}^{-}$, which remains unaffected by the black hole singularity. Next, we will turn our attention to $\mathcal{C}^{+}$. The outcome in this case will largely depend on whether the singularity is timelike or Kasner, so we will examine each of these scenarios separately.

\emph{Timelike singularities.}
As discussed in the preceding section, for timelike singularities, the future boundary $\Sigma_+$ approaches the inner horizon $r_-$ as the mean curvature becomes large. Then, \eqref{dcdtaue} evaluates to
\begin{align}\label{dcdtlkfeq22}
\begin{split}
\mathcal{F}&= 1: \qquad\qquad \lim_{t_b \rightarrow \infty}\frac{\dd \mathcal{C}^{+}}{\dd t_b} = \frac{V_{d-1}}{G_{\mathrm{N}}L} C(r_-)^{\frac{d-1}{2}}\sqrt{-D(r_-)} = 0 \,, \\
\mathcal{F}&= \abs{LK}: \quad \lim_{t_b \rightarrow \infty}\frac{\dd \mathcal{C}^{+}}{\dd t_b} = \left.\frac{V_{d-1}}{G_{\mathrm{N}}} \abs{\frac{C(r) D^{\prime}(r)+(d-1) D(r) C^{\prime}(r)}{2 \sqrt{B(r)D(r)}}} C(r)^{\frac{d-3}{2}}\right|_{r=r_{-}} \\
&\qquad\qquad\qquad\qquad\qquad\,\,\,\,\,  = 8 \pi \, T_- S_- \,,
\end{split}
\end{align}
where we used \eqref{TSFOR}. Thus, we reach the same conclusion as with $\mathcal{C}^{-}$: for $\mathcal{F}= 1$ we obtain a vanishing complexity rate since $\Sigma_+$ approaches a null surface, while for $\mathcal{F}= \abs{LK}$ the rate is proportional to $T\,S$, however, with the variables associated with the inner horizon.

\emph{Kasner singularities.}
Finally, we consider $\mathcal{C}^{+}$ in the presence of Kasner singularities. In the absence of an inner horizon, the future boundary $\Sigma_+$ approaches the spacelike singularity, $r \rightarrow 0$, so in this case we expect to be able to diagnose the singularity. 

To be more specific, we consider the Kasner metric\footnote{Note that we do not rescale $t$ and $r$ such that $c_r=c_t=1$ since the growth rate of complexity depends on the choice of the time coordinate and $V_{d-1}$ also depends on the choice of the spatial coordinate.}
\begin{align}\label{}
\begin{split}
\dd s^2 =  -c_r^2 \, \dd r^2 + c_t^2 \, r^{2 p_t} \dd t^2      + r^{2 p_x} \dd \vec x_{i}^{2}  \,, 
\end{split}
\end{align}
which, given our metric ansatz~\eqref{OURMETRIC}, implies
\begin{equation}
D(r)=-c_t^2 r^{2p_t}\,,\quad B(r)=-c_r^2\,,\quad C(r)=r^{2 p_x} \,.
\end{equation}
Substituting this expression into \eqref{dcdtaue} yields
\begin{equation}
\begin{aligned}
\label{eq:dcovtkasner}
\lim_{t_b \rightarrow \infty}\frac{\dd \mathcal{C}^{+}}{\dd t_b} \,=\, \frac{V_{d-1}}{G_{\mathrm{N}}L} \, \mathcal{F}(r) \, c_t \, r^{p_t+(d-1)p_x} \,.
\end{aligned}
\end{equation}
Now, for the two choices of $\mathcal{F}$ in (\ref{Fchoices}) we find:
\begin{align}\label{dcdtlkfeq33}
\begin{split}
\mathcal{F}&= 1: \qquad\qquad \lim_{t_b \rightarrow \infty}\frac{\dd \mathcal{C}^{+}}{\dd t_b} \,=\, \frac{V_{d-1}}{G_{\mathrm{N}}L} \, c_t \,\, r^{p_t+(d-1)p_x} \,\to\, 0 \,, \\
\mathcal{F}&= \abs{LK}: \quad \lim_{t_b \rightarrow \infty}\frac{\dd \mathcal{C}^{+}}{\dd t_b} \,=\, \frac{V_{d-1}}{G_{\mathrm{N}}}  \frac{c_t}{c_r} \, (p_t+(d-1)p_x) \,\, r^{p_t+(d-1)p_x-1}  \,=\, \frac{V_{d-1}}{G_{\mathrm{N}}}  \frac{c_t}{c_r}  \,,
\end{split}
\end{align}
where we used the Kasner condition, $p_t+(d-1)p_x=1$, and $\abs{K_{\Sigma_{+}}} = \frac{p_t+(d-1) p_x}{c_rr}$. Notably, the late time growth for $\mathcal{F}= \abs{LK}$ is constant, however, the explicit dependence on the Kasner exponents drops out. 

The above observation prompts us to explore more general choices of $\mathcal{F}$ with the goal of extracting more detailed information about the singularity. The main issue is that the combination of $p_t$ and $p_x$ in (\ref{dcdtlkfeq33}) exactly yields the linear Kasner identity $p_t+(d-1)p_x=1$, which does not include $p_\phi$, unlike the quadratic identity $p_\phi^2 + p_t^2 + (d-1)p_x^2 = 1$. One possibility would be to select an $\mathcal{F}$ with explicit dependence on the matter fields, e.g., $\mathcal{F}(\phi)$. However, this choice appears to be highly theory-dependent. Another option is to investigate other geometric invariants. Notably, we find that a simple choice of the form (\ref{Fchoices2}) already yields the desired result. Specifically, we find that
\begin{align}\label{KDCOMGRO}
\begin{split}
    \mathcal{F}= L\sqrt{K_{\mu \nu}K^{\mu \nu}}: \quad \lim_{t_b \rightarrow \infty}\frac{\dd \mathcal{C}^{+}}{\dd t_b} 
    &\,=\, \frac{V_{d-1}}{G_{\mathrm{N}}}  \frac{c_t}{c_r} \, \sqrt{p_t^2+(d-1)p_x^2} \,\, r^{p_t+(d-1)p_x-1}  \\
    &\,=\, \frac{V_{d-1}}{G_{\mathrm{N}}}  \frac{c_t}{c_r}\sqrt{1-p_\phi^2} \,,
\end{split}
\end{align}
where we used both Kasner identities, $p_t + (d-1)p_x = 1$ and $p_\phi^2 + p_t^2 + (d-1)p_x^2 = 1$, as well as the relation $K_{\mu \nu}K^{\mu \nu} = \frac{p_t^2 + (d-1)p_x^2}{c_r^2r^2}$. Note that the value of $p_\phi$, combined with these two identities, uniquely determines the full set of Kasner exponents. This indicates that, for the chosen $\mathcal{F}$, we can fully characterize the singularity geometry by examining the rate of complexity growth at late times.\footnote{By employing our results of Kasner components from equations \eqref{KAS1} and \eqref{KAS2}, we have confirmed $p_{\phi}^2<1$, thereby ensuring that \eqref{KDCOMGRO} is positive definite.}

A final comment is in order. For both $\mathcal{F}= \abs{LK}$ and $\mathcal{F}= L\sqrt{K_{\mu \nu}K^{\mu \nu}}$, we found a finite complexity rate in the late time regime. Similar to \cite{Jorstad:2023kmq}, one might speculate whether such rates can be expressed as proportional to $T\,S$. However, a quick calculation reveals that this is generally not the case, but rather a coincidence specific to the Schwarzschild singularity ($Q=0$ in our context). In hindsight, this should not come as a surprise. Unlike the previous scenarios (\eqref{dcdtlkfeq} and \eqref{dcdtlkfeq22}), in this case, $\Sigma_+$ does not approach any of the horizons, but instead the singularity itself. For the Schwarzschild case, the rate is proportional to $T\,S$ due to the absence of additional scales in the problem. However, this reasoning does not apply to a more generic Kasner singularity, resulting in more complex dependencies in the rate of complexity growth.

%
\subsection{Thermal $a$-function}

In this section, we explore another observable for studying the black hole interior, known as the thermal $a$-function, which was introduced in \cite{Caceres:2022smh} and further elaborated on in \cite{Caceres:2022hei,Caceres:2023zft}. Our goal is to characterize the endpoint of the $a$-function using near-singularity data, which we can access analytically through our exact backgrounds. Before doing so, let us briefly review the concept of `trans-IR' flows, defined holographically via black hole interiors.

The main concept of a `trans-IR' flow involves analytically continuing the conventional RG flow beyond its infrared (IR) fixed point to complex energies. For thermal states, the flow is characterized by the so-called thermal $a$-function, $a_T(z)$, which (i) measures the effective number of degrees of freedom at each energy scale and (ii) is therefore monotonically decreasing along the entire RG flow. This includes both the conventional RG flow, defined outside the horizon, and the `trans-IR' regime, defined inside the black hole. 

To define the $a$-function, we start with a black hole metric in domain-wall coordinates:
 \begin{equation}
ds^2 = e^{2A(u)}\left[-h(u)^2 dt^2 + d\vec{x}^2\right] + du^2,
\end{equation}
with $t \in \mathbb{R}$, $\vec{x} \in \mathbb{R}^{d-1}$, $u \geq 0$. As usual, $u$ represents an energy scale in the theory so that $u=0$ (the black hole horizon) corresponds to the IR of the theory, while $u\to \infty$ (the AdS boundary) corresponds to the UV \cite{Peet:1998wn,Agon:2014rda}.\footnote{For an examination of the $a$-function within the context of Lorentz violating RG flows, see \cite{Baggioli:2024vza}.} In these coordinates the thermal $a$-function reads \cite{Caceres:2022smh},
\begin{equation}
a_T(u) = \frac{\pi^{d/2}}{\Gamma\left(\frac{d}{2}\right)\ell_P^{d-1}}\left[\frac{h(u)}{A'(u)}\right]^{d-1},\label{aFunction0}
\end{equation}
where $\ell_P$ is the Planck length. To probe monotonicity, one simply differentiates with respect to $u$ and uses Einstein's equations to recast the result in terms of the stress tensor. Then, demanding that the bulk matter fields respect the null energy condition, one arrives at
\begin{equation}
    \frac{d a_T}{d u}\propto\frac{1}{A'(u)^d}\bigg(T^{z}_{z}-T^t_t\bigg)\geq 0\,.
\end{equation}

Notably, the above coordinates only cover the exterior of the black hole. To access the interior one can perform a dual analytic continuation of the time and radial coordinates, so that
\begin{equation}
    t=t_I-\text{sgn}(t_I)\frac{i \gamma}{2\mathcal{T}}\,,\qquad u=i\rho\,,
\end{equation}
where $\gamma$ is a half integer and $\mathcal{T}\equiv\frac{e^{A(0)}h'(0)}{2\pi}$.
To show monotonicity inside, we use the coordinates introduced in \eqref{ZCO}. Specifically, we perform 
the following coordinate transformation, $u\to u(z)$, and identifications:
\begin{eqnarray}
    \frac{d u}{dz}&=&-\frac{1}{z\sqrt{g(z)}}\,,\label{transzr}\\
    e^{2 A(u)}&=&\frac{1}{z^2}\label{coordtrans0}\,,\\
    h(u)^2&=&g(z)e^{-\chi(z)}\,,
    \label{coordtran01}
\end{eqnarray}
to obtain
 \begin{equation}
a_T(z) = \frac{\pi^{d/2}}{\Gamma\left(\frac{d}{2}\right)\ell_P^{d-1}} e^{-(d-1)\chi(z)/2}\,.\label{aFunctionSch}
\end{equation}
Moreover, by differentiating and using Einstein's equations, we obtain
\begin{equation}
    \frac{d a_T}{d z}\propto -g(z)\bigg(T^r_r-T^t_t\bigg)\leq 0\,,\label{derivar}
\end{equation}
demonstrating that $a_T(z)$ remains monotonic inside the black hole, i.e., for $z>z_h$.

Let us now analyze our specific hairy black holes. In the near-singularity limit ($z\rightarrow\infty$), reading $\chi(z)$ from \eqref{SLFOR}, we find that, up to the first subleading order,
\begin{align}\label{SLFOR2222}
\begin{split}
\chi(z) =  \frac{4}{(d-1)\d^2} \log z + 2 \log \left[ \frac{(d-2)(d-1)\d^2}{2+(d-2)(d-1)\d^2} Q^{\frac{2}{(d-2)(d-1)\d^2}}\right] \,.
\end{split}
\end{align}
This implies that the thermal $a$-function can be approximated as
\begin{align}\label{afunc}
\begin{split}
a_T(z\to\infty) \approx  \frac{\pi^{d/2}}{\Gamma\left(\frac{d}{2}\right) Q^{\frac{2}{(d-2)\d^2}} } \left[ \frac{2+(d-2)(d-1)\d^2}{(d-2)(d-1)\ell_p \d^2} \right]^{d-1} \, z^{\frac{-2}{\d^2}} \,.
\end{split}
\end{align}

A few remarks are in order. First, since $\d>0$, our thermal $a$-function approaches zero near the singularity, following a specific power law determined by $\d$. This behavior encompasses all solutions with either timelike or Kasner singularities, implying that the (classical) degrees of freedom freeze out at the singularities. This potentially leaves room for stringy or quantum degrees of freedom, which are not captured by this $a$-function, which can still be excited under such extreme conditions. Second, the decay is faster for timelike singularities, as these are associated with smaller values of $\d$ $(0<\d<\d_c)$. For Kasner singularities ($\d \geq \d_c$), our results are consistent with those in \cite{Caceres:2022smh}. Specifically, by recasting (\ref{afunc}) in terms of the Kasner exponents, we find that
\begin{align}\label{atherresul}
\begin{split}
a_T(z) \propto  z^{\frac{-(d-1)(d(1+p_t)-2)}{1-p_t}}\,,
\end{split}
\end{align}
where we have utilized \eqref{KAS1}-\eqref{KAS2}. Using the two Kasner relations, $p_t + (d-1)p_x = 1$ and $p_\phi^2 + p_t^2 + (d-1)p_x^2 = 1$, we can thus conclude that the power law in \eqref{atherresul} determines the complete set of Kasner exponents. Lastly, it is worth noting that \eqref{afunc} ensures the thermal $a$-function remains real-valued even into the trans-IR regime, given that $Q>0$ for our solutions.

\begin{figure}[]
 \centering
     \subfigure[$\d = {\d_c}/{3}$]
     {\includegraphics[width=4.83cm]{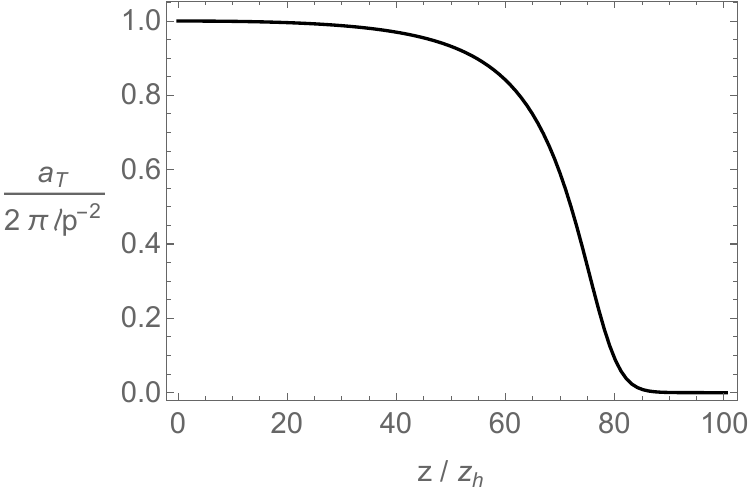} \label{}}
     \subfigure[$\d = \d_c$]
     {\includegraphics[width=4.83cm]{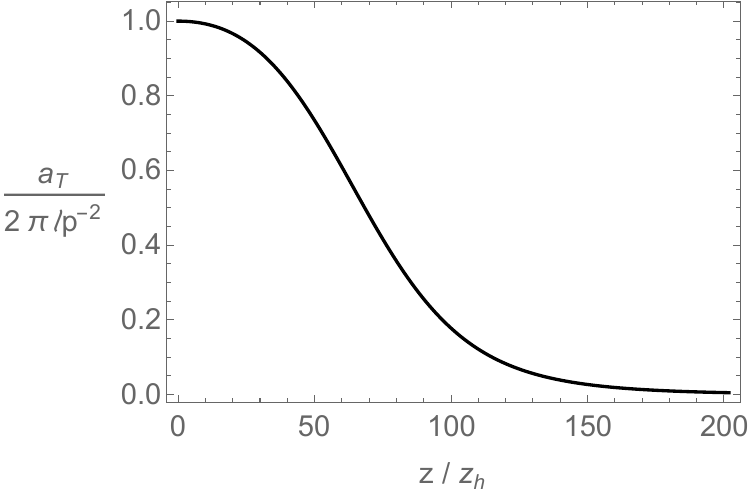} \label{}}
     \subfigure[$\d = 3 \d_c$]
     {\includegraphics[width=4.83cm]{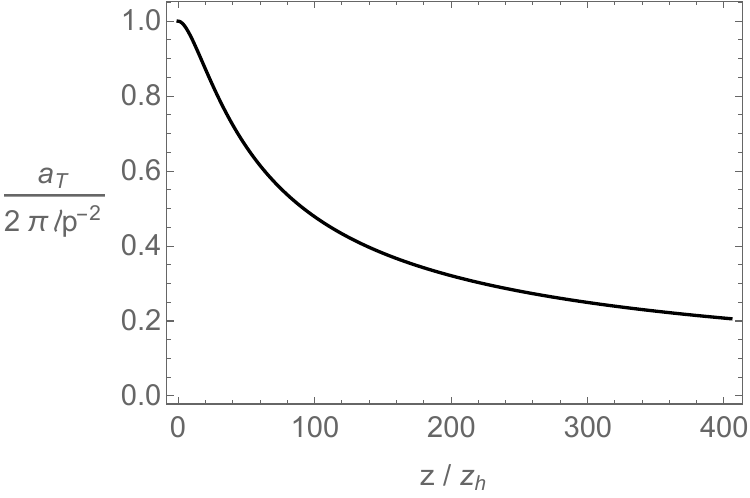} \label{}}
\caption{Thermal $a$-function for $d=3$ with $\d = \frac{1}{3}\d_c,\, \d_c,\, 3\d_c$ (left, center and right, respectively) where $\d_c = \sqrt{1/3}$. The left plot corresponds to a flow leading to a timelike singularity, while the center and right plots correspond to flows into Kasner singularities. \label{TAFIG}}
\end{figure}

We can also examine the full thermal $a$-function across the whole RG flow. For instance, for $d=3$ and $T/\mu=1$, Fig. \ref{TAFIG} depicts the thermal $a$-function from the AdS boundary to the singularity, exhibiting the expected monotonic behavior.
%

%
\section{The interior of black holes breaking translational invariance}
\label{sec:transbreaking}
In this section we study the interior geometry of black hole solutions that break translations on the dual theory.
We will consider both the case of explicit and spontaneous breaking.

It is worth noting that there exist numerical investigations concerning the analysis of black hole interiors with broken translations in holographic hairy black holes~\cite{Mansoori:2021wxf,Sword:2021pfm,Mirjalali:2022wrg}. In this paper
we construct and study analytical solutions of the interior of black hole geometries that break translations explicitly on the boundary theory. We compliment this analysis with a numerical simulation of the interior of black holes that instead break translations spontaneously.
To our knowledge, this work is the first analytic study of the interior of hairy black holes that break translations on the boundary theory.

In the following we focus on the Gubser-Rocha model in ($3+1$) dimensions~\cite{Gubser:2009qt}, i.e., we restrict our setup~\eqref{GGR}
to the case
\begin{equation}\label{}
\d = \d_c = \sqrt{\frac{2}{d(d-1)}} \,, \qquad d=3 \,,
\end{equation}
which brings the potentials~\eqref{MODELACTION} to the form
\begin{align}
\begin{split}
Z(\phi)=e^{-\frac{\phi}{\sqrt{3}}}  \,, \qquad    V(\phi)= 6 \cosh \frac{\phi}{\sqrt{3}} \,.
\end{split}
\end{align}
As shown in section~\ref{sec:interior}, this setup is the only case within
our model where, at low temperature, the singularity is of the Kasner type.

As we will discuss below, implementing the breaking of traslations in the Gubser-Rocha model is relevant for the description of strongly coupled condensed matter systems.
Thus, in order to break translational invariance on the dual theory, 
following~\cite{Andrade:2013gsa,Vegh:2013sk} we add to our model~\eqref{GGR}
the axion action
\begin{equation}\label{AXIONAC}
S_{\text{axion}} =  \int\dd^{4}x   \sqrt{-g}   \left[  -\frac{Y(\phi)}{2} \sum_{i=1}^{2}\partial \psi_{i}^2  \right] \,, \qquad  Y(\phi) = \left(1 - \eta \, e^{\phi} \right)^2\,.
\end{equation}
The axion fields $\psi_i$ are taken as
\begin{equation}
\label{eq:axionsol}
\psi_i = \beta x^i\,,
\end{equation}
which solves their equations of motion and results in the breaking of translations on the
boundary theory~\cite{Andrade:2013gsa,Vegh:2013sk},
with the axion charge $\beta$ denoting the strength of broken translations. A key feature of this model is that the geometry remains homogeneous, i.e., the metric and matter fields are functions of the radial coordinate only. Moreover, as we will see below, if the breaking of translations is explicit one finds analytic black hole solutions in the presence of axions~\eqref{eq:axionsol}.

The type of translational symmetry breaking implemented by the axion fields on the boundary theory can be regulated by the parameter $\eta$ in $Y(\phi)$ as follows:
\begin{align}\label{TWOTM}
\begin{split}
 \eta = 0  \quad\, (\text{Explicit breaking}) \,, \quad\quad 
 \eta = 1   \quad\, (\text{Spontaneous breaking}) \,,
\end{split}
\end{align}
together with the appropriate boundary conditions on 
the dilaton $\phi$. Specifically, the asymptotic solution of the axion fields
takes the form
\begin{align}\label{BCSSB}
\begin{split}
\psi_i \approx
\begin{cases}
\,\, \beta x^{i} \,+\, \psi_i^{(\Delta)} \, r^{-3} \,, \qquad\qquad\qquad\qquad\qquad\quad (\eta = 0) \\
\,\, \psi_i^{(\Delta)} \, r \,+\, \beta x^{i}  \qquad\text{with}\quad 
\phi  \approx  \left\langle\mathcal{O}\right\rangle \, r^{-2}  \,.  \qquad (\eta = 1)
\end{cases}
\end{split}
\end{align}
In accordance with the holographic dictionary, the leading term is interpreted as the source, while the subleading term corresponds to the expectation value of the dual operator. This implies that the
$\beta$ term serves as the source when $\eta=0$, whereas it represents the expectation value (vev) when $\eta=1$~\cite{Amoretti:2017frz,Jeong:2021zhz}. In other words, given that we set $\psi_i^{(\Delta)}=0$ in \eqref{BCSSB}, the translational symmetry is broken by the source (explicit breaking) when $\eta=0$, and 
instead by the vev (spontaneous breaking) when $\eta=1$.\footnote{Refer to section 2.2 in \cite{Jeong:2021zhz} for a more comprehensive and generic discussion on exploring the nature of broken translations within the Einstein-Maxwell-Dilaton with Axion model.}

It is noteworthy that holographic axion theories~\cite{Andrade:2013gsa,Vegh:2013sk} have served as valuable tools for investigating strongly coupled condensed matter systems~\cite{Zaanen:2015oix,Hartnoll:2016apf,Baggioli:2019rrs,Natsuume:2014sfa,Zaanen:2021llz,Liu_2012,Faulkner:2010da}, particularly aiding in the understanding of the strange metal phase and the associated high-$T_c$ superconductivity in cuprate materials and other strongly correlated systems. Notable and celebrated results have been attained through analyses involving conductivity~\cite{Davison:2013txa,Gouteraux:2014hca,Blauvelt:2017koq,Jeong:2018tua,PhysRevLett.120.171602,Ammon:2019wci,Ahn:2019lrh,Jeong:2021wiu,Baggioli:2022pyb,Balm:2022bju,Ahn:2023ciq}, transport coefficients~\cite{Davison:2014lua,Blake:2016jnn,Amoretti:2016cad,Blake:2017qgd,Baggioli:2017ojd,Ahn:2017kvc,Davison:2018ofp,Blake:2018leo,Jeong:2019zab,Arean:2020eus,Liu:2021qmt,Jeong:2021zhz,Wu:2021mkk,Jeong:2021zsv,Huh:2021ppg,Jeong:2022luo,Baggioli:2022uqb,Jeong:2023ynk,Ahn:2024aiw,Zhao:2023qms}, and the collective dynamics of strongly coupled phases~\cite{Baggioli:2014roa,Alberte:2015isw,Amoretti:2016bxs,Alberte:2017oqx,Amoretti:2017frz,Amoretti:2017axe,Alberte:2017cch,Andrade:2017cnc,Amoretti:2018tzw,Amoretti:2019cef,Ammon:2019wci,Baggioli:2019abx,Amoretti:2019kuf,Ammon:2019apj,Baggioli:2020edn,Amoretti:2021fch,Amoretti:2021lll,Wang:2021jfu,Zhong:2022mok,Bajec:2024jez}. Additionally, quantum information applications~\cite{RezaMohammadiMozaffar:2016lbo,Yekta:2020wup,Li:2019rpp,Zhou:2019xzc,Huang:2019zph,Jeong:2022zea,HosseiniMansoori:2022hok,Ahn:2024lkh,Babaei-Aghbolagh:2021ast} and the application of the AdS/Deep learning correspondence~\cite{Ahn:2024gjf,Ahn:2024lkh} have been explored.\footnote{For a comprehensive and up-to-date review of the holographic axion model and an extensive list of references, readers are directed to \cite{Baggioli:2021xuv}.}

Within this array of previous investigations, the Gubser-Rocha model~\cite{Gubser:2009qt} stands out as the most renowned and celebrated holographic model accountable for the characteristics of strange metals. Notably, the Gubser-Rocha model facilitates the achievement of linear-in-$T$ resistivity, attributable to the nature of the IR fixed point~\cite{Davison:2013txa,Gouteraux:2014hca,Anantua:2012nj,Jeong:2018tua,Jeong:2019zab,Jeong:2021wiu,Balm:2022bju,Ahn:2023ciq,Jeong:2023ynk,Ge:2023yom,Wang:2023rca}. It is worth mentioning that the model also exhibits characteristics relevant to high-$T_c$ cuprate superconductivity, such as Homes's law in high-$T_c$ superconductors~\cite{Jeong:2021wiu,Wang:2023rca}.\footnote{Moreover, explorations of the phase diagram employing fermionic spectral functions~\cite{Jeong:2019zab} or conductivity~\cite{Zhao:2023qms} have been conducted. In addition, some limitations in describing transport anomalies, such as the Hall angle, have also been discussed~\cite{Ahn:2023ciq,Ge:2023yom}.}

Despite the extensive scrutiny of the Gubser-Rocha model with broken translations, this paper marks the first analysis of its black hole interiors. Given the model's potential to delineate the characteristics of strange metals in holography, our work in this section could be likened to ``diving into holographic strange metals", as it is reminiscent of pioneering efforts in studying black hole interiors of holographic superconductors~\cite{Hartnoll:2020fhc}.

%
\subsection{Explicit symmetry breaking}
We begin by examining the scenario of explicit translational symmetry breaking, corresponding to $\eta=0$ in \eqref{AXIONAC}. This condition leads to the following
analytical background solutions in the ansatz \eqref{sol11}:
\begin{align}\label{GUBANS}
\begin{split}
f(r) &= r^2 \left(h(r)^{3}-\frac{r_h^3}{r^{3}} h(r_h)^{3}\right) - \frac{\beta^2}{2} \left( 1 - \left(\frac{r_h}{r}\right) \right)  \,, \qquad\,\,\, h(r) = 1+\frac{Q}{r} \,, \\
A_t(r) &= \sqrt{ 3 Q \, r_h \left( \frac{h(r_h)}{h(r)^2} - \frac{\beta^2}{2 r_h^2 h(r_h ) h(r)^2} \right)}  \left(1-\frac{r_h}{r}\right),   \qquad e^\phi = h(r)^{-\frac{\sqrt{3}}{2}}  \,.
\end{split}
\end{align}
We can extract the corresponding Hawking temperature ($T$) and chemical potential ($\mu$) as follows:
\begin{align}\label{TDV}
\begin{split}
T &=  r_{h} \, \frac{6(1+\tQ)^2 - \tilde{\beta}^2}{8 \pi (1+\tQ)^{3/2}} \,,  \qquad \mu = r_{h} \,  \sqrt{  3\tQ (1+\tQ)  \left( 1 - \frac{\tilde{\beta}^2}{2(1+\tQ)^2}  \right) }  \,,
\end{split}
\end{align}
where $\tQ \equiv Q/r_h$ and $\tilde{\beta} \equiv \beta/r_h$. 
As before, the underlying UV conformal symmetry of our holographic theory
implies that physically distinct solutions are functions of dimensionless ratios of the different scales. These are $T$, $\mu$, and $\beta$. We choose to normalize the temperature and strength of breaking of translations $
\beta$ in terms of the chemical potential, which allows to compare our results to those that obtain at fixed chemical potential. 
The relevant ratios can be read as
\begin{align}\label{}
\frac{T}{\mu} =   \frac{6(1+\tQ)^{2}-\tilde{\beta}^{2}}{4\sqrt{6}\pi\sqrt{\tQ(1+\tQ)^{2}(2(1+\tQ)^{2}-\tilde{\beta}^{2})}} \,, \qquad 
\frac{\beta}{\mu} =   \sqrt{\frac{2(1+\tQ)\tilde{\beta}^{2}}{3\tQ(2(1+\tQ)^2-\tilde{\beta}^2)}} \,.
\end{align}
Physically meaningful solutions exist only in the following range:
\begin{align}\label{PCC}
\frac{T}{\mu} \,,\,   \frac{\beta}{\mu} \geq 0 \,, \quad\longrightarrow\quad  \tilde{\beta}^2 < 2 (1 + \tQ)^2 \,,
\end{align}
where $({T}/{\mu} \,,\,   {\beta}/{\mu}) \rightarrow \infty$ as $\tilde{\beta}^2 \rightarrow 2 (1 + \tQ)^2 $. It is worth noting that 
with $\tilde\beta$ in the allowed range, any real value of
$T/\mu$ and $\beta/\mu$ can be reached, including the extremal zero temperature
case.\footnote{It was shown that the Gubser-Rocha model remains well defined at low $T$ even with finite $\beta$~\cite{Davison:2013txa,Gouteraux:2014hca,Jeong:2018tua}. For instance, at low $T$, the entropy density and butterfly velocity are approximated as ${s}/{\mu^2} \approx \frac{16\pi^2}{3\sqrt{3}}\, {T}/{\mu}$ and $v_B^2 \approx 16\pi^2 \, \left({T}/{\mu}\right)^2$, respectively.}

\paragraph{Kasner singularity with broken translations.}
As in Section \ref{sec23}, we study the singularities in the $z$-coordinate \eqref{ZCO}, employing the same singularity limit ($z\rightarrow\infty$) expressed in \eqref{ZCO23}. Substituting \eqref{ZCO23} into \eqref{GUBANS}, we can derive the approximate solution in the singularity limit as follows:
\begin{align}\label{NSGR1}
\begin{split}
g(z) &= \frac{\tQ^3}{32} \left[ \tilde{\beta}^2 - 2 \left( (1+\tQ)^3 - \tQ^3 \right) \right] r_h^6 \, z^6  =: g_c \, z^6\,, \\
\chi(z) &= 6 \log z  \,, \quad
\phi(z) = -2\sqrt{3} \log z  \,,
\end{split}
\end{align}
and one can check that for $\tilde\beta$ within its allowed range~\eqref{PCC}, $g_c$ is always negative. As we show below, this implies that the singularity is of the spacelike Kasner type.

Subsequently, by applying the coordinate transformation
\begin{align}\label{}
z = \tau^{-{1}/{3}} \,,
\end{align}
the metric near the singularity becomes
\begin{align}\label{GRKS}
\begin{split}
\dd s^2 &= \frac{1}{9 \, g_c} \dd \tau^2  - g_c \tau^{2 p_t} \dd t^2  + \tau^{2 p_x} \dd \vec x_{i}^{2}  \,, \quad \phi(\tau) = -\sqrt{2} p_{\phi} \log \tau \,,
\end{split}
\end{align}
with the Kasner exponents
\begin{align}\label{KCC}
\begin{split}
p_t = p_x = \frac{1}{3} \,, \quad 
p_\phi = -\sqrt{\frac{2}{3}} \,,
\end{split}
\end{align}
satisfying the Kasner condition:
\begin{align}\label{eq:kasnercondd4}
\begin{split}
p_t + 2 p_x = p_\phi^2 + p_t^2 + 2 p_x^2 = 1 \,.
\end{split}
\end{align}
Since $g_c$ is negative for $\tilde\beta$ in its physical range~\eqref{PCC}, the singularity is spacelike.
Moreover, it is also interesting that the Kasner exponents \eqref{KCC} are independent of $\beta$, a unique feature not observed in previous 
examples of Kasner singularities featured by black holes that break
translations~\cite{Mansoori:2021wxf}.

\paragraph{No inner horizon.} 
One can also check that these black holes do not posses an inner horizon.
Specifically, considering \eqref{GUBANS},
a would-be inner horizon would be given by 
\begin{align}\label{}
  f(r) = 0  \quad\longrightarrow\quad  r = r_{I} \equiv \frac{r_h}{2} \left( \sqrt{2 \tilde{\beta}^2 - 3(1+\tQ)^2} -1 - 3 \tQ \right) \,,
\end{align}
which yields
\begin{align}\label{}
  0 \leq r_I  \leq r_h \,, \quad\longrightarrow\quad \sqrt{2+6\tQ(1+\tQ)} \leq \tilde{\beta} \leq \sqrt{6}(1+\tQ) \,,
\end{align}
but, as expected, this condition does not hold for $\tilde\beta$ in its allowed range~\eqref{PCC}.

To summarize our analysis, we have found that the black hole geometries~\eqref{GUBANS} always end in a spacelike Kasner singularity and do not feature an inner horizon.
It turns out that the implementation of homogeneous breaking of translations via the addition of axions to the Gubser-Rocha model does not change the character of the singularity. Indeed we have seen in section~\ref{ssec:singularities} that
in the absence of axions the Gubser-Rocha black holes also feature a Kasner spacelike singularity and no inner-horizon.

\paragraph{Holographic flows to singularities.}

We shall now plot and briefly discuss instances of the geometries~\eqref{GUBANS} realizing the holographic flows towards spacelike Kasner singularities.
We will set $T/\mu=1$ for all the examples we discuss.

\begin{figure}[]
 \centering
     \subfigure[$g(z)$ vs. $z/z_h$]
     {\includegraphics[width=7.3cm]{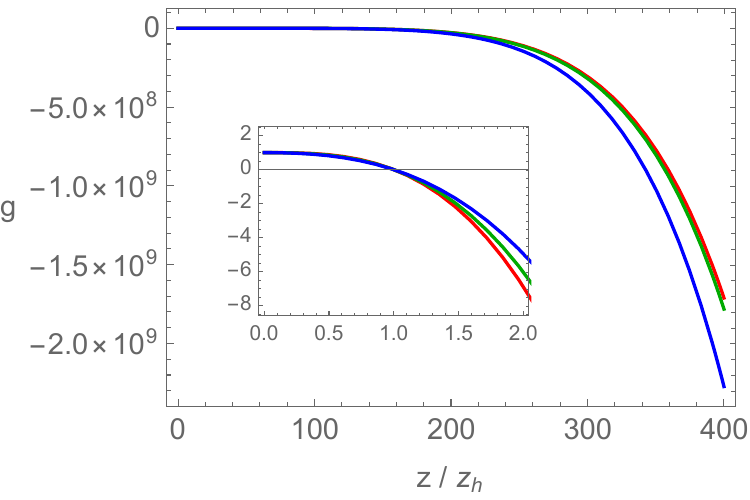} \label{}}
     \subfigure[$A_t(z)$ vs. $z/z_h$]
     {\includegraphics[width=6.7cm]{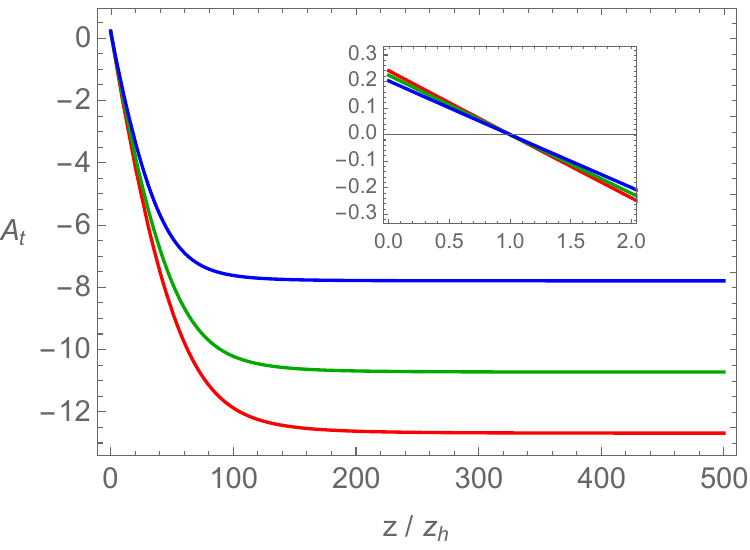} \label{}}
\caption{Holographic flows of the blackening factor $g(z)$ and gauge field $A_t(z)$ for $T/\mu=1$ and $\beta/\mu = 0, 3, 5$ (red, green, blue). The insets show the near-horizon behavior.}\label{GRfig1}
\end{figure}

\begin{figure}[]
 \centering
\subfigure[$z \, \dd (\log g_{tt}')/\dd z$ vs. $z/z_h$]
     {\includegraphics[width=4.83cm]{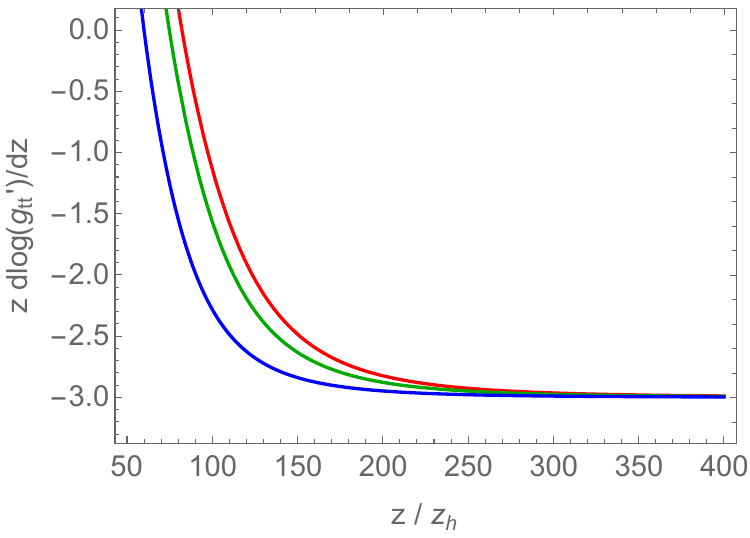} \label{}}
\subfigure[$z \, \dd \chi/\dd z$ vs. $z/z_h$]
     {\includegraphics[width=4.83cm]{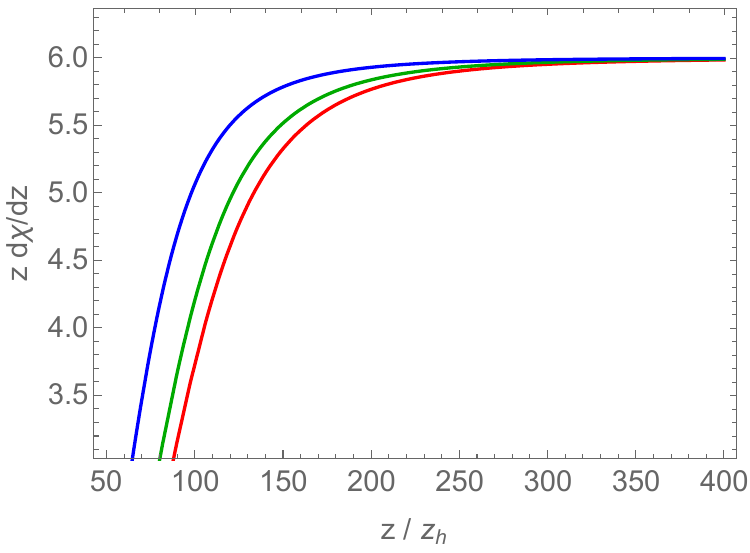} \label{}}
\subfigure[$z \, \dd \phi/\dd z$ vs. $z/z_h$]
     {\includegraphics[width=4.83cm]{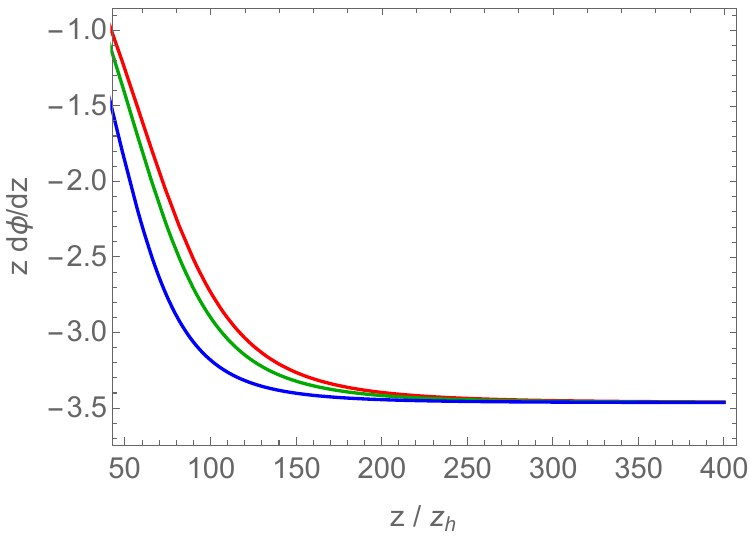} \label{}}
\caption{The functions $z \, \dd X/\dd z$ where $X = \left\{ \log g_{tt}',\, \chi,\, \phi \right\}$ for $T/\mu=1$ and  $\beta/\mu = 0, 3, 5$ (red, green, blue). The saturated value near the singularity ($z\rightarrow\infty$) is \eqref{NSVGR}.}\label{GRfig2}
\end{figure}

First, in Fig.~\ref{GRfig1} we plot the blackening factor $g(z)$ and gauge
field $A_t(z)$ for three different values of $\beta/\mu$. As expected, we observe the absence of an inner horizon. Additionally, the gauge field, which vanishes at the horizon, quickly approaches a constant towards the singularity.
Next, to further explore the behavior near the singularity, we examine the functions defined in Eq.~\eqref{THF}. As in section~\ref{ssec:singularities} these functions should approach a constant towards the Kasner singularity. Indeed, after plugging in the
near singularity solution~\eqref{NSGR1}, we find
\begin{align}\label{NSVGR}
\begin{split}
z \frac{\dd}{\dd z} \log (g_{tt}') \approx -3 \,, \qquad
z \frac{\dd}{\dd z} \chi \approx 6 \,, \qquad z \frac{\dd}{\dd z} \phi \approx -2\sqrt{3} \,,
\end{split}
\end{align}
which are related to the Kasner exponents \eqref{KCC} as we will see below in
Eq.~\eqref{RELKLAT}.
In Fig. \ref{GRfig2} we display these functions inside the event horizon, finding they all approach the value 
\eqref{NSVGR} towards the singularity. Notably, this constant value is independent of the value of $\beta$ since the Kasner exponents \eqref{KCC} are unaffected by $\beta$.

\begin{figure}[]
 \centering
     \subfigure[$\beta/\mu = 0, 3, 5$ (red, green, blue)]
     {\includegraphics[width=7.1cm]{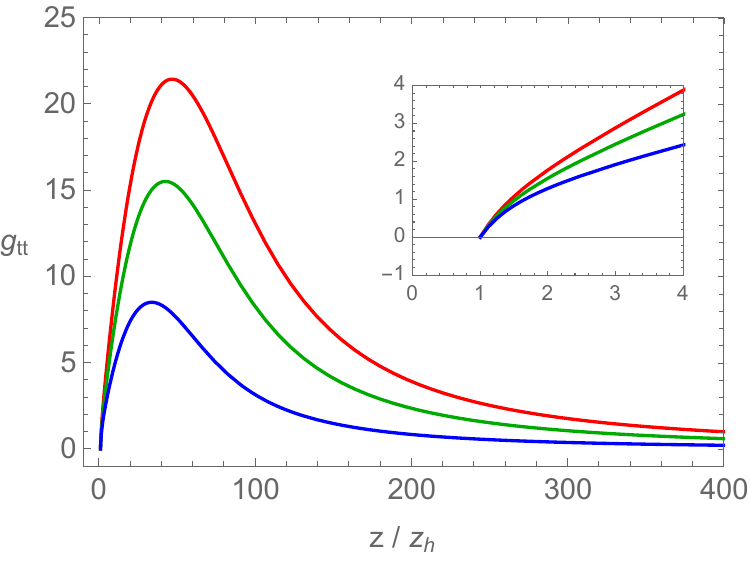} \label{GRfig3a}}
     \subfigure[$\beta/\mu = 10, 150, 300$ (purple, gray, black)]
     {\includegraphics[width=7.1cm]{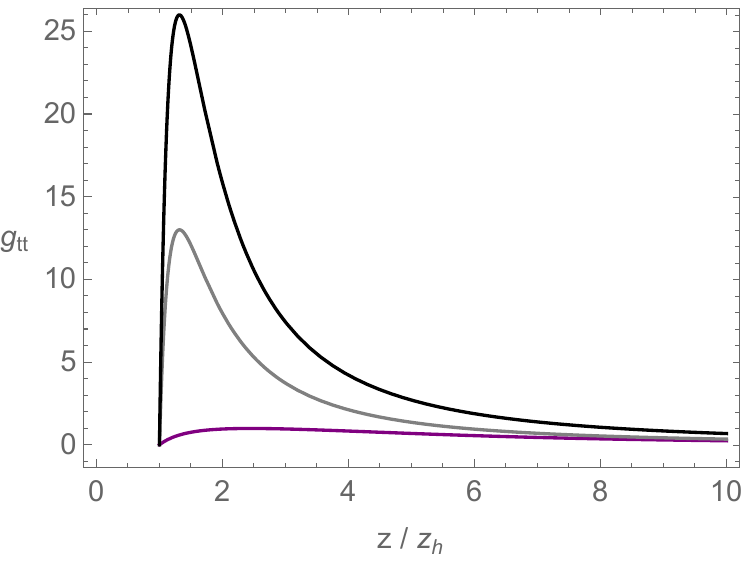} \label{GRfig3b}}
\caption{Plot of $g_{tt}$ vs. $z/z_h$ for $T/\mu=1$ and $\beta/\mu = 0, 3, 5, 10, 150, 300$ (red, green, blue, purple, gray, black). The inset shows the near horizon behavior.}\label{GRfig3}
\end{figure}

Finally, in Fig. \ref{GRfig3} we present the plot of $g_{tt}(z)$ inside the event horizon. 
We observe the collapse of the Einstein-Rosen bridge even at finite $\beta$, along with the non-trivial effect of $\beta$ on $g_{tt}$ -- its maximum exhibits non-monotonic behavior with increasing $\beta$, decreasing initially (see from red to blue in Fig. \ref{GRfig3a}) then increasing again (see from purple to black in Fig. \ref{GRfig3b}).\\

To summarize, we have found that the Gubser-Rocha model with explicit translational symmetry breaking features a spacelike Kasner singularity \eqref{GRKS} and no inner horizon. Notably, the Kasner exponents~\eqref{KCC} are independent of $\beta$ and thus agree with those found when translational symmetry is respected.
It is worth checking if these features also hold when the breaking of translations is spontaneous. We will do that in the remaining of this section.

%
\subsection{Spontaneous symmetry breaking}
Next, by numerically solving the equations of motion resulting from the action (\ref{GGR}, \ref{AXIONAC}) with $\eta=1$, we study the interior of black holes that break translations spontaneously on the boundary theory.

\begin{figure}[]
 \centering
     {\includegraphics[width=7.1cm]{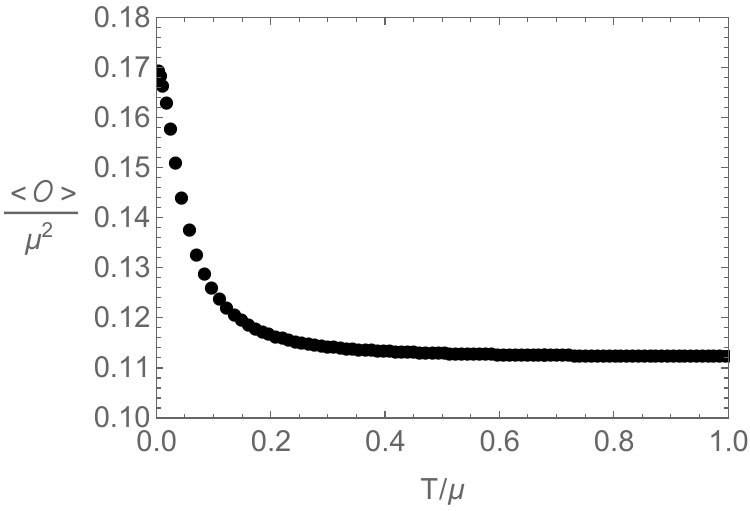}}
\caption{Plot of the scalar condensate vs. temperature when $\beta/\mu=0.01$.}\label{SSBFIG1}
\end{figure}

\begin{figure}[]
 \centering
     \subfigure[$g(z)$ vs. $z/z_h$]
     {\includegraphics[width=7.5cm]{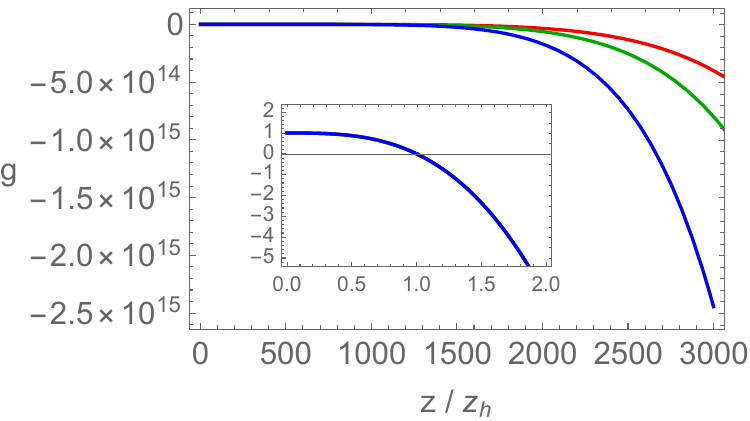} \label{}}
     \subfigure[$A_t(z)$ vs. $z/z_h$]
     {\includegraphics[width=6.7cm]{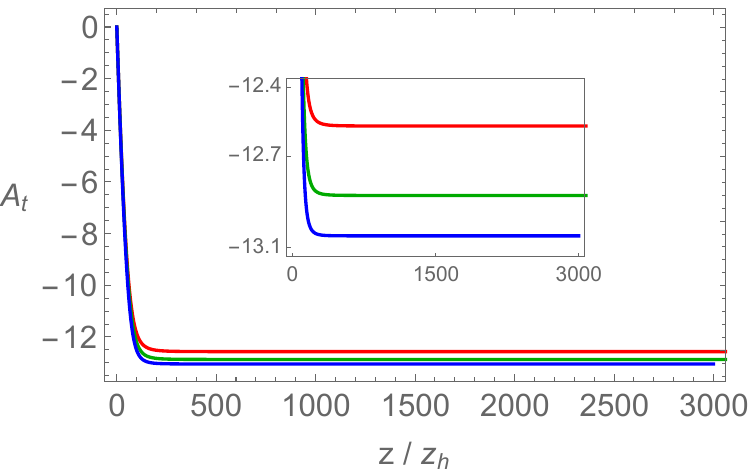} \label{}}
\caption{Holographic flows of the blackening factor $g(z)$ and gauge field $A_t(z)$ at $\beta/\mu = 0.01, 3, 5$ (red, green, blue). The inset in the left panel shows the near horizon behavior, where all data appear nearly identical in that regime: the deviation among the data is, at most, of the order of $10^{-4}$ outside the event horizon.}\label{SSBFIG2}
\end{figure}

\begin{figure}[]
 \centering
\subfigure[$z \, \dd (\log g_{tt}')/\dd z$ vs. $z/z_h$]
     {\includegraphics[width=4.83cm]{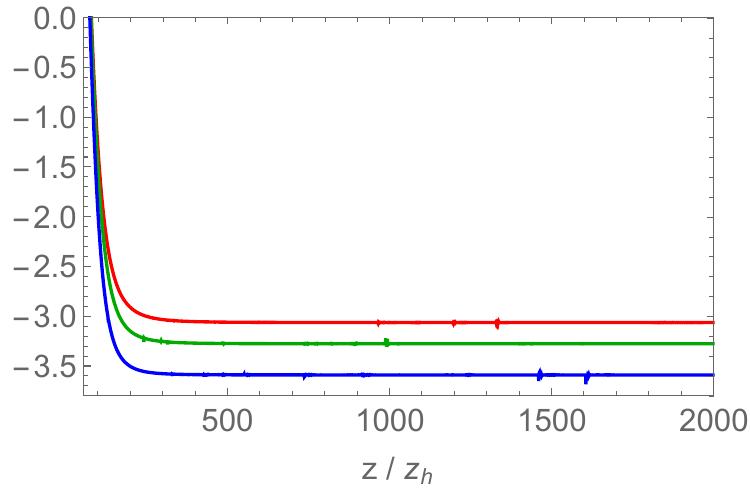} \label{}}
\subfigure[$z \, \dd \chi/\dd z$ vs. $z/z_h$]
     {\includegraphics[width=4.53cm]{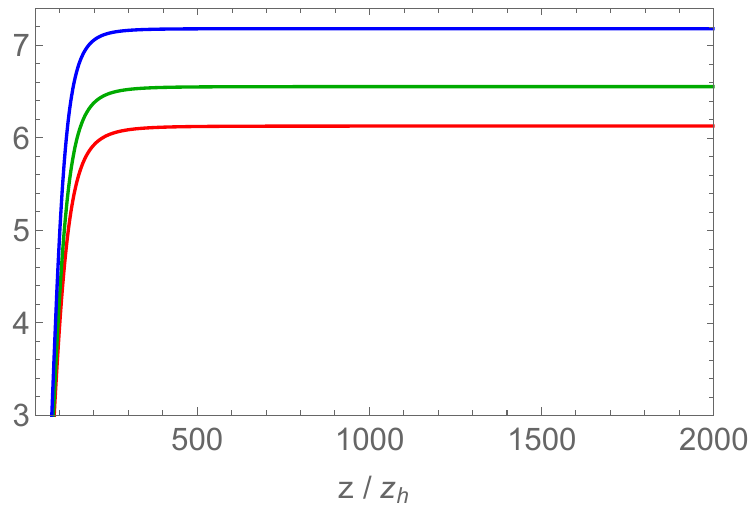} \label{}}
\subfigure[$z \, \dd \phi/\dd z$ vs. $z/z_h$]
     {\includegraphics[width=4.83cm]{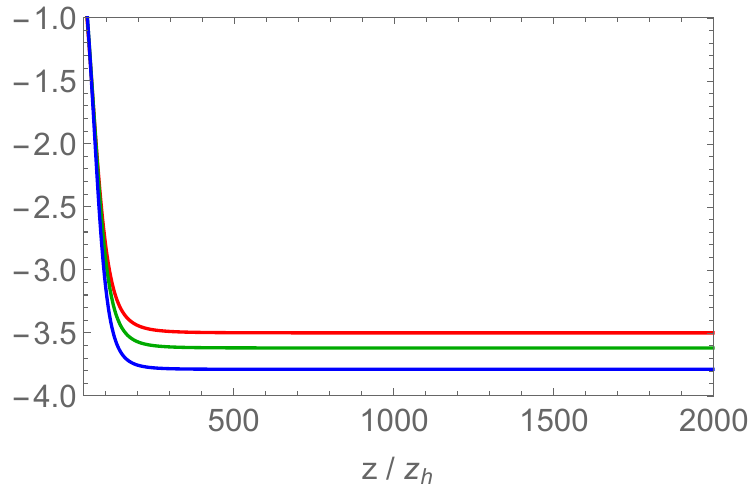} \label{}}
\caption{The functions $z \, \dd X/\dd z$ when $X = \left\{ \log g_{tt}',\, \chi,\, \phi \right\}$ at $\beta/\mu = 0.01, 3, 5$ (red, green, blue).}\label{SSBFIG3}
\end{figure}

\begin{figure}[]
 \centering
\subfigure[$p_t$ vs. $\beta/\mu$]
     {\includegraphics[width=4.73cm]{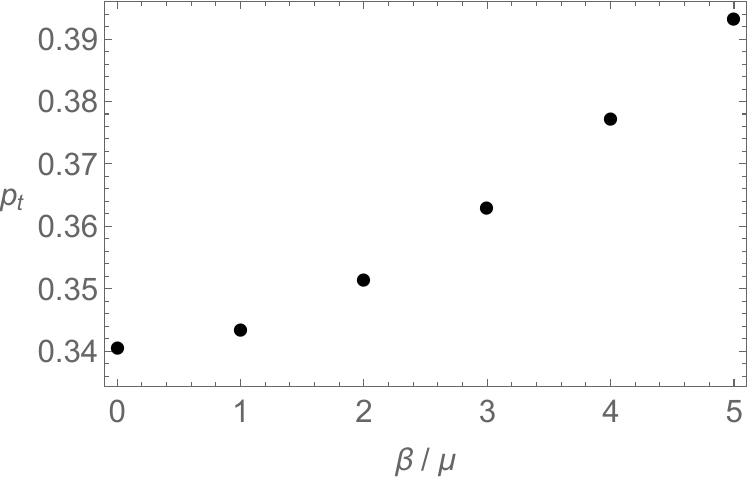} \label{}}
\subfigure[$p_x$ vs. $\beta/\mu$]
     {\includegraphics[width=4.73cm]{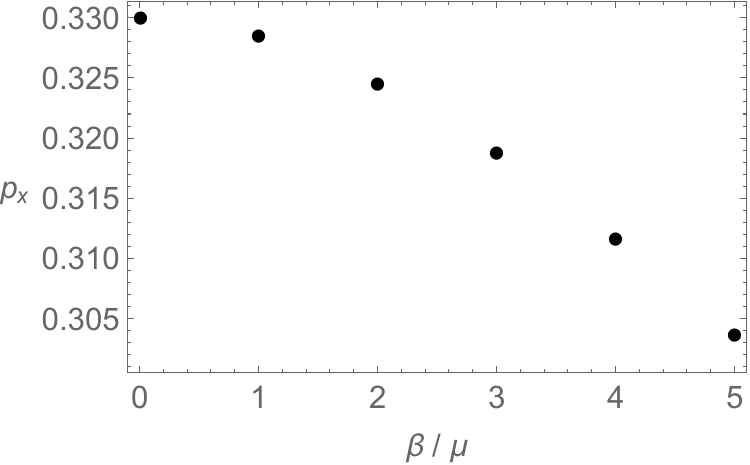} \label{}}
\subfigure[$p_\phi$ vs. $\beta/\mu$]
     {\includegraphics[width=4.93cm]{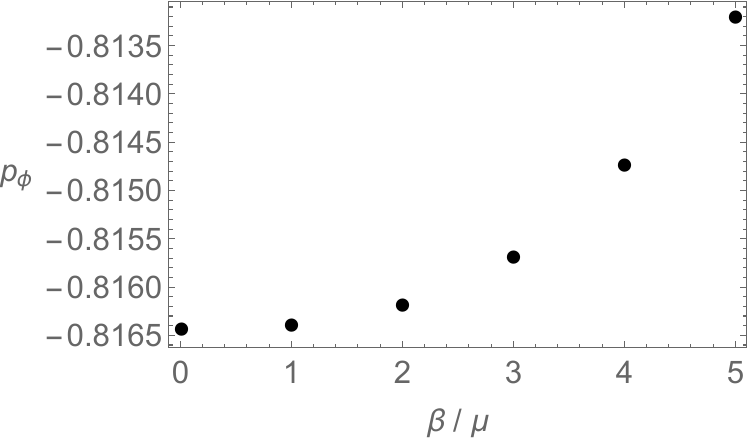} \label{}}
\caption{Kasner exponents ($p_t,\,p_x,\,p_\phi$) vs. $\beta/\mu$.}\label{SSBFIG5}
\end{figure}

\begin{figure}[]
 \centering
     \subfigure[$p_t + 2p_x$]
     {\includegraphics[width=6.5cm]{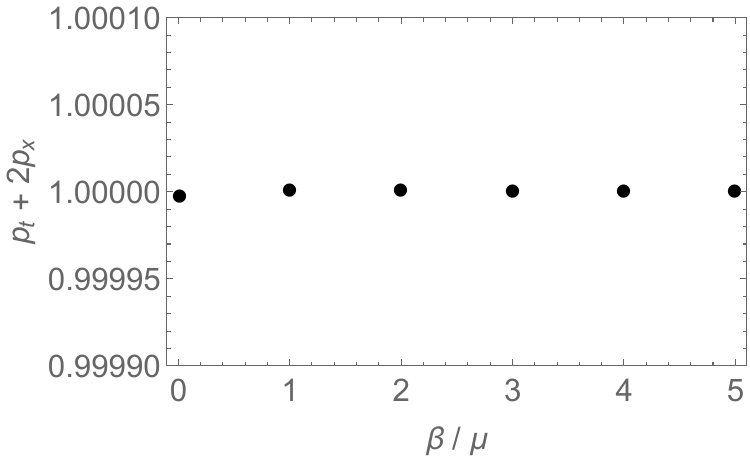} \label{}}
\qquad
     \subfigure[$p_\phi^2 + p_t^2 + 2p_x^2$]
     {\includegraphics[width=6.5cm]{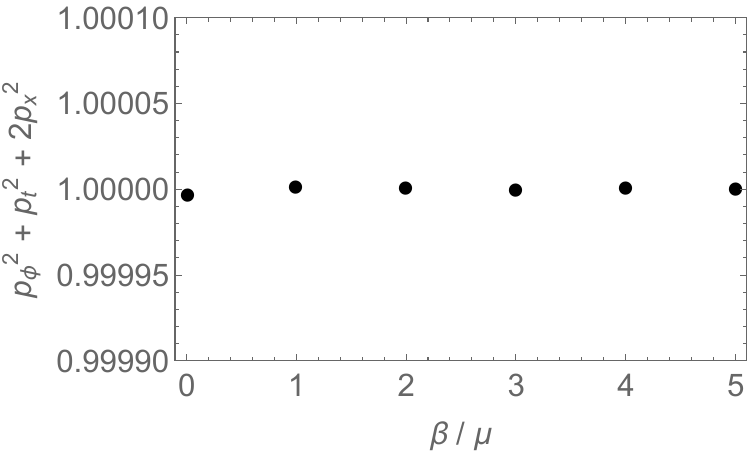} \label{}}
\caption{Kasner condition ($p_t + 2p_x=p_\phi^2 + p_t^2 + 2p_x^2=1$) when the translations are broken spontaneously.}\label{SSBFIG6}
\end{figure}

First, in Fig. \ref{SSBFIG1} we plot the scalar condensate $\left\langle\mathcal{O}\right\rangle$ as a function of temperature for solutions 
fulfilling \eqref{BCSSB}.
Notice that these solutions do not feature a critical temperature as the order parameter is nonzero for any temperature.\footnote{These results are consistent
with previous literature; see for instance~Fig. 2 in \cite{Amoretti:2017frz}. Moreover, as discussed in \cite{Amoretti:2017frz} these geometries correspond to unstable phases where translations are broken spontaneously at non-zero strain.}
Then, in Figs.~\ref{SSBFIG2}-\ref{SSBFIG6} we present plots characterizing the interior of these black hole solutions that break translations spontaneously.

By comparing these geometries breaking translations spontaneously to those  corresponding to explicit breaking analyzed in the previous subsection, we can
summarize the effect of the type of translation symmetry breaking on the black hole interior as follows
\begin{itemize}
\item{There is no inner horizon for both type of symmetry breaking: cfr. Fig. \ref{GRfig1}(a) vs. Fig. \ref{SSBFIG2}(a). In both cases, the geometry ends in a Kasner spacelike singularity.}

\item{Unlike the explicit breaking case, when translations are broken spontaneously the Kasner exponents depend on $\beta$ (i.e., the strength of broken translations): cfr. Fig. \ref{GRfig2} vs. Fig. \ref{SSBFIG3}.}
\end{itemize}
Our latter finding can also be understood as follows.
By considering the near-singularity behavior with generic coefficients ($\alpha_1,\,\alpha_2,\,\alpha_3$) given by
\begin{align}\label{GMANS}
\begin{split}
g(z) = g_c \,\, z^{\alpha_1} \,, \qquad
\chi(z) = \alpha_2 \log z  \,, \qquad
\phi(z) =  -\alpha_3 \log z  \,, 
\end{split}
\end{align}
along with the coordinate transformation
\begin{align}\label{}
\begin{split}
 z = \tau^{-\frac{2}{\alpha_1}} \,,
\end{split}
\end{align}
we can write the near-singularity metric as
\begin{align}\label{}
\begin{split}
\dd s^2 =   \frac{1}{g_c} \frac{4}{\alpha_1^2} \dd \tau^2  - g_c \tau^{2 p_t} \dd t^2  + \tau^{2 p_x} \dd \vec x_{i}^{2}  \,, \qquad \phi(\tau) = -\sqrt{2} p_{\phi} \log \tau \,,
\end{split}
\end{align}
where
\begin{align}\label{GMANS2}
\begin{split}
p_t = \frac{2-\alpha_1+\alpha_2}{\alpha_1}\,,\qquad p_x = \frac{2}{\alpha_1} \,, \qquad 
p_\phi = - \frac{\sqrt{2}\alpha_3}{\alpha_1} \,.
\end{split}
\end{align}
Then, using \eqref{GMANS} together with \eqref{GMANS2}, we find the following 
relation between $z \, \dd X/\dd z$, where $X = \left\{ \log g_{tt}',\, \chi,\, \phi \right\}$, and the Kasner exponents:
\begin{align}\label{RELKLAT}
\begin{split}
z \frac{\dd}{\dd z} \log (g_{tt}') &= -3 + \alpha_1 - \alpha_2 = -1 - \frac{2p_t}{p_x} \,, \\
z \frac{\dd}{\dd z} \chi &= \alpha_2 = \frac{2(1+p_t-p_x)}{p_x} \,, \\
z \frac{\dd}{\dd z} \phi &= -\alpha_3 = \frac{\sqrt{2} p_\phi}{p_x} \,.
\end{split}
\end{align}
Applying this relation to our numerical results in Fig. \ref{SSBFIG3}, we find that the Kasner exponents depend on $\beta$ when translations are spontaneously broken: see Fig. \ref{SSBFIG5}. 
Nevertheless, as we verify in Fig.~\ref{SSBFIG6}, the Kasner exponents fulfill the Kasner conditions~\eqref{eq:kasnercondd4}.

Notice that for the models we have analyzed in this section, the details of the singularity are sensitive to the type of symmetry breaking, while the near horizon IR geometry is not. Indeed, the Gubser-Rocha based models \eqref{TWOTM} feature the same IR geometry both in the case of explicit and spontaneous breaking of translations; the near horizon geometry at extremality is conformal to
AdS$_2 \times \R^{d-1}$~\cite{Amoretti:2017frz}. On the other hand, as we have seen above, while the singularity type is Kasner in both cases, the Kasner exponents are sensitive to the strength of translation symmetry breaking $\beta$ only in the case of spontaneous breaking.

\begin{figure}[]
 \centering
      {\includegraphics[width=7.1cm]{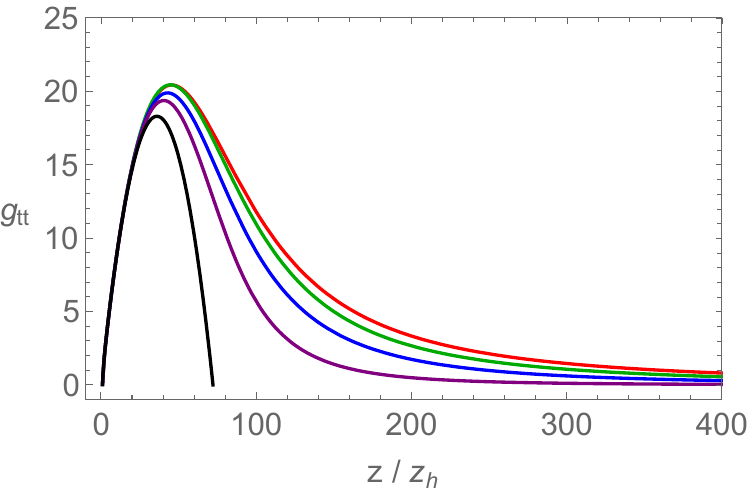} \label{}}
\caption{Plot of $g_{tt}$ vs. $z/z_h$ for $\beta/\mu = 0.01, 3, 5, 10, 300$ (red, green, blue, purple, black).}\label{SSBFIG4}
\end{figure}

We close this section with some additional observations 
\begin{itemize}
\item{As $\beta$ increases (from red to blue), the saturated value of $A_t$ is suppressed for explicit breaking while enhanced for spontaneous breaking: cfr. Fig. \ref{GRfig1}(b) vs. Fig. \ref{SSBFIG2}(b).}
\item{Unlike the explicit breaking case, when translations are broken spontaneously the maximum of $g_{tt}$ shows monotonic behavior with increasing $\beta$: cfr. Fig. \ref{GRfig3} vs. Fig. \ref{SSBFIG4}.}
\end{itemize}

One final remark is in order. We have found that the numerical stability of our code for computing the geometry inside the event horizon deteriorates as the value of $\beta/\mu$ increases. For instance, in the black data presented in Fig. \ref{SSBFIG4}, we determine the inner horizon geometry up to $z/z_h\approx70$. Nevertheless, within the same parameter range utilized for the explicit breaking case (with $\beta/\mu$ up to 300 in Fig. \ref{GRfig3}), we can conclude that the maximum value of $g_{tt}$ in the spontaneous case demonstrates a monotonic behavior.

%
\section{Conclusions}\label{SEC4}

In this work, we have investigated the interior geometry of a family of AdS hairy black holes, which are essential for applications of gauge/gravity dualities to QCD and condensed matter systems. Our analytical approach has allowed us to clearly characterize the singularities, offering valuable insights into observables in the dual field theory that can probe and diagnose these singularities.

The geometries we have studied comprise a one-parameter ($\delta$) family of asymptotically AdS$_{d+1}$ hairy black holes, which emerge as solutions of a particular Einstein-Maxwell-Dilaton model. For a specific value of this parameter, $\delta_c$, given in~\eqref{CRITICAL}, the geometry corresponds to the well-known Gubser-Rocha background~\cite{Gubser:2009qt}, which has been a workhorse in dual descriptions of strongly coupled phases of matter.

After thoroughly reviewing this family of black hole solutions, we conducted an in-depth analysis of the singularities behind their horizons. Our findings indicate that, depending on the value of $\delta$, these geometries exhibit either timelike or Kasner singularities. Notably, we identified geometries with a Cauchy horizon in the presence of a scalar deformation. This observation prompted us to generalize previous theorems that prohibit the existence of inner horizons in geometries with a nontrivial scalar~\cite{Hartnoll:2020rwq,Cai:2020wrp,An:2021plu}.\footnote{See also \cite{Chew:2023upu} for the case of spherically symmetric charged black holes using the double-null formalism.} We complemented our singularity analysis with a detailed characterization of the entire solution flow from the AdS boundary to the singularity at the end of space.

After characterizing the interior of the asymptotic AdS black holes, we analyzed observables in the dual theory that can access the geometry behind the horizon. Specifically, we focused on the Complexity = Anything proposal \cite{Jorstad:2023kmq} and the thermal $a$-function \cite{Caceres:2022smh}.

First, we extended prior analyses of holographic complexity to the general metric ansatz describing our black hole solutions, thereby establishing explicit connections between complexity and the structure of singularities. Notably, we demonstrated that for geometries with a Kasner singularity and no inner horizon, a new variant of the proposals introduced by Jørstad, Myers, and Ruan results in hypersurfaces that (i) extend all the way to the singularity and (ii) accurately characterize the linear growth of complexity through the specific Kasner exponents of the singularity. This new prescription involves evaluating the observable (\ref{CANYE}) with the function $\mathcal{F}$ given in (\ref{Fchoices2}) on a surface of large constant mean curvature, found by extremizing the functional (\ref{CCMC}).

The second observable we investigated to diagnose the black hole singularity is the thermal $a$-function. We found that it approaches zero near the singularity, exhibiting a specific power-law behavior that we computed analytically. This power is fully determined by the Kasner exponents~\cite{Caceres:2022smh}. The novel result of our analysis is that, through our exact RG flow solutions, we were able to express this behavior completely analytically in terms of the parameters of the model (\ref{atherresul}), using the relations we found earlier in (\ref{KAS1})-\eqref{KAS2}.

In the final section of this work, we examined the interior geometry of asymptotically AdS black holes that break translational symmetry in the boundary theory. Specifically, we explored the effects of adding axion fields to the same Einstein-Maxwell-dilaton model, resulting in homogeneous geometries that, depending on the boundary conditions, either explicitly or spontaneously break translations in the dual theory~\cite{Gouteraux:2014hca,Jeong:2018tua}.

We found that the nature of boundary symmetry breaking ---whether explicit or spontaneous--- can significantly influence the physics and characteristics of the singularity, potentially more so than the near-horizon IR geometry of the background. Notably, for the solutions we examined, the near-horizon IR geometry is conformally AdS$_2 \times \mathbb{R}^{d-1}$ in both symmetry-breaking scenarios. As for the interior geometries, both types of symmetry breaking result in Kasner singularities and the absence of an inner horizon. In the case of explicit symmetry breaking, the Kasner exponents remain unaffected by the strength of translation breaking. Conversely, in scenarios of spontaneous symmetry breaking, these exponents vary with the strength of the symmetry breaking.\footnote{It is worth noting that for the case of spontaneous symmetry case, we do not find the characteristic Josephson oscillations found in holographic superconducting models. This is due to the fact that the scalar fields present in our models are neutral \cite{Mansoori:2021wxf}.}

Several promising directions for future research are worth highlighting. While our analysis has focused on black hole solutions within the Einstein-Maxwell-Dilaton model \eqref{GGR}, including scenarios with broken translations, expanding this investigation to a broader range of AdS black hole geometries could yield significant insights. Notably, exploring inhomogeneous geometries that realize holographic lattices \cite{Kachru:2009xf,Nakamura:2009tf,Donos:2011bh,Horowitz:2012ky,Horowitz:2012gs,Donos:2013eha,Donos:2014yya}, with disorder \cite{Lucas:2014zea,Hartnoll:2014cua,Arean:2014oaa,Araujo:2016jlf} or broken isotropy \cite{Mateos:2011ix,Giataganas:2017koz,Jeong:2017rxg,Arefeva:2018hyo,Gursoy:2018ydr,Hoyos:2020zeg,Gursoy:2020kjd}, which are often used in applications to QCD and condensed matter systems, would be particularly enlightening. Additionally, investigating stringy $1/\alpha'$ corrections or semiclassical $1/N$ corrections near black hole singularities represents another intriguing direction. These could potentially be addressed through higher curvature gravity models \cite{Bueno:2024fzg,Caceres:2024edr} or holographic braneworld frameworks \cite{Emparan:2020znc,Emparan:2022ijy,Panella:2023lsi,Feng:2024uia,Climent:2024nuj,Panella:2024sor} respectively. We hope to come back to some of these topics in the near future.

%
\acknowledgments
It is a pleasure to thank {Yongjun Ahn, Matteo Baggioli, Elena Cáceres, Javier Carballo, Blaise Goutéraux, Hong-Yue Jiang, Yu-Xiao Liu, Ángel Murcia, Gerben Oling, Ayan Patra and Shan-Ming Ruan} for valuable discussions and correspondence.  
DA, HSJ and JFP are supported by the Spanish MINECO ‘Centro de Excelencia Severo Ochoa' program under grant SEV-2012-0249, the Comunidad de Madrid ‘Atracci\'on de Talento’ program (ATCAM) grant 2020-T1/TIC-20495, the Spanish Research Agency via grants CEX2020-001007-S and PID2021-123017NB-I00, funded by MCIN/AEI/10.13039/501100011033, and ERDF A way of making Europe. 
LCQ acknowledges the financial support provided by the scholarship granted by the Chinese Scholarship Council (CSC).
All authors contributed equally to this paper and should be considered as co-first authors.

%
\appendix
\section{Singularities in the $r$-coordinate}\label{appa1}

In this section, we study the singularities of our model \eqref{GGR}-\eqref{MODELACTION} in the $r$-coordinate. For this purpose we initially expand our metric \eqref{OURMETRIC}-\eqref{sol11int} in the
near-singularity limit $r\rightarrow0$, arriving at
\begin{equation}\label{}
\dd s^2 \approx -D(r)\dd t^2+B(r)\dd r^2+C(r)\dd \vec x_{i}^{2} \,,
\end{equation}
where
\begin{align}\label{RCMT1}
\begin{split}
D(r) &=  Q^\frac{4}{(d-2)\left( 2 + (d-2)(d-1)\d^2 \right)} r^{2-d+\frac{4(d-2)}{2+(d-2)(d-1)\d^2}} \left( r^{d+\frac{4-4d}{2+(d-2)(d-1)\d^2}} - r_h^{d+\frac{4-4d}{2+(d-2)(d-1)\d^2}} \right) \,, \\  
B(r) &= Q^{-\frac{4}{2+(d-2)(d-1)\d^2}} \frac{ r^{d-2-\frac{4}{2+(d-2)(d-1)\d^2}} }{  r^{d+\frac{4-4d}{2+(d-2)(d-1)\d^2}} - r_h^{d+\frac{4-4d}{2+(d-2)(d-1)\d^2}} }   \,, \\ 
C(r) &=  Q^\frac{4}{(d-2)\left( 2 + (d-2)(d-1)\d^2 \right)} r^{2-\frac{4}{2+(d-2)(d-1)\d^2}}  \,.
\end{split}
\end{align}
Then, it is evident that depending on the power of $r$ in the functions $D(r)$ and $B(r)$ above, we can categorize the near-singularity geometry into three classes, similar to what we did in the $z$-coordinate case:
\begin{align}\label{}
\begin{split}
\begin{cases}
\,\, d+\frac{4-4d}{2+(d-2)(d-1)\d^2} > 0  \qquad\longrightarrow\qquad \d > \d_c \,, \qquad\quad\, (\text{Class I})  \\
\,\, d+\frac{4-4d}{2+(d-2)(d-1)\d^2} = 0  \qquad\longrightarrow\qquad \d = \d_c \,,  \qquad\quad\, (\text{Class II}) \\
\,\, d+\frac{4-4d}{2+(d-2)(d-1)\d^2} < 0 \qquad\longrightarrow\qquad 0< \d < \d_c \,. \quad\,\, (\text{Class III}) 
\end{cases}
\end{split}
\end{align}
Note that Class II cannot be examined within the metric \eqref{RCMT1}; instead, it needs to be analyzed separately. Specifically, when we impose $\d = \d_c$ into \eqref{OURMETRIC}-\eqref{sol11int}, the corresponding metric of Class II is obtained as
\begin{align}\label{RCMT2}
\begin{split}
D(r) &=  Q^{\frac{d}{1-d}} r^{\frac{d-2}{d-1}} \left( Q^{\frac{d}{d-2}} - r_h^{d} \left( 1+ Q r_h^{2-d} \right)^{\frac{d}{d-2}} \right)  \,, \\  
B(r) &= \frac{ Q^{\frac{d}{(d-2)(d-1)}} r^{d-3+\frac{1}{1-d}} }{ Q^{\frac{d}{d-2}} - r_h^d \left( 1 + Q r_h^{2-d} \right)^{\frac{d}{d-2}}  }   \,, \\ 
C(r) &= Q^{\frac{d}{(d-2)(d-1)}} r^{\frac{d-2}{d-1}}   \,.
\end{split}
\end{align}
In what follows, we identify the singularities for each class.\\

\paragraph{Class I ($\d > \d_c$): Kasner singularity.}
For Class I, applying the coordinate transformation
\begin{align}\label{}
\begin{split}
 r = \tau^{\frac{4+2(d-2)(d-1)\d^2}{(d-2)(2+d(d-1)\d^2)}} \,,
\end{split}
\end{align}
the metric \eqref{RCMT1} becomes
\begin{align}\label{}
\begin{split}
\dd s^2 \approx  -\dd \tau^2  + \tau^{2 p_t} \dd t^2  + \tau^{2 p_x} \dd \vec x_{i}^{2}  \,, \qquad \phi(\tau) = -\sqrt{2} p_{\phi} \log \tau \,,
\end{split}
\end{align}
where we omit the coefficients for simplicity in this section. We obtain identical Kasner exponents as described in \eqref{KAS1}:
\begin{align}\label{}
\begin{split}
p_t = \frac{2-(d-2)(d-1)\d^2}{2+d(d-1)\d^2} \,, \qquad 
p_x = \frac{2(d-1)\d^2}{2+d(d-1)\d^2} \,, \qquad 
p_\phi = -\frac{2\sqrt{2}(d-1)\d}{2+d(d-1)\d^2} \,,
\end{split}
\end{align}
that satisfy the Kasner condition
\begin{align}\label{eq:kasnercondapp}
\begin{split}
p_t + (d-1)p_x = p_\phi^2 + p_t^2 + (d-1)p_x^2 = 1 \,.
\end{split}
\end{align}

\paragraph{Class II ($\d = \d_c$): Kasner singularity.}
For Class II, with the transformation
\begin{align}\label{}
\begin{split}
 r = \tau^{\frac{2(d-1)}{d(d-2)}} \,,
\end{split}
\end{align}
the metric \eqref{RCMT2} becomes
\begin{align}\label{}
\begin{split}
\dd s^2 \approx  -\dd \tau^2  + \tau^{2 p_t} \dd t^2  + \tau^{2 p_x} \dd \vec x_{i}^{2}  \,, \qquad \phi(\tau) = -\sqrt{2} p_{\phi} \log \tau \,,
\end{split}
\end{align}
with the Kasner exponents \eqref{KAS2} found for Class II in the main text, namely
\begin{align}\label{}
\begin{split}
p_t = p_x = \frac{1}{d} \,, \qquad 
p_\phi = -\sqrt{\frac{d-1}{d}} \,,
\end{split}
\end{align}
which of course satisfy the Kasner condition~\eqref{eq:kasnercondapp}.

\paragraph{Class III ($0< \d < \d_c$): timelike singularity.}
Finally, for Class III geometries, through the coordinate transformation
\begin{align}\label{}
\begin{split}
 r = \tau^{\frac{2+(d-2)(d-1)\d^2}{2(d-2)}} \,,
\end{split}
\end{align}
our metric \eqref{RCMT1} becomes
\begin{align}\label{}
\begin{split}
\dd s^2 \approx  \dd \tau^2  - \tau^{2 \tilde{p}_t} \dd t^2  + \tau^{2 \tilde{p}_x} \dd \vec x_{i}^{2}  \,, \qquad \phi(\tau) = -\sqrt{2} \tilde{p}_{\phi} \log \tau \,,
\end{split}
\end{align}
which has a timelike singularity.
As expected we find the same exponents as in \eqref{TLSCOMP}:
\begin{align}\label{}
\begin{split}
\tilde{p}_t = \tilde{p}_x = \frac{(d-1)\d^2}{2} \,, \qquad 
\tilde{p}_\phi = -\frac{(d-1)\d}{\sqrt{2}}\,,
\end{split}
\end{align}
and one can check that they do not satisfy the Kasner condition. Instead one has
\begin{align}\label{}
\begin{split}
\tilde{p}_t + (d-1)\tilde{p}_x = \frac{d(d-1)\d^2}{2}  \,, \qquad \tilde{p}_\phi^2 + \tilde{p}_t^2 + (d-1)\tilde{p}_x^2 =  \frac{(d-1)^2 (2+d \d^2) \d^2}{4} \,,
\end{split}
\end{align}
which is of course~\eqref{eq:class3kas}.
Similarly, the cases $\d=0$ (and $Q\neq0$) and $Q=0$, corresponding to the timelike and Schwarzschild singularities, can be straightforwardly obtained in the $r$-coordinate.

In essence, this section demonstrates that singularities within our model, analyzed within the $z$-coordinate \eqref{FINRESUM}, can be consistently identified using the $r$-coordinate.

%
\section{Two branches of solutions for gapped geometries ($\d>\d_c$)}\label{appB}

In this section, we study further the gapped geometries 
that constitute the solutions of our model for $\d>\d_c$.
In particular, we aim to determine which of the two existent branches of solutions is the thermodynamically stable one.

\paragraph{Free energy density in the generic dimension.}
For this purpose, it is instructive to find the free energy density ($\mathcal{W}$) within our model (\ref{GGR})-(\ref{MODELACTION}), which can be determined through the thermodynamic pressure ($\mathcal{P}$) according to the expression:
\begin{align}\label{FE1}
\begin{split}
\mathcal{W} = -\mathcal{P} = -\frac{s T + \mu \rho}{d} \,.
\end{split}
\end{align}
Here, in the second equality above, we used both the Gibbs-Duhem relation $\epsilon + \mathcal{P} = s T + \mu \rho$ and the equation of state $\epsilon = (d-1) \mathcal{P}$ involving the thermodynamic energy $\epsilon$. Note that this equation of state follows from the fact that the stress tensor is traceless when conformal symmetry is unbroken.\footnote{However, it is also worth noting that external sources, such as magnetic fields, can modify the equation of state of our models, as discussed in \cite{Caldarelli:2016nni}.}

Then, by incorporating the expressions for the temperature ($T$) and chemical potential ($\mu$) from \eqref{TmuFOR}, the entropy ($s$) from \eqref{enforre}, and the charge density ($\rho$) of our model \cite{Gouteraux:2014hca,Jeong:2018tua}
\begin{align}
\begin{split}
\rho = \frac{A_t' Z}{(BD)^{\frac{1}{2}}C^{\frac{-(d-1)}{2}}}\Bigr|_{r\rightarrow r_{h}} = \frac{2\sqrt{(d-2)(d-1)\tQ} \left(1+\tQ\right)^{\frac{2(d-1)}{(d-2) \left( 2+(d-2)(d-1)\d^2 \right)}} r_h^{d-1}}{\sqrt{2+(d-2)(d-1)\d^2}} \,,
\end{split}
\end{align}
we can compute the free energy density \eqref{FE1} as follows:
\begin{align}\label{FE2}
\begin{split}
\mathcal{W} = - \left(1+\tQ \right)^{\frac{4(d-1)}{(d-2)\left(2+(d-2)(d-1)\d^2\right)}} r_h^{d} \,.
\end{split}
\end{align}

\paragraph{Free energy density and holographic renormalization.}
In this section, for simplicity, we focus on the $d=3$ scenario, where \eqref{FE2} simplifies to 
\begin{align}\label{FE3}
\begin{split}
\mathcal{W} = - \left(1+\tQ \right)^{\frac{4}{1+\d^2}} r_h^{3} \,.
\end{split}
\end{align}
It is worth noting that for the case of $d=3$, \eqref{FE3} can also be derived through the standard holographic renormalization procedure, as outlined in \cite{Caldarelli:2016nni}, which we briefly summarize below. 

Firstly, we begin by determining the regularized action of our model, denoted as $S_{\text{reg}}$, which is expressed as: 
\begin{align}\label{RNOSA1}
\begin{split}
S_{\text{reg}} = \lim_{\epsilon\rightarrow0} \left(S_{\text{bulk}} + S_{\text{GH}} + S_{\text{ct}}\right) \,,
\end{split}
\end{align}
where $S_{\text{bulk}}$ corresponds to $S$ in \eqref{GGR}, and the standard Gibbons-Hawking term $S_{\text{GH}}$ and counterterms $S_{\text{ct}}$ take the form 
\begin{align}\label{}
\begin{split}
S_{\text{GH}} = \int_{z=\epsilon} \dd^3 x \sqrt{-\gamma} \,\,2 K \,, \qquad
S_{\text{ct}} = -\int_{z=\epsilon} \dd^3 x \sqrt{-\gamma} \,\,\left( 4 + R[\gamma] + \frac{\phi^2}{2} \right) \,.
\end{split}
\end{align}
Here, $\epsilon$ denotes the UV cutoff, $\gamma_{ij}$ represents the induced metric on the radial cutoff, and $K$ stands for the trace of the extrinsic curvature.
Moreover, it is noted in \cite{Caldarelli:2016nni} that an additional boundary term ($S_\mathcal{F}$) must be added to the renormalized on-shell action \eqref{RNOSA1} to yield \eqref{FE3} (i.e., to attain the traceless stress tensor). This additional term is defined as
\begin{align}\label{}
\begin{split}
S_{\mathcal{F}} = \int_{z=\epsilon} \dd^3 x \sqrt{-\gamma} \,\,\left( J_{\mathcal{F}} \, \phi_{(0)} + \mathcal{F}(\phi_{(0)}) \right) \,, \qquad J_{\mathcal{F}} = -\phi_{(1)} - \mathcal{F}'(\phi_{(0)}) \,,
\end{split}
\end{align}
where $\phi_{(0)}$ is the leading coefficient of the scalar field near the AdS boundary, and $\phi_{(1)}$ denotes the sub-leading coefficient. The function $\mathcal{F}$ corresponds to a multi-trace deformation of the Neumann theory~\cite{Papadimitriou:2007sj,Witten:2001ua,Berkooz:2002ug,Mueck:2002gm}, determined via the analysis of Ward identities.\footnote{This extra boundary term, as highlighted in~\cite{Caldarelli:2016nni}, serves to impose more general boundary conditions on $\phi$. Essentially, it keeps a function $J_{\mathcal{F}}$ constant at the boundary.} 
For our model, it takes the form
\begin{align}\label{}
\begin{split}
\mathcal{F}(\phi_{(0)}) = \frac{1-\d^2}{12\d} \phi_{(0)}^3 \,.
\end{split}
\end{align}
Consequently, the modified renormalized on-shell action is expressed as:
\begin{align}\label{RNOSA2}
\begin{split}
S'_{\text{reg}} = \lim_{\epsilon\rightarrow0} \left(S_{\text{bulk}} + S_{\text{GH}} + S_{\text{ct}} + S_{\mathcal{F}}\right) \,.
\end{split}
\end{align}
The grand canonical potential $\Omega$ can then be obtained as:
\begin{align}\label{}
\begin{split}
\Omega = - T S'_{\text{reg}} \,,
\end{split}
\end{align}
and subsequently, the corresponding free energy density is derived as:
\begin{align}\label{FE4}
\begin{split}
\mathcal{W} = \frac{\Omega}{\mathcal{V}} \,, 
\end{split}
\end{align}
where $\mathcal{V}$ is the (formally infinite) spatial volume. One can also verify that \eqref{FE4} yields 
\eqref{FE3}, and that \eqref{RNOSA2} produces the traceless stress tensor.

\paragraph{Thermodynamically stable solutions in gapped geometries.}
To examine the thermodynamically stable solutions using the free energy density \eqref{FE3}, we first render it into a dimensionless quantity as
\begin{align}\label{FE5}
\begin{split}
\frac{\mathcal{W}}{\mu^3} = - \frac{\left(1+\tQ \right)^{\frac{1+3\d^2}{1+\d^2}} \left(1+\d^2\right)^{3/2}}{8\tQ^{3/2}} \,.
\end{split}
\end{align}
In addition, using \eqref{TmuFOR}, we obtain
\begin{align}\label{APPBFOR1}
\begin{split}
\frac{T}{\mu} = \frac{3-\tQ+3(1+\tQ)\d^2}{8\pi \sqrt{\tQ(1+\d^2)}} \,,
\end{split}
\end{align}
From this expression we can solve for $\tQ$ in terms of $T/\mu$ as
\begin{align}\label{QsFOR}
\begin{split}
\tQ = \tQ_{\pm} \equiv \frac{9(1+\d^2)}{3(1-3\d^2) + 8\pi \frac{T}{\mu} \left( 4\pi \frac{T}{\mu} \pm \sqrt{16\pi^2 \left(\frac{T}{\mu}\right)^2 + 3\left(1-3\d^2\right)} \right) } \,,
\end{split}
\end{align}
where $\d>\d_c$, i.e., for the gapped geometries.
Note that at a given $T/\mu$, there can exist two branches of solutions ($\tQ_{\pm}$), which converge when
\begin{align}\label{CRITICALTfor}
\begin{split}
\tQ_{+}=\tQ_{-}: \qquad \frac{T}{\mu} = \frac{T_c}{\mu} \equiv \frac{\sqrt{3(3\d^2-1)}}{4\pi} \,.
\end{split}
\end{align}

\begin{figure}[]
 \centering
     {\includegraphics[width=6.5cm]{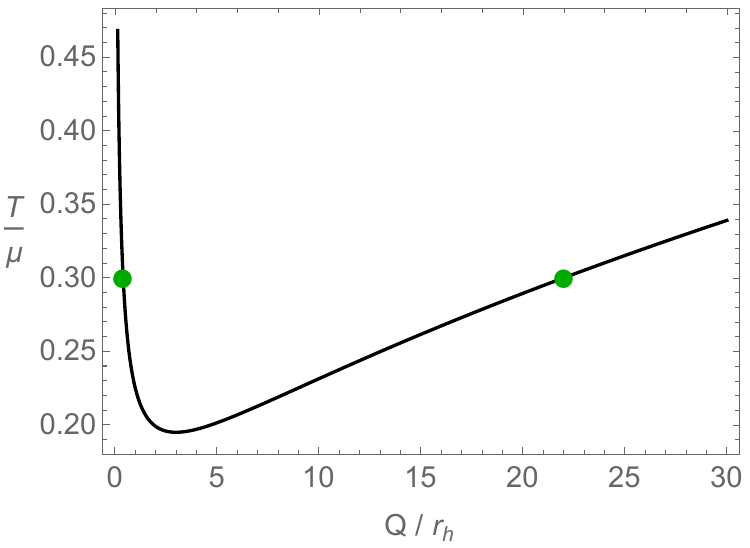} \label{}}
\qquad
     {\includegraphics[width=6.5cm]{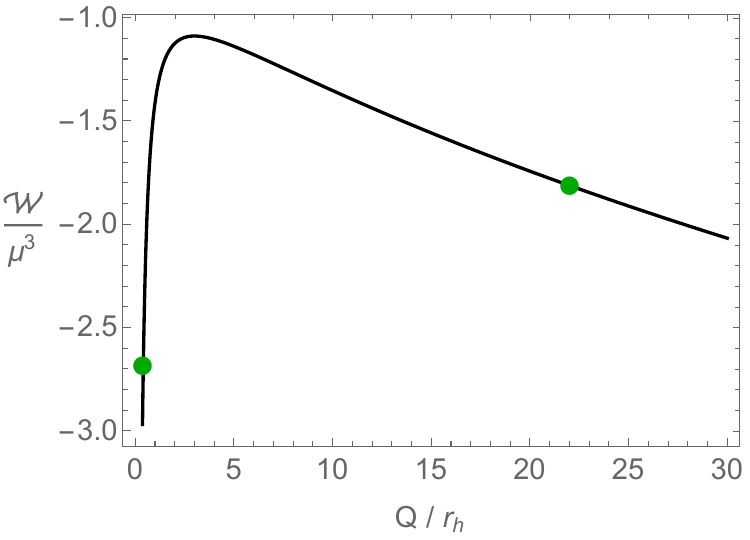} \label{}}
\caption{Two branches of solutions of gapped geometries when $\d=1$. The left panel shows the existence of two solutions 
at given $T/\mu$, e.g.,  $\left(\tQ_{+},\,\tQ_{-}\right) \approx (0.4,\, 22)$, denoted as green dots, when $T/\mu=0.3$, where $\tQ\equiv Q/r_h$ in \eqref{tildeQval}. The critical temperature \eqref{CRITICALTfor} is $T_c/\mu \approx 0.19$. 
We plot the free energy of both branches in the right panel. It is evident that the solution with $\tQ_{+}$ is the stable one when $T/\mu=0.3$.}\label{appBfig1}
\end{figure}

We illustrate the two branches of gapped geometries in Fig. \ref{appBfig1}.
Since multiple solutions arise when $\d>\d_c$, we can now determine the thermodynamically stable solution by analyzing the free energy density \eqref{FE5}. For instance, considering the parameters used in the left panel of Fig. \ref{appBfig1}, we find that $\tQ_{+}$ represents the stable solution, as depicted in the right panel of Fig. \ref{appBfig1}. 

\begin{figure}[]
 \centering
     {\includegraphics[width=6.5cm]{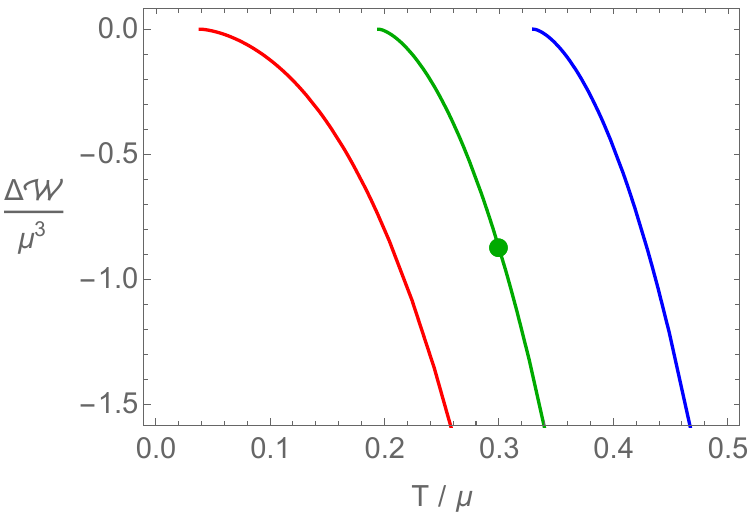} \label{}}
\caption{The difference in free energy density \eqref{DIFFED} for $\d=0.6, 1, 1.5$ (red, green, blue). The critical temperature \eqref{CRITICALTfor} corresponds to the point where ${\Delta\mathcal{W}}/{\mu^3}=0$. The green dot depicted in the plot corresponds to the same data point represented by the greed dots in Fig. \ref{appBfig1}.}\label{appBfig2}
\end{figure}

To further elaborate on the thermodynamically stable solution, it is beneficial to introduce the difference in the free energy density as  
\begin{align}\label{DIFFED}
\begin{split}
\frac{\Delta\mathcal{W}}{\mu^3} \equiv \frac{\mathcal{W}(\tQ_{+}) - \mathcal{W}(\tQ_{-})}{\mu^3}  \,.
\end{split}
\end{align}
This signifies that when ${\Delta\mathcal{W}}/{\mu^3}<0$, the solution characterized by $\tQ_{+}$ is the stable solution. By plotting ${\Delta\mathcal{W}}/{\mu^3}$ across various values of $\d$ and $T/\mu$, we find that for the gapped geometries, $\tQ_{+}$ indeed represents the stable solution: see Fig. \ref{appBfig2}.

In principle, one can plug \eqref{QsFOR} into \eqref{FE5} to derive the analytic expression for \eqref{DIFFED}. However, due to its complexity and lack of clarity, we refrain from presenting it here. Instead, we display its plot in Fig. \ref{appBfig2} and provide its asymptotic form in the vicinity of $T_c$, i.e., when ${T} = {T_c} + {\Delta T}$ as 
\begin{align}\label{DIFFED22}
\begin{split}
\frac{\Delta\mathcal{W}}{\mu^3}  \,=\, \underbrace{
-\frac{2^{\frac{9+13\d^2}{2(1+\d^2)}}}{3^{\frac{9}{4}}} \pi^{\frac{3}{2}} \left(3\d^2-1\right)^{\frac{3-5\d^2}{4(1+\d^2)}} \left(3\d^2+1\right)^{\frac{\d^2-1}{\d^2+1}} 
}_{ <\, 0 } \,\left(\frac{\Delta T}{\mu}\right)^{\frac{3}{2}} \,+\, \cdots \,,
\end{split}
\end{align}
where $\cdots$ denotes the sub-leading corrections of ${\Delta T}/{\mu}$.

The negative sign of ${\Delta\mathcal{W}}/{\mu^3}$ in \eqref{DIFFED22} indicates that $\tQ_{+}$ serves as the stable solution near $T_c$ for any $\d$ value within gapped geometries, consistent with Fig. \ref{appBfig2}. Beyond the temperature range depicted in Fig. \ref{appBfig2}, we have also numerically verified that ${\Delta\mathcal{W}}/{\mu^3}$ remains a monotonically decreasing function even in the high $T/\mu$ regime, such as up to $T/\mu = 10^5$. This implies that the solution characterized by $\tQ_{+}$ is the thermodynamically stable solution.

\bibliographystyle{JHEP}

\providecommand{\href}[2]{#2}\begingroup\raggedright\endgroup

\end{document}